\documentclass[runningheads]{llncs}
\usepackage[linesnumbered,boxed,noend]{algorithm2e}
\usepackage{wrapfig}
\usepackage[version=latest]{pgf}
\usepackage{tikz}
\usetikzlibrary{arrows}
\usepackage{tkz-graph}
\usepackage{mathpartir}
\usepackage{stmaryrd}
\expandafter\def\csname opt@stmaryrd.sty\endcsname
{only,shortleftarrow,shortrightarrow}
\usepackage{listings}
\usepackage{proof}

\usepackage{multirow}

\usepackage{lineno}

\usepackage{xcolor}
\usepackage{colortbl}

\DeclareMathAlphabet{\mathpzc}{OT1}{pzc}{m}{it}
\usepackage{amssymb}
\usepackage{mathtools}

\usepackage{hhline}
\usepackage{color, colortbl}
\definecolor{LightCyan}{rgb}{0.88,1,1}
\usepackage{pifont}
\usepackage{extpfeil}
\newcommand{\cmark}{\ding{51}}%
\newcommand{\xmark}{\ding{55}}%
\newcolumntype{C}[1]{>{\centering\let\newline\\\arraybackslash\hspace{0pt}}m{#1}}

\usepackage{textcomp}

\newcommand{\bools}{\ensuremath{\mathbb{B}}}
\newcommand{\nats}{\ensuremath{\mathbb{N}}}

\newcommand{\gaps}{\ensuremath{\mathbb{G}}}

\newcommand{\supscript}[2]{{#1}^{\set{#2}}}

\newcommand{\set}[1]{\left\{{#1}\right\}}
\newcommand{\sqset}[1]{\left<{#1}\right>}

\newcommand{\setcomp}[2]{\set{#1 ~|~#2 }}
\newcommand{\tuple}[1]{\left({#1}\right)}

\newcommand{\pFunctionsOf}[2]{{\mathtt{partialFunctions}}\tuple{{#1},{#2}}}

\newcommand{\tFunctionsOf}[2]{{\mathtt{totalFunctions}}\tuple{{#1},{#2}}}

\newcommand{\substSet}[2]{{#1}{\left[{#2}\right]}}
\newcommand{\mapSubst}[3]{{#1}{\left[{#2}\leftarrow{#3}\right]}}

\newcommand{\multisetsOf}[1]{\mathcal{M}\left({#1}\right)}

\newcommand{\val}{\mathtt{val}}

\newcommand{\encodingOf}[3]{\mathtt{enc}({#1},{#2},{#3})}

\newcommand{\acc}{\mathpzc{acc}}
\newcommand{\envAcc}{{\mathpzc{env}}}
\newcommand{\encTuple}{\tuple{\boolMp,\acc,\envAcc}}
\newcommand{\pencTuple}{\tuple{\boolMp',\acc',\envAcc'}}

\newcommand{\gapSet}{\mathpzc{G}}
\newcommand{\gapElem}{\mathpzc{g}}

\newcommand{\apreceq}{\preceq}

\renewcommand{\iff}{\Longleftrightarrow}

\newcommand{\isRegistred}[3]{\mathtt{isReg}(#1,#2,#3)}

\newcommand{\eval}[2]{\llbracket#1\rrbracket_{#2}}

\newcommand{\suff}{\mathtt{Suff}}
\newcommand{\usuff}{\mathtt{UnrSuff}}

\newcommand{\bVarSet}{\mathtt{B}}
\newcommand{\phVarSet}{\mathtt{V}}

\newcommand{\phvar}{\mathtt{v}}
\newcommand{\ophvar}{\mathtt{w}}

\newcommand{\bvar}{\mathtt{b}}
\newcommand{\var}{\mathtt{var}}

\newcommand{\seqVal}{\mathtt{s}}

\newcommand{\seqSet}{\mathtt{S}}

\newcommand{\grammarOr}{\;\mbox{\makebox[0mm][l]{\raisebox{-0.6ex}{\textbar}}%
    \raisebox{0.6ex}{\textbar}}\;}
\newcommand{\program}{\mathtt{prg}}
\newcommand{\programTuple}{\tuple{\bVarSet,\phVarSet,\tasks}}

\newcommand{\stmt}{\mathtt{stmt}}

\newcommand{\newp}{\mathtt{newPhaser()}}

\newcommand{\asynch}[2]{\mathtt{asynch(#1,#2)}}
\newcommand{\dereg}[1]{#1\mathtt{.drop()}}
\newcommand{\sig}[1]{#1.\mathtt{signal()}}
\newcommand{\wait}[1]{#1.\texttt{wait()}}
\newcommand{\nextt}[1]{#1.\mathtt{next()}}
\newcommand{\nextblock}[2]{#1.\mathtt{next()\{#2\}}}
\newcommand{\while}[2]{\mathtt{while(#1)\set{ #2 }}}

\newcommand{\cond}{\mathtt{cond}}

\newcommand{\true}{\mathtt{true}}
\newcommand{\false}{\mathtt{false}}
\newcommand{\ndet}{\mathtt{ndet()}}
\newcommand{\ifp}[2]{\mathtt{if(#1) \set{  #2 }}}
\newcommand{\assert}[1]{\mathtt{assert(#1)}}
\newcommand{\assign}[2]{\mathtt{#1:=#2}}
\newcommand{\exit}{\mathtt{exit}}
\newcommand{\task}{\mathtt{task}}
\newcommand{\tasks}{\mathtt{T}}
\newcommand{\parametersOf}[1]{\mathtt{paramOf}(#1)}

\newcommand{\boolMp}{\mathpzc{bv}}
\newcommand{\pcMp}{\mathpzc{seq}}

\newcommand{\nreg}{\mathtt{nreg}}
\newcommand{\relMp}{\mathpzc{phase}}

\newcommand{\relgapMp}{\mathpzc{gap}}
\newcommand{\envMp}{\mathpzc{egap}}

\newcommand{\conctask}{\mathtt{concretizeTask}}
\newcommand{\concvar}{\mathtt{concretizeVar}}
\newcommand{\concphaser}{\mathtt{concretizePhaser}}
\newcommand{\concseq}{\mathtt{concretizeSeq}}

\newcommand{\conc}{\mathtt{concTasksPhasers}}
\newcommand{\cnc}{\mathtt{conc}}

\newcommand{\plusOf}[1]{(#1)^+}

\newcommand{\cConf}{c}

\newcommand{\cConfInit}{\cConf_{init}}
\newcommand{\sConf}{\phi}
\newcommand{\osConf}{\psi}

\newcommand{\sConfSet}{\Phi}

\newcommand{\sConfSetBad}{\Phi_{bad}}

\newcommand{\raceErrors}{\mathtt{RaceErrors}}
\newcommand{\assertErrors}{\mathtt{AssertErrors}}
\newcommand{\registerErrors}{\mathtt{RegisterErrors}}
\newcommand{\deadErrors}{\mathtt{DeadlockErrors}}

\newcommand{\sConfSetAssert}[2]{\mathtt{badCstrs}_{\mathtt{assert}}}
\newcommand{\sConfSetRace}[2]{\mathtt{badCstrs}_{\mathtt{race}}}
\newcommand{\sConfSetRuntime}[2]{\mathtt{badCstrs}_{\mathtt{runtime}}}
\newcommand{\sConfSetDeadlock}[2]{\mathtt{badCstrs}_{\mathtt{deadlock}}}

\newcommand{\entailedBy}{\sqsubseteq}

\newcommand{\cConfTuple}{\tuple{\thidSet,\phidSet,\boolMp,\pcMp,\relMp}}

\newcommand{\cConfTupleInd}[1]{\tuple{\thidSet_{#1},\phidSet_{#1},\boolMp_{#1},\pcMp_{#1},\relMp_{#1}}}

\newcommand{\sConfTuple}{\tuple{\thidSet,\phidSet,\boolMp,\pcMp,\relgapMp,\envMp}}
\newcommand{\psConfTuple}{\tuple{\thidSet',\phidSet',\boolMp',\pcMp',\levMp',\relgapMp',\envMp'}}
\newcommand{\sConfTupleInd}[1]{\tuple{\thidSet_{#1},\phidSet_{#1},\boolMp_{#1},\pcMp_{#1},\relgapMp_{#1},\envMp_{#1}}}

\newcommand{\denotationOf}[1]{[\![{#1}]\!]}

\newcommand{\thid}{t}
\newcommand{\othid}{u}

\newcommand{\thidSet}{\mathpzc{T}}
\newcommand{\othidSet}{\mathpzc{U}}
\newcommand{\uniquely}{\mathpzc{Un}}
\newcommand{\fresh}{\mathpzc{F}}

\newcommand{\phid}{p}
\newcommand{\ophid}{q}

\newcommand{\phidSet}{\mathpzc{P}}

\newcommand{\waitVal}[2]{w_{#1}^{#2}}
\newcommand{\sigVal}[2]{s_{#1}^{#2}}

\newcommand{\pwaitVal}[2]{{w'}_{#1}^{#2}}
\newcommand{\psigVal}[2]{{s'}_{#1}^{#2}}

\newcommand{\lev}{\mathpzc{l}}

\newcommand{\levMp}{\mathpzc{level}}

\newcommand{\reducesTo}[2]{\xrightarrow[#2]{#1}}

\newcommand{\xrightarrowdbl}[2][]{%
  \xrightarrow[#1]{#2}\mathrel{\mkern-14mu}\rightarrow
}

\newcommand{\preReducesTo}[2]{\xrightarrowdbl[#2]{#1}}

\newcommand{\taskMp}{\tau}
\newcommand{\phaserMp}{\pi}

\newcommand{\sConfSetWorking}{\mathtt{Working}}
\newcommand{\sConfSetVisited}{\mathtt{Visited}}

\newcommand{\waitMode}{\mathrm{\textsc{Wait}}}
\newcommand{\sigMode}{\mathrm{\textsc{Sig}}}
\newcommand{\sigWaitMode}{\mathrm{\textsc{Sig\_Wait}}}

\usetikzlibrary{%
  arrows,%
  shapes,%
  positioning,%
  petri,%
  automata,%
  backgrounds%
}

\newcommand{\headOf}[1]{\mathtt{hd}(#1)}
\newcommand{\tailOf}[1]{\mathtt{tl}(#1)}

\newcommand{\movesto}[1]{\stackrel{#1}{\longrightarrow}}

\makeatletter
\def\slashedarrowfill@#1#2#3#4#5{%
  $\m@th\thickmuskip0mu\medmuskip\thickmuskip\thinmuskip\thickmuskip
   \relax#5#1\mkern-7mu%
   \cleaders\hbox{$#5\mkern-2mu#2\mkern-2mu$}\hfill
   \mathclap{#3}\mathclap{#2}%
   \cleaders\hbox{$#5\mkern-2mu#2\mkern-2mu$}\hfill
   \mkern-7mu#4$%
}

\def\rightslashedarrowfilla@{%
  \slashedarrowfill@\relbar\relbar{\raisebox{1.2pt}{$\scriptscriptstyle\diagup$}}\rightarrow}
\newcommand\xslashedrightarrowa[2][]{%
  \ext@arrow 0055{\rightslashedarrowfilla@}{#1}{#2}}

\def\rightslashedarrowfillb@{%
  \slashedarrowfill@\relbar\relbar/\rightarrow}
\newcommand\xslashedrightarrowb[2][]{%
  \ext@arrow 0055{\rightslashedarrowfillb@}{#1}{#2}}

\def\rightslashedarrowfillc@{%
  \slashedarrowfill@\relbar\relbar{\raisebox{.12em}{\tiny/}}\rightarrow}
\newcommand\xslashedrightarrowc[2][]{%
  \ext@arrow 0055{\rightslashedarrowfillc@}{#1}{#2}}

\newcommand{\notmovesto}[2]{\xslashedrightarrowc{#1}{#2}}

\newcommand{\domainOf}[1]{\mathtt{dom}(#1)}

\newcommand{\reach}[2]{\mathtt{reach}(#1,#2)}
\newcommand{\breach}[3]{\mathtt{reach}_{#3}(#1,#2)}

\newcommand{\bpreceq}{\unlhd}

\newcommand{\aepreceq}{{{~}_\forall\!\!\preceq_\exists~}}

\newcommand{\eapreceq}{{{~}_\exists\!\!\preceq_\forall~}}
\newcommand{\neapreceq}{{{~}_\exists\!\!\not\preceq_\forall~}}

\newcommand{\epreceq}{\sqsubseteq_e}

\newcommand{\jnoti}[3]{{#1}_{#2}^{\neg{#3}}}

\newcommand{\wqo}{\ensuremath{\mathcal{WQO}}}

\newcommand{\lwbw}{\mathtt{lw}}
\newcommand{\upbw}{\mathtt{uw}}

\newcommand{\lwbs}{\mathtt{ls}}
\newcommand{\upbs}{\mathtt{us}}

\newcommand{\gaptuple}{\tuple{\lwbw,\lwbs,\upbw,\upbs}}

\newcommand{\uwmin}{\mathtt{uwmin}}
\newcommand{\usmin}{\mathtt{usmin}}

\newcommand{\lwmax}{\mathtt{lwmax}}
\newcommand{\lsmax}{\mathtt{lsmax}}

\newcommand{\elwbw}{\mathtt{ew}}
\newcommand{\elwbs}{\mathtt{es}}

\newcommand{\indexOf}[2]{{#2}_{[{#1}]}}

\begin{document}

\title{On Reachability in Parameterized Phaser Programs}


\author{Zeinab Ganjei\inst{1} \and
Ahmed Rezine\inst{1} \and
Ludovic Henrio\inst{2}\and
Petru Eles\inst{1}\and
Zebo Peng\inst{1} 
}
\authorrunning{Z. Ganjei et al.}
%
\institute{Linköping University, Sweden \and
Univ Lyon, EnsL, UCBL, CNRS, Inria,  LIP, F-69342, LYON Cedex 07, France
\email{\{zeinab.ganjei,ahmed.rezine,petru.eles,zebo.peng\}@liu.se}\\
\email{ludovic.henrio@ens-lyon.fr}}

\maketitle
\begin{abstract}
  We address the problem of statically checking safety properties
  (such as assertions or deadlocks) for parameterized
  {\em phaser programs}.
  Phasers embody a non-trivial and modern synchronization construct
  used to orchestrate executions of parallel tasks.
  This generic construct supports dynamic parallelism with runtime
  registrations and deregistrations of spawned tasks. It generalizes
  many synchronization patterns such as collective and point-to-point
  schemes.
  For instance, phasers can enforce barriers or producer-consumer
  synchronization patterns among all or subsets of the running tasks.
  %
  %
  We consider in this work programs that may generate arbitrarily many tasks and phasers.
  We study different formulations of the verification problem and
  propose an exact procedure that is guaranteed to terminate for some
  reachability problems even in the presence of unbounded phases and
  arbitrarily many spawned tasks.
  In addition, we prove  undecidability results for several problems
  on which our procedure cannot be guaranteed to terminate.
  %

\end{abstract}

\section{Introduction}
\label{sec:intro}


We focus on the parameterized verification problem of parallel
programs that adopt the phasers construct for synchronization
\cite{JDVW:phasers:2008}. This coordination construct unifies
collective and point-to-point synchronization.
Parameterized verification is particularly relevant for mainstream
parallel programs as the number of interdependent tasks in many
applications, from scientific computing to web services or e-banking,
may not be known apriori.
Parameterized verification of phaser programs is a
challenging problem due to the arbitrary numbers of involved tasks and
phasers.
In this work, we address this problem and provide an exact symbolic
verification procedure. 
We identify parameterized problems for which our procedure is
guaranteed to terminate and prove the undecidability of several
variants on which our procedure cannot be guaranteed to terminate in
general.

Phasers build on the clock construct from the X10 programming language
\cite{X10:2005} and are implemented in Habanero Java
\cite{cave2011habanero}. They can be added to any parallel programming
language with a shared address space.
%
%
Conceptually, phasers are synchronization entities to which tasks can
be registered or unregistered. Registered tasks may act as producers,
consumers, or both.
Tasks can individually issue \texttt{signal}, \texttt{wait}, and
\texttt{next} commands to a phaser they are registered
to.
Intuitively, a \texttt{signal} command is used to inform other tasks
registered to the same phaser that the issuing task is done with its
current phase. It increments the \emph{signal} value associated to the
issuing task on the given phaser.
The \texttt{wait} command on the
other hand checks whether all \emph{signal} values in the phaser are
greater than the number of \texttt{wait}s issued by this task, i.e.
all registered tasks have passed the issuing task's \emph{wait}
phase.
It then increments the \emph{wait} value associated to the
task on the phaser.
As a result, the \texttt{wait} command might block the issuing
task until other tasks issue enough \texttt{signal}s.
The
\texttt{next} command consists in a \texttt{signal} followed by a
\texttt{wait}.
The \texttt{next} command may be associated to a sequence of
statements that are to be executed in isolation by one of the registered
tasks participating in the command.
A program that does not use this feature of
the next statement is said to be {\em non-atomic}.
A task deregisters from a phaser by issuing a \texttt{drop} command on it.
%
%

The dynamic synchronization allowed by the  construct
suits applications that need dynamic
load balancing (e.g, for solving non-uniform
problems with unpredictable load estimates \cite{dlb}).
Dynamic behavior is enabled by the possible runtime creation of tasks and phasers
and their registration/de-registration.
Moreover, the spawned tasks can work in different phases, adding
flexibility to the synchronization pattern.
The generality of the construct makes it also
interesting from a theoretical perspective, as many language
constructs can be expressed using phasers.
For example, synchronization barriers of Single Program Multiple Data
programs, the Bulk Synchronous Parallel computation
model~\cite{Valiant1990}, or promises and futures
constructs~\cite{SDE:BoerCJ07} can be expressed using phasers.

This paper provides general (un)decidability results that we believe will guide
verification of other synchronization constructs.
We identify combinations of features (e.g., unbounded differences between signal
and wait phases, atomic statements) and properties to be checked (e.g., assertions, deadlocks)
for which the parameterized verification problem becomes undecidable.
These help identify synchronization constructs with enough expressivity to
result in undecidable parameterized verification problems.
We also provide a symbolic verification procedure that terminates
even on fragments with arbitrary phases and numbers of spawned tasks.
We get back to possible implications in the conclusion.
We summarize our contributions:
\begin{itemize}
\item We show an operational model for phaser programs based on
  \cite{JDVW:phasers:2008,cave2011habanero,dynamicverif,fmcad17}.
\item We propose an exact symbolic verification procedure for checking
  reachability  of  sets  of   configurations  for  non-atomic  phaser
  programs even when arbitrarily many tasks and phasers may be generated.
\item We prove undecidability results for several reachability problems. 
  %
\item We show termination of our procedure when checking assertions
  for non-atomic programs even when arbitrary many tasks may be spawned.
\item We show termination of our procedure when checking
  deadlock-freedom and assertions for non-atomic programs in which the
  difference between $signal$ and $wait$ phases is bounded, even when
  arbitrary many tasks may be spawned.
\end{itemize}

\paragraph{Related work.}
The closest work to ours is \cite{fmcad17}, which is the only work on
automatic and static formal verification of phaser programs. The work
in \cite{fmcad17} does not consider the parameterized case.
For instance, this work can decide whether some program assertion
is violated even in the presence of arbitray many tasks with arbitrary large
phaser gaps. This is well beyond \cite{fmcad17} and requires a more complex
symbolic representation with a deeper termination argument.
%
The work of \cite{dynamicverif} considers the dynamic verification of phaser 
programs and can therefore only be used to detect deadlocks at runtime. The work in 
\cite{Anderson:jpf:hj:2014} uses Java Path Finder \cite{havelund:jpf:2000}
to explore all execution paths of the application. It is however also restricted
to work on one concrete input at a time.
%
A more general description of the phasers mechanism 
of Habanero Java has also been formalized in Coq~\cite{COGUMBREIRO201750}.

\paragraph{Outline.} 
We describe the phasers construct in
Sec.~\ref{sec:example}.
We then formally introduce the construct and show the associated
general reachability problem to be undecidable in
Sec.~\ref{sec:language}.
We describe in Sec.~\ref{sec:symbolic} our symbolic representation and
state some of its non-trivial properties.
We use the representation in Sec.~\ref{sec:proc} to define a
verification procedure and establish decidability results.
We refine our undecidability results in Sec.~\ref{sec:limits} and
summarize our findings in Sec.~\ref{sec:conc}.

%


\renewcommand\tabcolsep{3pt}

\begin{figure*}
  \begin{tabular}{c}
\begin{minipage}{.39\textwidth}
\begin{lstlisting}[basicstyle=\scriptsize\ttfamily]
   bool a, done;  
   main(){
    done = $\false$;  
    p= newPhaser($\sigWaitMode$);
    c= newPhaser($\sigWaitMode$);
    while($\ndet$){
     asynch(Prod,p:$\sigMode$,c:$\waitMode$);   
     asynch(Cons,p:$\waitMode$,c:$\sigMode$);
    }
    p.drop();
    c.drop();
   }
\end{lstlisting}
\end{minipage}
  \end{tabular}
  \begin{tabular}{cc}
\begin{minipage}{.28\textwidth}
{
   \begin{lstlisting}[firstnumber=14,basicstyle=\scriptsize\ttfamily]
   Prod(p:$\sigMode$,c:$\waitMode$)
   {     
     while($\neg$done)
     {      
       p.signal();
       c.wait();
       assert(a);
       a = $\false$;
     };      
     p.drop();
     c.drop();
   }   
  \end{lstlisting}
  }
\end{minipage}
&
\begin{minipage}{.29\textwidth}
{
  \begin{lstlisting}[firstnumber=27,basicstyle=\scriptsize\ttfamily]
    Cons(p:$\waitMode$,c:$\sigMode$)
    {
     while($\neg$done){
       p.wait();       
       if($\ndet$)
        done = $\true$;                 
       a = $\true$;
       c.signal();
     };
     p.drop();
     c.drop();
    }   
  \end{lstlisting}
  }
\end{minipage}
\end{tabular}
\renewcommand\tabcolsep{6pt}
\caption{An unbounded number of producers and consumers are
  synchronized using two phasers. In this construction, each consumer
  requires all producers to be ahead of it (wrt. the \texttt{p}
  phaser) in order for it to consume their respective products. At the
  same time, each consumer needs to be ahead of all producers
  (wrt. the \texttt{c} phaser) in order for the producers to be able
  to move to the next phase and produce new items. }
\label{fig:aggregators}
\end{figure*}

\section{Motivating example}
\label{sec:example}

The program listed in Fig.~(\ref{fig:aggregators}) uses Boolean shared
variables $\bVarSet=\set{\mathtt{a},\mathtt{done}}$.
The \texttt{main} task creates two phasers (line 4-5).  When creating
a phaser, the task gets automatically registered to it. The main task
also creates an unbounded number of other task instances (lines 7-8).
When a task $\thid$ is registered to a phaser $\phid$, a pair 
$(\waitVal{\thid}{\phid},\sigVal{\thid}{\phid})$ in
$\nats^2$ can be associated to the couple $(\thid,\phid)$.
%
The pair represents the individual
{\em wait} and {\em signal} phases of task $\thid$ on phaser $\phid$.

Registration of a task to a phaser can occur in one of three modes:
$\sigWaitMode$, $\waitMode$ and $\sigMode$.
In $\sigWaitMode$ mode, a task may issue
both $\mathtt{signal}$ and $\mathtt{wait}$ commands.
In $\waitMode$ mode, a task may only issue
$\mathtt{wait}$ commands on the phaser. 
Finally, when registered in $\sigMode$ mode,  
a task may only issue
$\mathtt{signal}$ commands. 
Issuing a $\mathtt{signal}$ command by a task
on a phaser results in the task
incrementing its signal phase associated to the phaser. This command
is non-blocking.
On the other-hand, issuing a $\mathtt{wait}$ command by a task
on a phaser
$\phid$ will block until {\bf all} tasks registered to $\phid$ 
get signal values on $\phid$ that are strictly
larger than the wait value of the issuing task on the same phaser. 
In this case, the wait phase of the issuing task is incremented.
Intuitively, a signal command allows the issuing task to state that
other tasks need not wait for it to complete its 
signal phase. 
In retrospect, a $\mathtt{wait}$ command allows a task to make sure all
registered tasks have moved past its wait phase. 
Upon creation of a phaser, wait and signal phases are initialized to
$0$ (except in $\waitMode$ mode where no signal phase is associated to
the task in order to not block other waiters).
The only other way a task may get registered to a phaser is if 
an already registered task spawns and registers it in the same mode 
(or in $\waitMode$ or $\sigMode$ if the registrar is registered in
$\sigWaitMode$).
In this case, wait and signal phases of the newly registered task are
initialized to those of the registrar. 
Tasks are therefore dynamically registered (e.g., lines 7-8).
They can also dynamically deregister themselves (e.g.,
line 10-11).

In this example, an unbounded number of producers and consumers are
synchronized using two phasers. Consumers require producers to be
ahead of them (wrt. the phaser they point to with $\mathtt{p}$) in
order for them to consume their products.
At the same time, consumers need to be ahead of all producers
(wrt. the phaser pointed to with $\mathtt{c}$) in order for these to
produce their items.
It should be clear that phasers can be used as barriers for
synchronizing dynamic subsets of concurrent tasks.
Observe that tasks need not, in general, proceed in a lock step
fashion. The difference between the largest signal value and the
smallest wait value can be arbitrarily large (several signals
before waits catch up). Tasks
have then more flexibility to proceed at their own speeds.

We are interested in checking: (a) control reachability as in
assertions (e.g., line 20), race conditions (e.g., mutual exclusion
of lines 20 and 33 ) or registration errors (e.g., signaling a dropped
phaser), and (b) plain reachability as in deadlocks (e.g., a producer
at line 19 and a consumer at line 30 with equal phases waiting for each other). 
Intuitively, both problems concern themselves with the reachability of
target sets of program configurations. The difference is that control
state reachability defines the targets with the states of the tasks
(their control locations and whether they are registered to some
phasers).
Plain reachability can, in addition, constrain values of the phases in
the target configurations (e.g., requiring equality between wait and
signal values for deadlocks).
Observe that control state reachability depends on the values of the actual
phases, but these values are not used to define the target sets.
For example, assertions are expressed as predicates over Boolean
variables (e.g., line 20).  Validity of such assertions may depend
on respecting phasers synchronizations.
%


\section{Phaser programs and reachability}
\label{sec:language}
We define the syntax and semantics of a core phaser programs language.
We make sure the simplified language
presented here is representative of the general purpose languages
using phasers so that our results have a practical impact.
A phaser program $\program=\programTuple$ involves a set $\tasks$ of
tasks including a unique ``main'' task
$\mathtt{main()\{\stmt\}}$. Arbitrary many instances of each task
might be spawned during a program execution. All task instances share
a set $ \bVarSet$ of Boolean variables and make use of a set
$\phVarSet$ of phaser variables that are local to individual task
instances. Arbitrary many phasers might also be generated during
program execution. Syntax of programs is as follows.


{
  \footnotesize
$$
\begin{array}{lcl}
  \program & ::= & \mathtt{bool}~~ \bvar_1, \ldots, \bvar_{p};  \\
  & & \task_1(\phvar_{1_1},\ldots,\phvar_{k_1}) \set{\stmt_1}; \\
  & & \ldots \\
  & & \task_n(\phvar_{1_n},\ldots,\phvar_{k_n}) \set{\stmt_n};  \\
  \\
  \stmt & ::= & ~~ \phvar = \newp;  
  \grammarOr \asynch{\task}{\phvar_1,\ldots,\phvar_k};
  \grammarOr \dereg{\phvar}; ~ \grammarOr \sig{\phvar}; \\
  && ~\grammarOr \wait{\phvar}; \grammarOr\nextt{\phvar};
    \grammarOr \nextblock{\phvar}{\stmt}; \grammarOr \bvar:=\cond; 
    \grammarOr\assert{\cond};    
\\
  &&
  \grammarOr \while{\cond}{\stmt};
  \grammarOr \stmt~\stmt \grammarOr \exit;
  \\
  \\
  \cond & ::= &  \ndet \grammarOr  \true
  \grammarOr \false 
  \grammarOr \bvar
  \grammarOr \cond \vee \cond 
  \grammarOr \cond \wedge \cond
  \grammarOr \neg\cond 
\end{array}
$$
}


%

Initially, a unique task instance starts
executing the $\mathtt{main()\{\stmt\}}$ task.
 A phaser can recall a pair of values (i.e., wait and signal) for each
 task instance registered to it. A task instance can create a new
 phaser with $\phvar=\newp$, get registered to it (i.e., gets zero as
 wait and signal values associated to the new phaser) and refer to the
 phaser with its local variable $\phvar$.  We
 simplify the presentation by assuming all registrations to be in
 $\sigWaitMode$ mode.
Including the other modes is a matter of depriving $\waitMode$-registered tasks of a
signal value (to ensure they do not block other registered tasks) and of 
ensuring issued commands respect registration modes.
We use $\phVarSet$ for the union of all local phaser
variables.
A task $\task(\phvar_{1},\ldots,\phvar_{k}) \set{\stmt}$ in $\tasks$
takes the phaser variables $\phvar_1, \ldots \phvar_k$ as
parameters (write $\parametersOf{\task}$ to mean these parameters).
%
%
A task instance can spawn another task instance with
$\asynch{\task}{\phvar_1,\ldots,\phvar_n}$. The issuing task instance
registers the spawned task to the
phasers pointed to by $\phvar_1,\ldots,\phvar_n$,  with its own wait and signal values.
Spawner and Spawnee execute concurrently.
%
%
A task instance can deregister itself from a phaser
with $\dereg{\phvar}$.

A task instance can issue signal or wait commands on a
phaser referenced by $\phvar$ and on which it is registered. A wait
command on a phaser blocks until the wait value of the task instance
executing the wait on the phaser is strictly smaller than the signal
value of all task instances registered to the phaser. In other words, $\wait{\phvar}$ 
 blocks if $\phvar$ points to a phaser such that 
at least one of the signal values stored by the phaser is
equal to the wait 
value of the task that tries to perform the wait.
A signal command does not block. It only increments the signal value
of the task instance executing the signal command on the phaser.
$\nextt{\phvar}$ is syntactic sugar for a signal followed by a
wait.
Moreover, $\nextblock{\phvar}{\stmt}$ is similar to
$\nextt{\phvar}$ but the block of code $\stmt$ is executed atomically
by exactly one of the tasks participating in the synchronization
before all tasks continue the execution that follows the
barrier. $\nextblock{\phvar}{\stmt}$ thus requires all tasks to be
synchronized on exactly the same statement and is less flexible.
Abscence of a $\nextblock{\phvar}{\stmt}$ makes a program {\em non-atomic}.

Note that assignment of phaser variables is excluded from the syntax;
additionally, we restrict task creation
$\asynch{\task}{\phvar_1,\ldots,\phvar_n}$ and require that parameter
variables $\phvar_i$ are all different. This prevents two variables
from pointing to the same phaser and avoids the need to deal with aliasing:
we can reason on the single variable in a process that points to a
phaser. Extending our work to deal with aliasing is easy but would
require heavier notations.

We will need the notions of configurations, partial configurations and
inclusion in order to define the reachability problems we consider in
this work.
We introduce them in the following and assume a phaser program $\program=\programTuple$.

\paragraph{\bf Configurations.}
Configurations of a phaser program describe valuations of its variables,
control sequences of its tasks and registration details to the
phasers. 
\noindent\emph{Control sequences.}
We define the set $\suff$ of control sequences of $\program$ to be
the set of suffixes of all sequences $\stmt$
appearing in some statement $\task(\ldots)\set{\stmt}$.
In addition, we define $\usuff$ to be the smallest set containing $\suff$ in
addition to the suffixes of all (i)
$\seqVal_1;\while{\cond}{\seqVal_1};\seqVal_2$ if 
$\while{\cond}{\seqVal_1};\seqVal_2$ is in $\usuff$, and of all
(ii)  $\seqVal_1;\seqVal_2$ if
$\ifp{\cond}{\seqVal_1};\seqVal_2$ is in $\usuff$, and of all (iii)
$\seqVal_1;\nextblock{\phvar}{};\seqVal_2$ if
$\nextblock{\phvar}{\seqVal_1};\seqVal_2$ in $\usuff$, and finally of all (iv)
$\sig{\phvar};\wait{\phvar}{};\seqVal_2$ if
$\nextblock{\phvar}{};\seqVal_2$ is in $\usuff$.
%
%
%
%
We write $\headOf{\seqVal}$ and $\tailOf{\seqVal}$ to respectively
mean the head and the tail of a sequence $\seqVal$.

\noindent\emph{Partial configurations.}
Partial configurations allow the characterization of sets of
configurations by partially stating some of their common
characteristics.
A \emph {partial configuration} $\cConf$ of
$\program=\tuple{\bVarSet,\phVarSet,\tasks}$ is a tuple
$\tuple{\thidSet,\phidSet,\boolMp,\pcMp,\relMp}$ where:
\begin{itemize}
\item $\thidSet$ is a finite set  of
  task identifiers. We let $\thid,\othid$  range over the values in $\thidSet$.
\item $\phidSet$ is a finite set of phaser identifiers. We let
  $\phid,\ophid$  range over the values in
  $\phidSet$.
\item $\boolMp:\bVarSet \to \supscript{\bools}{*}$ fixes the values of
  some of the shared variables.\footnote{For any set $S$,
    $\supscript{S}{a,b,...}$ denotes ${S}\cup\{a,b,...\}$.}
\item $\pcMp:\thidSet \to \supscript{\usuff}{*}$  fixes the control sequences of some of the tasks. 
\item $\relMp:\thidSet \to
  \pFunctionsOf{\phidSet}{\supscript{\phVarSet}{-,*} \times
    \left(\nats^2\cup\set{(*,*),\nreg}\right)}$
  is a mapping that associates to each task $\thid$ in $\thidSet$ a
  partial mapping stating which phasers are known by the task and with which registration values. 
\end{itemize}

Intuitively, partial configurations are used to state some facts
about the valuations of variables and the control sequences of tasks and their 
registrations. Partial configurations leave some details
unconstrained using partial mappings or the symbol $*$.
For instance, if $\boolMp(\bvar)=*$ in a partial configuration 
$\tuple{\thidSet,\phidSet,\boolMp,\pcMp,\relMp}$, then the  
partial configuration does not constrain the value
of the shared variable $\bvar$. 
Moreover, a partial configuration does not constrain
the relation between a task $\thid$ and a phaser $\phid$
when $\relMp(\thid)(\phid)$ is undefined.
Instead, when the partial mapping $\relMp(\thid)$ is defined
on phaser $\phid$,  it associates 
a pair $\relMp(\thid)(\phid)=(\var,\val)$ to $\phid$.
If $\var\in\supscript{\phVarSet}{-,*}$ is a variable
$\phvar\in\phVarSet$ then the task $\thid$ in $\thidSet$ uses its
variable $\phvar$ to refer to the phaser $\phid$ in $\phidSet$\footnote{The uniqueness of this variable is due to the absence of aliasing discussed above}.  If
$\var$ is the symbol $-$ then the task $\thid$ does not refer to
$\phvar$ with any of its variables in $\phVarSet$.  If $\var$ is the
symbol $*$, then the task might or might not refer to $\phid$.
The value $\val$ in $\relMp(\thid)(\phid)=(\var,\val)$ is either
the value $\nreg$ or a pair
$(\waitVal{}{},\sigVal{}{})$.  The value $\nreg$ 
means the task $\thid$ is not registered to phaser
$\phid$. The pair $(\waitVal{}{},\sigVal{}{})$
belongs to
$(\nats\times\nats)\cup\set{(*,*)}$. In this case,
task $\thid$ is registered to phaser $\phid$ with a symbolic wait phase
$\waitVal{}{}$ and a symbolic signal phase $\sigVal{}{}$. The value $*$
means that the wait phase $\waitVal{}{}$ (resp. signal phase
$\sigVal{}{}$) can be any value in $\nats$. For
instance, $\relMp(\thid)(\phid)=(\phvar,\nreg)$ means variable $\phvar$ of the task $\thid$
refers to phaser $\phid$  but the task is not
registered to $\phid$. On the other hand,
$\relMp(\thid)(\phid)=(-,(*,*))$ means the task $\thid$ does not refer to
$\phid$ but is registered to
it with a arbitrary wait and signal phases. 

\noindent\emph{Concrete configurations}. A concrete configuration (or
configuration for short) is a partial configuration
$\tuple{\thidSet,\phidSet,\boolMp,\pcMp,\relMp}$
where $\relMp(\thid)$ is total for each $\thid\in\thidSet$
and where the symbol $*$ does not appear in any range.
It is a tuple
$\tuple{\thidSet,\phidSet,\boolMp,\pcMp,\relMp}$ where 
$\boolMp:\bVarSet \to \bools$, $\pcMp:\thidSet \to
\usuff$, and $\relMp:\thidSet \to
\tFunctionsOf{\phidSet}{\supscript{\phVarSet}{-} \times
  \left(\left(\nats\times\nats\right)\cup\set{\nreg}\right)}$.
For a concrete configuration $\tuple{\thidSet,\phidSet,\boolMp,\pcMp,\relMp}$,
we write $\isRegistred{\relMp}{\thid}{\phid}$ to mean the predicate
$\relMp(\thid)(\phid)\not\in\left(\supscript{\phVarSet}{-}\times\set{\nreg}\right)$.
The predicate $\isRegistred{\relMp}{\thid}{\phid}$ captures whether the task $\thid$
is registered to phaser $\phid$ according to the mapping $\relMp$.
%


\noindent \emph{Inclusion of configurations}.  A configuration
$\cConf'=\tuple{\thidSet',\phidSet',\boolMp',\pcMp',\relMp'}$
\emph{includes} a partial configuration
$\cConf=\tuple{\thidSet,\phidSet,\boolMp,\pcMp,\relMp}$ if renaming and deleting
tasks and phasers from $\cConf'$  can give a
configuration that ``matches'' $\cConf$.
More formally, $\cConf'$ includes $\cConf$ if
$\left((\boolMp(\bvar)\neq\boolMp'(\bvar))\implies (\boolMp(\bvar)=*)\right)$ for each $\bvar\in\bVarSet$
and there are injections $\tau:\thidSet\to\thidSet'$ and 
$\pi:\phidSet\to\phidSet'$ s.t. for each
$\thid\in\thidSet$ and  $\phid\in\phidSet$:
(1) $((\pcMp(\thid)\neq\pcMp'(\tau(\thid)))\implies (\pcMp(\thid)=*))$, and 
either (2.a) $\relMp(\thid)(\phid)$ is undefined, or (2.b)
$\relMp(\thid)(\phid)=(var, val)$ and
$\relMp'(\taskMp(\thid))(\phaserMp(\phid))=(var', val')$ with
$((var\neq var')\implies (var=*))$ and either $(val=val'=\nreg)$ or
$val=(\waitVal{}{},\sigVal{}{})$ and
$val'=(\pwaitVal{}{},\psigVal{}{})$ with $((\waitVal{}{}\neq
\pwaitVal{}{}) \implies (\waitVal{}{}=*))$ and $((\sigVal{}{}\neq
\psigVal{}{}) \implies (\sigVal{}{}=*))$.


\paragraph{\bf Semantics and reachability.}
Given a program $\program=\programTuple$, the main task
$\mathtt{main()\{\stmt\}}$ starts executing $\stmt$ from an initial configuration
$\cConfInit=(\thidSet_{init},\phidSet_{init},\boolMp_{init},\pcMp_{init},\relMp_{init})$
where $\thidSet_{init}$ is a singleton, $\phidSet_{init}$ is empty,
$\boolMp_{init}$ sends all shared variables to $\false$ and
$\pcMp_{init}$ associates $\stmt$ to the unique
task in $\thidSet_{init}$.
%
%
We write $\cConf \reducesTo{\thid}{\stmt} \cConf'$ to mean a task
$\thid$ in $\cConf$ can fire statement $\stmt$ and result in
configuration $\cConf'$. See Fig.~\ref{fig:semantics} for a
description of the operational semantics. We write $\cConf
\reducesTo{}{\stmt} \cConf'$ if $\cConf \reducesTo{\thid}{\stmt}
\cConf'$ for some task $\thid$, and $\cConf \reducesTo{}{} \cConf'$ if
$\cConf \reducesTo{}{\stmt} \cConf'$ for some statement $\stmt$. We
also write $\reducesTo{}{\stmt}^+$ for the transitive closure of
$\reducesTo{}{\stmt}$ and let ${\reducesTo{}{}}^*$ be the reflexive
transitive closure of $\reducesTo{}{}$.  Fig.~\ref{fig:semantics-err}
identifies erroneous configurations.

%

\begin{figure} 
  \begin{center}
    \scriptsize

    \[
    \begin{array}{c}

\infer[\mathtt{(newPhaser)}]
      {       
        \cConfTuple 
        \reducesTo{\thid}{\phvar:=\newp}\tuple{\thidSet,\phidSet',\boolMp,
          \mapSubst{\pcMp}{\thid}{\tailOf{\pcMp(\thid)}},
          \relMp''}        
      }
      {
        \begin{array}{c}
        \headOf{\pcMp(\thid)}=\phvar:=\newp \quad 
        \phid\not\in\phidSet \\
        \phidSet'=\phidSet\cup\set{\phid} \quad 
        \relMp'=\mapSubst{\relMp}{\thid}{\set{\phid \leftarrow (\phvar,(0,0))}} \\
        \relMp''=\substSet{\relMp'}{\setcomp{\othid\leftarrow \mapSubst{\relMp'(\othid)}{\phid}{(-,\nreg)}}{\othid \in \thidSet\setminus\set{\thid}}}
        \end{array}
      } 
      \\
      \\
\infer[\mathtt{(asynch)}]
      {
        \cConfTuple \reducesTo{\thid}{\asynch{\task}{\phvar_1,\ldots \phvar_k}\{\stmt\}}{}
        \tuple{\thidSet',\phidSet,\boolMp,
          \substSet{\pcMp'}{\set{\othid\leftarrow\stmt}\cup\set{\thid\leftarrow\tailOf{\pcMp(\thid)}}},
          \relMp' }
      }
      {
        \begin{array}{c}
        \headOf{\pcMp(\thid)}=\asynch{\task}{\phvar_1,\ldots \phvar_k}\{\stmt\} \quad 
        \parametersOf{\task}=(\ophvar_1,\ldots \ophvar_k) \quad  \othid\not\in\thidSet \quad
        \thidSet'=\thidSet\cup\set{\othid} 
        \\
        \textrm{for each } i:1\leq i\leq \mathtt{k}.~\relMp(\thid)(\phid_i)=
        (\phvar_i,(\waitVal{i}{},\sigVal{i}{})) \quad 
         \textrm{for each } i,j:1\leq i,j\leq \mathtt{k}.~i\neq j\Rightarrow\phvar_i\neq\phvar_j \\        
         \relMp'=
         \mapSubst{\relMp}{\othid}{
           \setcomp{\phid_i\leftarrow(\ophvar_i,(\waitVal{i}{},\sigVal{i}{}))}
                   {1\leq i\leq \mathtt{k}} 
                   \cup
           \setcomp{\phid\leftarrow(-,\nreg)}
                   {\phid \not\in\setcomp{p_i}{1 \leq i \leq \mathtt{k}}}
         }
        \end{array}
      }       
      \\
      \\      
                
\infer[\mathtt{(signal)}]
      {
        \cConfTuple \reducesTo{\thid}{\sig{\phvar}}{}
        \tuple{\thidSet,\phidSet,\boolMp,\mapSubst{\pcMp}{\thid}{\tailOf{\pcMp(\thid)}},\relMp'}
      }
      {
        \begin{array}{c}
        \headOf{\pcMp(\thid)}=\sig{\phvar}  \quad 
        \relMp(\thid)(\phid)=(\phvar,(\waitVal{}{},\sigVal{}{}))  
        \quad
        \relMp'=\mapSubst{\relMp}{\thid}                 
                         {\set{\phid \leftarrow (\phvar,(\waitVal{}{},1+\sigVal{}{}))}}
        \end{array}
      }
      \\
      \\
\infer[\mathtt{(wait)}]
      {
        \cConfTuple \reducesTo{\thid}{\wait{\phvar}}{}             
        \tuple{\thidSet,\phidSet,\boolMp,\mapSubst{\pcMp}{\thid}{\tailOf{\pcMp(\thid)}},\relMp'}
      }
      {
        \begin{array}{c}
        \headOf{\pcMp(\thid)}=\wait{\phvar} \quad 
        \relMp(\thid)(\phid)=(\phvar,(\waitVal{\thid}{},\sigVal{\thid}{})) \\
        \forall 
        \othid\in\thidSet, var\in\supscript{\phVarSet}{-}.~\left(\relMp(\othid)(\phid)=(var,(\waitVal{\othid}{},\sigVal{\othid}{}))
        \Rightarrow \waitVal{\thid}{} < \sigVal{\othid}{} \right) \\
        \relMp'=\mapSubst{\relMp}{\thid}                 
                         {\set{\phid \leftarrow (\phvar,(1+\waitVal{\thid}{},\sigVal{\thid}{}))}}
        \end{array}
      }
      \\
      \\
\infer[\mathtt{(next)}]
      {
        \cConfTuple \reducesTo{\thid}{\nextt{\phvar}}{}
        \tuple{\thidSet,\phidSet,\boolMp,\pcMp',\relMp}
      }
      {
        \begin{array}{c}
        \headOf{\pcMp(\thid)}=\nextt{\phvar} \quad 
        \pcMp'=\mapSubst{\pcMp}{\thid}{\sig{\phvar};\wait{\phvar};\tailOf{\pcMp(\thid)}}
        \end{array}
      }
      \\
      \\
\infer[\mathtt{(next\{stmt\})}]
      {
        \cConfTuple 
        \reducesTo{\thid}{\nextblock{\phvar}{\stmt}}{}\tuple{\thidSet,\phidSet,\boolMp,\pcMp',\relMp}
      }
      {
        \begin{array}{c}
	\othidSet=\setcomp{\othid}{\isRegistred{\relMp}{\othid}{\phid}} \quad 
	\forall \othid\in \othidSet.\,\headOf{\pcMp(\othid)}=\nextblock{\phvar}{\stmt}
        \quad 	\thid\in\othidSet\\
        \pcMp'=\substSet{\mapSubst{\pcMp}{\thid}{\stmt;\nextt{\phvar};\tailOf{\pcMp(\thid)}}}
	                {\setcomp{\othid\leftarrow\nextt{\phvar};\tailOf{\pcMp(\othid)}}
                          {\othid\in\othidSet\setminus 
	                    \set{\thid}}}
        \end{array}
      }
      \\
      \\      
\infer[{\mathtt{(assign)}}]
      {
        \cConfTuple \reducesTo{\thid}{\assign{b}{\cond}}{}   
        \tuple{\thidSet,\phidSet,\mapSubst{\boolMp}{b}{\cond},\mapSubst{\pcMp}{\thid}{\tailOf{\pcMp(\thid)}},\relMp}
      }
      {
        \begin{array}{c}
          \headOf{\pcMp(\thid)}=\assign{b}{\cond}
        \end{array}
      }
      \\
      \\  
\infer[\left(\begin{array}{c}\mathtt{assertion:ok}\end{array}\right)]
      {
        \cConfTuple \reducesTo{\thid}{\assert{\cond}}{}
        \tuple{\thidSet,\phidSet,\boolMp,\mapSubst{\pcMp}{\thid}{\tailOf{\pcMp(\thid)}},\relMp}
      }
      {
        \begin{array}{c}
        \headOf{\pcMp(\thid)}=\assert{\cond}  \quad 
        \eval{\cond}{\boolMp}=\true
        \end{array}
      }   
      \\
      \\
\infer[\left(\begin{array}{c}\mathtt{selection:then}\end{array}\right)]
      {
        \cConfTuple \reducesTo{\thid}{\ifp{\cond}{\stmt}}{}
        \tuple{\thidSet,\phidSet,\boolMp,\mapSubst{\pcMp}{\thid}{\stmt;\tailOf{\pcMp(\thid)}},\relMp}
      }
      {
        \begin{array}{c}
        \headOf{\pcMp(\thid)}=\ifp{\cond}{\stmt}  \quad 
        \eval{\cond}{\boolMp}=\true
        \end{array}
      }   
      \\
      \\
\infer[\left(\begin{array}{c}\mathtt{selection:else }\end{array}\right)]
      {
        \cConfTuple \reducesTo{\thid}{\ifp{\cond}{\stmt}}{}
        \tuple{\thidSet,\phidSet,\boolMp,\mapSubst{\pcMp}{\thid}{\tailOf{\pcMp(\thid)}},\relMp}
      }
      {
        \begin{array}{c}
        \headOf{\pcMp(\thid)}=\ifp{\cond}{\stmt}  \quad 
        \eval{\cond}{\boolMp}=\false
        \end{array}
      }   
      \\
      \\
\infer[\left(\begin{array}{c}\mathtt{while:unroll}\end{array}\right)]
      {
        \cConfTuple \reducesTo{\thid}{\while{\cond}{\stmt}}{}
        \tuple{\thidSet,\phidSet,\boolMp,\mapSubst{\pcMp}{\thid}{\stmt;\pcMp(\thid)},\relMp}
      }
      {
        \begin{array}{c}
        \headOf{\pcMp(\thid)}=\while{\cond}{\stmt}  \quad 
        \eval{\cond}{\boolMp}=\true
        \end{array}
      }   
      \\
      \\
\infer[\left(\begin{array}{c}\mathtt{while:exit}\end{array}\right)]
      {
        \cConfTuple \reducesTo{\thid}{\while{\cond}{\stmt}}{}
        \tuple{\thidSet,\phidSet,\boolMp,\mapSubst{\pcMp}{\thid}{\tailOf{\pcMp(\thid)}},\relMp}
      }
      {
        \begin{array}{c}
        \headOf{\pcMp(\thid)}=\while{\cond}{\stmt}  \quad 
        \eval{\cond}{\boolMp}=\false
        \end{array}
      }   
      \\
      \\
\infer[\mathtt{(drop)}]
      {
        \cConfTuple \reducesTo{\thid}{\dereg{\phvar}}{}
        \tuple{\thidSet,\phidSet,\boolMp,\mapSubst{\pcMp}{\thid}{\tailOf{\pcMp(\thid)}},\relMp'}
      }
      {
        \begin{array}{c}
        \headOf{\pcMp(\thid)}=\dereg{\phvar} \quad 
        \relMp(\thid)(\phid)=(\phvar,(\waitVal{}{},\sigVal{}{}))  \quad 
        \relMp'=\substSet{\relMp}{
          \substSet{\thid\leftarrow\relMp(\thid)}{\phid \leftarrow (\phvar,\nreg)}}
        \end{array}
      }
      \\
      \\      
\infer[\mathtt{(exit)}]
      {
        \cConfTuple 
        \reducesTo{\thid}{\exit}{}      
        \tuple{\thidSet\setminus\set{\thid},\phidSet,\boolMp,\pcMp',\relMp'}
      }
      {
        \begin{array}{c}
        \headOf{\pcMp(\thid)}=\exit \quad 
        \pcMp'=\pcMp\setminus\set{\thid} \quad 
        \relMp'=\relMp\setminus\set{\thid} 
        \end{array}
      }
    \end{array}
    \]
  \end{center}

  \caption{Operational semantics of phaser statements without
    errors. Each transition corresponds to a task $\thid\in\thidSet$
    executing a statement from a configuration $\cConfTuple$. For
    instance, the $\mathtt{drop}$ transition corresponds to a task
    $\thid$ executing $\dereg{\phvar}$ when registered to phaser
    $\phid\in\phidSet$ (with phases $(\waitVal{}{},\sigVal{}{})$) and
    referring to it with variable $\phvar$. The result is the same
    configuration where task $\thid$ moves to its next statement
    without being registered to $\phid$ (albeit still refering to
    $\phid$ with $\phvar$).}
  \label{fig:semantics}
\end{figure}

\begin{figure}[!t]
  \begin{center}
    \scriptsize

    \[
    \begin{array}{c}

\infer[\left(\begin{array}{c}\mathtt{assertion:errors}\end{array}\right)]
      {
        \cConfTuple \in \assertErrors
      }
      {
        \begin{array}{c}
        \headOf{\pcMp(\thid)}=\assert{\cond}  \quad 
        \eval{\cond}{\boolMp}=\false
        \end{array}
      }
      \\
      \\
\infer[\left(\begin{array}{c}\mathtt{race:errors}\end{array}\right)]
      {
        \cConfTuple \in \raceErrors
      }
      {
        \begin{array}{c}
          \headOf{\pcMp(\thid)}=\assign{b'}{\cond'}  \quad
          \textrm{(}b' \textrm{ coincides with } b \textrm{ or appears in } \cond \textrm{)} \\
        \headOf{\pcMp(\othid)}\in\set{\assign{b}{\cond},
          \ifp{\cond}{\stmt}, \while{\cond}{\stmt},\assert{\cond}} 
        \end{array}
      }   
      \\
      \\
\infer[\mathtt{(registration:errors)}]
      {
        \cConfTuple \in \registerErrors
      }
      {
        \begin{array}{c}
          \thid\in\thidSet \quad \phvar\in\phVarSet
          \quad \phid\in\phidSet \quad \relMp(\thid)(\phid)=(\phvar,\nreg) \\   
          \headOf{\pcMp(\thid)}\in\set{
            \asynch{\task}{\ldots \phvar \ldots}\{\stmt\},
            \sig{\phvar},
            \wait{\phvar},
            \dereg{\phvar}
          } 
        \end{array}
      }
      \\
      \\
\infer[\mathtt{(deadlock:errors)}]
      {
        \cConfTuple \in \deadErrors
      }
      {
        \begin{array}{l}
          \thid_0,\ldots\thid_n\in\thidSet \quad \phvar_0,\ldots \phvar_n\in\phVarSet
          \quad var_0,\ldots var_n\in\supscript{\phVarSet}{-} \quad
          \phid_0,\ldots\phid_n\in\phidSet \\ 
		\sigVal{0}{}\ldots\sigVal{n}{}\in\nats\quad \sigVal{0}{\prime}\ldots \sigVal{n}{\prime}\in\nats
\quad \waitVal{0}{}\ldots \waitVal{n}{}\in\nats \quad \waitVal{0}{\prime}\ldots \waitVal{n}{\prime}\in\nats\\
          \forall i:0\leq i \leq n.
          \left\{
          \begin{array}{l}       
          \headOf{\pcMp(\thid_i)}=\wait{\phvar_i} \\		  
          \relMp(\thid_i)(\phid_i)=(\phvar_i,(\waitVal{i}{},\sigVal{i}{})) \\
          \relMp(\thid_i)(\phid_{(i+1)\%(n+1)})=(var_{i},(\waitVal{i}{\prime},\sigVal{i}{\prime}))\\
          \sigVal{i}{\prime} = \waitVal{(i+1)\%(n+1)}{}
          \end{array}\right.
        \end{array}
      }
    \end{array}
    \]
  \end{center}

  \caption{Definition of error configurations. Starting from 
    $\cConfTuple$, error configurations are obtained
    when tasks execute the above statements under certain
    conditions. For instance, a deadlock is obtained if tasks in
    a subset $\set{\thid_0,\ldots,\thid_n}\subseteq\thidSet$ form a cycle
    where each $\thid_{i}$ blocks (with its signal phase
    $\sigVal{i}{\prime}$) the wait being executed by
    $\thid_{(i+1)\%(n+1)}$ on phaser $\phid_{(i+1)\%(n+1)}$ (with wait
    phase $\waitVal{(i+1)\%(n+1)}{}$).}
  \label{fig:semantics-err}
\end{figure} 

We are interested in the reachability of sets of
configurations (i.e., checking safety properties).
We differentiate between two reachability problems depending on whether
the target sets of configurations constrain the registration phases or not.
The \emph{plain reachability} problem  constrains the registration phases
of the target configurations.
%
The \emph{control reachability} problem only  constrains registration status, control sequences, or variable values. 
We will see that decidability of the two problems can be different.
The two problems are defined in the following.

\noindent\emph{Plain reachability.} First, we define equivalent
configurations.
A configuration
$\cConf=\tuple{\thidSet,\phidSet,\boolMp,\pcMp,\relMp}$ is
equivalent to configuration 
$\cConf'=\tuple{\thidSet',\phidSet',\boolMp',\pcMp',\relMp'}$
if  $\boolMp=\boolMp'$ and there are bijections
$\taskMp:\thidSet\to\thidSet'$ and $\phaserMp:\phidSet\to\phidSet'$ such that,
for all $\thid\in\thidSet$, $\phid\in\phidSet$ and
$var\in\supscript{\phVarSet}{-}$,
$\pcMp(\thid)=\pcMp'(\tau(\thid))$
 and there are some integers $(k_\phid)_{\phid\in\phidSet}$ such that
 $\relMp(\thid)(\phid)=(var,(w,s))$ iff
 $\relMp'(\taskMp(\thid))(\phaserMp(\phid))=(var,(w+k_\phid,s+k_\phid))$. We
 write $\cConf\sim\cConf'$ to mean that $\cConf$ and $\cConf'$ are 
 equivalent. 
 Intuitively, equivalent configurations simulate each other.
%
  We can establish the following: 
  
  \begin{lemma}[Equivalence]
    Assume two configurations $\cConf_1$ and $\cConf_2$. If
    $\cConf_1\reducesTo{}{}\cConf_2$ and $\cConf_1'\sim\cConf_1$
    then there is a configuration $\cConf_2'$ s.t. $\cConf_2'\sim\cConf_2$ and $\cConf_1'\reducesTo{}{}\cConf_2'$.
    \end{lemma}
Observe that if the wait value of a task $\thid$ on a phaser $\phid$
is equal to the signal of a task $\thid'$ on the same phaser $\phid$
in some configuration $\cConf$, then this is also the case, up to a
renaming of the phasers and tasks, in all equivalent configurations.
This is particularly relevant for defining deadlock configurations
where a number of tasks are waiting for each other.
The plain reachability problem is given a program and a target partial
configuration and asks whether a
configuration (equivalent to a configuration) that includes the
target partial configuration is reachable.
More formally, given a program $\program$ 
and a partial configuration $\cConf$, let $\cConfInit$ be
the  initial configuration of $\program$, then
 $\reach{\program}{\cConf}$ if and only if 
$\cConfInit \reducesTo{}{}^* \cConf_1$ for $\cConf_1\sim\cConf_2$ and
$\cConf_2$  includes $\cConf$.

\begin{definition}[Plain reachability]
  For a program $\program$ and a partial configuration $\cConf$,
  decide whether
  $\reach{\program}{\cConf}$ holds. 
\end{definition}

\noindent\emph{Control reachability.}
A partial configuration
$\cConf=\tuple{\thidSet,\phidSet,\boolMp,\pcMp,\relMp}$ is said to be
a \emph{control partial configuration} if for all $\thid\in\thidSet$  and
$\phid\in\phidSet$, either
$\relMp(\thid)(\phid)$ is undefined or 
$\relMp(\thid)(\phid)\in(\supscript{\phVarSet}{-,*} \times \set{\tuple{*,*},\nreg})$.
Intuitively, control partial configurations
do not constrain phase values.
%
%
%
%
They are enough to characterize, for example, configurations
where an assertion is violated (see Fig.~\ref{fig:semantics-err}).
%

\begin{definition}[Control reachability]\label{def:control}
  For a program $\program$ and a \underline{control} partial configuration $\cConf$,
  decide whether $\reach{\program}{\cConf}$ holds.
\end{definition}

Observe that plain reachability is at least as hard to answer as control reachability
since any control partial configuration is also a partial configuration.
It turns out the control reachability problem is undecidable for programs resulting
in arbitrarily many tasks and phasers as stated by the theorem below.
This is proven by reduction of the state reachability problem for 2-counter Minsky machines.
A 2-counter Minsky machine $\tuple{S,\set{x_1,x_2},\Delta,s_0,s_F}$
has a finite
set $S$ of states, two counters $\set{x_1,x_2}$ with values in
$\nats$, an initial state $s_0$ and a final state  $s_F$.
Transitions may increment, decrement or test a counter.
For example $\tuple{s_0,\textrm{test($x_1$)},s_F}$ takes the machine
from $s_0$ to $s_F$ if the counter $x_1$ is zero. 

\begin{theorem}[Minsky machines \cite{minsky}]
  Checking whether $s_F$ is reachable from configuration $(s_0,0,0)$
  for 2-counter machines is undecidable in general.
\end{theorem}

\begin{theorem}
  \label{thm:control:general}
 Control reachability is undecidable in general. 
\end{theorem}
\paragraph{Proof sktech.}
  State reachability of an arbitrary 2-counters
  Minsky machine is encoded as the control reachability problem of a phaser
  program (captured in Fig. \ref{fig:thm-control-gen}).
  %
  %
  The phaser program has three tasks $\mathtt{main}$, $\mathtt{xUnit}$ and $\mathtt{yUnit}$.
  It uses Boolean shared variables to encode the state $s\in S$
  and to pass information between different
  task instances. 
  The phaser program builds two chains, one with $\mathtt{xUnit}$ instances for the $x$-counter,
  and one with $\mathtt{yUnit}$ instances for the $y$-counter.
  Each chain alternates a phaser and a task and encodes the values
  of its counter with its length.
  The idea is to have the phaser program simulate all transitions of
  the counter machine, i.e., increments, decrements and tests for
  zero.
  Answering state reachability of the counter machine amounts to
  checking whether there are reachable configurations where the
  boolean variables encoding the counter machine can evaluate to the
  target machine state $s_F$. This can be captured with a control partial
  configuration.
  \begin{figure}
  \begin{center}
  \begin{tabular}{ccc}
\begin{minipage}{.5\textwidth}
{
\begin{lstlisting}
   bool s1, s2,...,sF; 
   bool xInc,xDec,yInc,yDec,temp; 
   main(){
      $\mathtt{xPh}$ = newPhaser();
      $\mathtt{yPh}$ = newPhaser();     
      while(true){  
         temp = false;
         //(${\color{olive}\mathtt{q_i}}$:inc(x):${\color{olive}\mathtt{q_j}}$)
         if(ndet() $\wedge$ si){
            xInc=$\true$;
            xPh.signal();
            xPh.wait();  
            if(!temp){ 
               asynch(xTask,xPh); 
            }
            else{
               temp = $\false$;
            }
            xInc = $\false$; 
            
            si$=\false$;
            sj$=\true$;
         }       
         //(${\color{olive}\mathtt{q_i}}$:dec(x):${\color{olive}\mathtt{q_j}}$)
         if(ndet() $\wedge$ sj){
            $\mathtt{xDec}=\true$;
            xPh.signal();
            xPh.wait();
            while(!temp){};
            temp =$\false$;
            xDec = $\false$;
            si$=\false$;
            sj$=\true$;
         }  
         //(${\color{olive}\mathtt{q_i}}$:test(x):${\color{olive}\mathtt{q_j}}$)
         if(ndet() $\wedge$ sj){
            $\mathtt{xPh}$.signal(); 
            $\mathtt{xPh}$.wait(); 
            si$=\false$;
            sj$=\true$;
         }
\end{lstlisting}
}
\end{minipage}
&
\begin{minipage}{.5\textwidth}
{
\begin{lstlisting}[firstnumber=42] 
      //(${\color{olive}\mathtt{q_i}}$:inc(y):${\color{olive}\mathtt{q_j}}$)
      if(ndet() $\wedge$ si){
            ...
      }
      //(${\color{olive}\mathtt{q_i}}$:dec(y):${\color{olive}\mathtt{q_j}}$)
      if(ndet() $\wedge$ si){    
         ...     
      } 
      //(${\color{olive}\mathtt{q_i}}$:test(y):${\color{olive}\mathtt{q_j}}$)
      if(ndet() $\wedge$ si){    
         ...     
      }
   } 
     
   //***** xTask *****
   xTask(parent){
      child=newPhaser();     		
      while(true){
        child.signal();
        child.wait();	     	
        if($\mathtt{xInc}$){
            temp=$\true$; 
            asynch(xTask,child);
            parent.signal();  
            while(temp){};
        }  
       if(xDec){
            temp=$\true$; 
            exit(); 
       }
       if(temp){
         parent.signal(); 
         while(temp){}; 
       }   		
     }
   } 
   
   //**** yTask ****
   yTask(parent){
      ...
   }
\end{lstlisting}
}
\end{minipage}

\end{tabular}
\end{center}
  \caption[Encoding a 2-counter Minsky machine for the proof of Theorem \ref{thm:control:general}]{For the proof of Theorem \ref{thm:control:general}. Encoding a 2-counter Minsky machine with the counters $\set{x,y}$ using two task-phaser-chains 
    with the lengths of the chains capturing the values of each counter.
 The messages among the tasks are used to orchestrate the simulation  and are transmitted in the Boolean variables \texttt{temp, xInc, xDec, yInc,} and \texttt{yDec}. For instance,  $\mathtt{xInc}$ stands for incrementing the counter $x$. } 
\label{fig:thm-control-gen}
\end{figure}

  %
  %


\section{A gap-based symbolic representation}
\label{sec:symbolic}

The symbolic representation we propose builds on the following
intuitions.
First, observe the language semantics impose, for each phaser, the
invariant that signal values are always larger or equal to wait
values.
We can therefore assume this fact in our symbolic representation.
In addition, our reachability problems from
Sec.~\ref{sec:language} are defined in terms of reachability of
equivalence classes, not of individual configurations.
This is because configurations violating considered properties (see
Fig. \ref{fig:semantics-err}) are not defined in terms of concrete
phase values but rather in terms of relations among them (in addition
to the registration status, control sequences and variable values).
Finally, we observe that if a wait is enabled with smaller gaps on a
given phaser, then it will be enabled with larger ones.
We therefore propose to track the gaps of the differences between
ignal and wait values wrt. to an existentially quantified level (per
phaser) that lies between wait and signal values of all registered
tasks (to the considered phaser).

We formally define our symbolic representation and
describe a corresponding entailment relation.
We also establish a desirable property (namely that of
being a well-quasi-ordering) on some classes of representations.
This property is crucial for the decidability of certain
reachability problems (see. Sec.~\ref{sec:proc}). 

\paragraph{Named gaps.}
A {\em named gap} is associated to a task-phaser pair.
It consists in a tuple $(var,val)$ in
$\gaps=\left(
{\supscript{\phVarSet}{-,*} \times \left(
    \left(\nats^4
    \cup 
    \left(\nats^2\times\set{\infty}^2\right)\right)
    \cup
    \set{\nreg}\right)}
\right)$. 
Like for partial configurations in Sec.~\ref{sec:language},
$var\in\supscript{\phVarSet}{-,*}$ constrains variable values.
The $val$ value describes task registration to the phaser.
If registered, then $val$ is a 4-tuple $(\lwbw,\lwbs,\upbw,\upbs)$.
This intuitively captures, together with some level $l$ common to
all tasks registered to the considered phaser, all concrete wait and signal values $(\waitVal{}{},\sigVal{}{})$
satisfying $\lwbw \leq
(l-\waitVal{}{}) \leq \upbw$ and $\lwbs\leq (\sigVal{}{}-l)\leq\upbs$.
A named gap $(var,(\lwbw,\lwbs,\upbw,\upbs))$ is said to be
\emph{free} if $\upbw=\upbs=\infty$.
It is said to be \emph{$B$-gap-bounded}, for $B\in\nats$, if both $\upbw
\leq B$ and $\upbs \leq B$ hold.
A set $\gapSet\subseteq\gaps$ is said to be free (resp. $B$-gap-bounded) 
if all its named gaps are free (resp. $B$-gap-bounded).
The set $\gapSet$ is said to be \emph{$B$-good} if each one
of its named gaps is either free or $B$-gap-bounded.
Finally, $\gapSet$ is said to be \emph{good} if it is $B$-good for
some $B\in\nats$.
%
%
%
Given a set $\gapSet$ of named gaps, we define the partial order
$\bpreceq$ on $\gapSet$, and write $(var,val) \bpreceq
(var',val')$, to mean
(i) $\left(var\neq
var'\Rightarrow var=*\right)$, and (ii) $(val = \nreg) \iff (val' =
\nreg)$, and (iii) if $val=(\lwbw,\lwbs,\upbw,\upbs)$ and
$val'=(\lwbw',\lwbs',\upbw',\upbs')$ then $\lwbw \leq \lwbw'$, $\lwbs \leq \lwbs'$,
$\upbw' \leq \upbw$ and $\upbs' \leq \upbs$.
Intuitively, named gaps are used in the definition of constraints in
order to capture relations (i.e., reference, registration and possible
phases) of tasks and phasers.
The partial order $(var,val) \bpreceq (var',val')$ ensures the
relations allowed by $(var',val')$ are also allowed by $(var,val)$.

\paragraph{Constraints. }
A constraint $\sConf$ of $\program=\programTuple$ is a tuple
$\sConfTuple$ that denotes a possibly infinite set of
configurations. Intuitively, $\thidSet$ and $\phidSet$ respectively
represent a minimal set of tasks and phasers that are required in any configuration
denoted by the constraint. In addition:
\begin{itemize}
\item $\boolMp:\bVarSet\to\supscript{\bools}{*}$ and
  $\pcMp:\thidSet\to\supscript{\usuff}{*}$ respectively represent,
  like for partial configurations, a
  valuation of the Boolean variables and a mapping of tasks to their
  control sequences.
\item $\relgapMp:\thidSet\to \tFunctionsOf{\phidSet}{\gaps}$
  constrains relations between $\thidSet$-tasks and 
  $\phidSet$-phasers by associating to each task $\thid$ a mapping
  $\relgapMp(\thid)$ that defines for each phaser $\phid$ a named gap
  $(var,val)\in\gaps$ capturing the relation of $\thid$ and $\phid$.
\item $\envMp:\phidSet\to \nats^2$ associates \emph{lower bounds}
  $(\elwbw,\elwbs)$ on gaps of tasks that are registered to 
  $\phidSet$-phasers but which are not explicitly captured by
  $\thidSet$. This is described further in the deonotations of a
  constraint below.
\end{itemize}

We write $\isRegistred{\relgapMp}{\thid}{\phid}$ to mean the task $\thid$ is registered
to the phaser $\phid$, i.e., 
$\relgapMp(\thid)(\phid)\not\in (\supscript{\phVarSet}{-,*}\times\set{\nreg})$.
A constraint $\sConf$ is said to be free (resp. $B$-gap-bounded or $B$-good) if the set
$\gapSet=\setcomp{\relgapMp(\thid)(\phid)}{\thid\in\thidSet,
  \phid\in\phidSet}$ is free (resp. $B$-gap-bounded or $B$-good).
The dimension of a constraint is the number of its phasers (i.e.,
$|\phidSet|$).
A set of constraints $\sConfSet$ is said to be free, $B$-gap-bounded, $B$-good
or $K$-dimension-bounded if each of its constraints are. 

\noindent{\bf Denotations.}
We write $\cConf\models\sConf$ to mean constraint
$\sConf=\sConfTupleInd{\sConf}$
denotes configuration
$\cConf=\tuple{\thidSet_\cConf,\phidSet_\cConf,\boolMp_\cConf,\pcMp_\cConf,\relMp_\cConf}$.
Intuitively, the configuration $\cConf$ should have at least as many tasks (captured by a surjection
$\taskMp$ from a subset $\thidSet_\cConf^1$ of $\thidSet_\cConf$ to $\thidSet_\sConf$) and phasers
(captured by a bijection $\phaserMp$ from a subset $\phidSet_\cConf^1$ of $\phidSet_\cConf$ to
$\phidSet_\sConf$).
Constraints on the tasks and phasers in $\thidSet_\cConf^1$ and $\phidSet_\cConf^1$ ensure
target configurations are reachable.
Additional constraints on the tasks in $\thidSet_\cConf^2=\thidSet_\cConf\setminus\thidSet_\cConf^1$
ensure this reachability is not blocked by tasks not captured by $\thidSet_\sConf$.
More formally:

\begin{enumerate}
\item for each $\bvar\in\bVarSet$, $(\boolMp_\sConf(\bvar)\neq\boolMp_\cConf(\bvar))\implies(\boolMp_\sConf(\bvar)=*)$, and 
\item $\thidSet_\cConf$ and $\phidSet_\cConf$ can be written as
  $\thidSet_\cConf=\thidSet_\cConf^1\uplus \thidSet_\cConf^2$ and
    $\phidSet_\cConf=\phidSet_\cConf^1\uplus \phidSet_\cConf^2$, with
\item $\taskMp:\thidSet_\cConf^1\to\thidSet_\sConf$ is a surjection and
  $\phaserMp:\phidSet_\cConf^1\to\phidSet_\sConf$ is a bijection, and
\item for $\thid_\cConf \in \thidSet_\cConf^1$ with
  $\thid_\sConf=\taskMp(\thid_\cConf)$,
  $(\pcMp_\sConf(\thid_\sConf)\neq\pcMp_\cConf(\thid_\cConf))\implies(\pcMp_\sConf(\thid_\sConf)=*)$,
  and
\item for each $\phid_\sConf=\phaserMp(\phid_\cConf)$, there is a
  natural level $\lev:0\leq\lev$ such that:
  \begin{enumerate}
  \item if $\thid_\cConf \in \thidSet_\cConf^1$ with $\thid_\sConf=\taskMp(\thid_\cConf)$,
    $\relMp_\cConf(\thid_\cConf)(\phid_\cConf)=(var_\cConf,val_\cConf)$
    and $\relgapMp_\sConf(\thid_\sConf)(\phid_\sConf)=(var_\sConf,val_\sConf)$, then it is the case that:
    \begin{enumerate}
    \item $(var_\cConf \neq var_\sConf) \implies (var_\sConf=*)$, and
    \item $(val_\cConf=\nreg) \iff (val_\sConf=\nreg)$, and
    \item if $(val_\cConf=(\waitVal{}{},\sigVal{}{}))$ and
      $(val_\sConf=(\lwbw,\lwbs,\upbw,\upbs))$ then $\lwbw \leq \lev -
      \waitVal{}{} \leq \upbw$ and $\lwbs \leq \sigVal{}{} - \lev \leq
      \upbs$.
    \end{enumerate}    
  \item if $\thid_\cConf\in\thidSet_\cConf^2$, then for each
    $\phid_\sConf=\phaserMp(\phid_\cConf)$ with
    $\relMp_\cConf(\thid_\cConf)(\phid_\cConf)=(var_\cConf,(\waitVal{}{},\sigVal{}{}))$ 
    and $\envMp(\phid_\sConf)=(\elwbw,\elwbs)$, we have: 
    $(\elwbs\leq \sigVal{}{}-\lev)$ and $(\elwbw\leq
    \lev-\waitVal{}{})$
  \end{enumerate}
\end{enumerate}

We say in this case that $\taskMp$ and $\phaserMp$ witness the
denotation of $\cConf$ by $\sConf$.
Intuitively, for each phaser, the bounds given by $\relgapMp$
constrain the values of the phases belonging to tasks captured by
$\thidSet_\sConf$ (i.e., those in $\thidSet_\cConf^1$) and registered
to the given phaser.
This is done with respect to some non-negative level, one
per phaser.
The same level is used to constrain phases of tasks registered to the
phaser but not captured by $\thidSet_\sConf$ (i.e., those in
$\thidSet_\cConf^2$). For these tasks, lower bounds are enough as we only
want to ensure they do not block executions to target sets of
configurations. We write $\denotationOf{\sConf}$ for
$\setcomp{\cConf}{\cConf\models\sConf}$.

{\bf Entailment.}
We write $\sConf_a\entailedBy\sConf_b$
to mean 
$\sConf_a=\sConfTupleInd{a}$
is entailed by 
$\sConf_b=\sConfTupleInd{b}$.
This will ensure that configurations denoted by $\sConf_b$
are also denoted by $\sConf_a$.
Intuitively, $\sConf_b$ should have at least as many tasks (captured by a surjection
$\taskMp$ from a subset $\thidSet_b^1$ of $\thidSet_b$ to $\thidSet_a$) and phasers
(captured by a bijection $\phaserMp$ from a subset $\phidSet_b^1$ of $\phidSet_b$ to
$\phidSet_a$).
Conditions on tasks and phasers in $\thidSet_b^1$ and $\phidSet_b^1$ ensure
the conditions in $\sConf_a$ are met.
Additional conditions on the tasks in $\thidSet_b^2=\thidSet_b\setminus\thidSet_b^1$
ensure at least the $\envMp_a$ conditions in $\sConf_a$ are met. 
More formally:

\begin{enumerate}
  \item $(\boolMp_a(\bvar)\neq\boolMp_b(\bvar))\implies(\boolMp_a(\bvar)=*)$, for each $\bvar\in\bVarSet$ and 
  \item $\thidSet_b$ and $\phidSet_b$ can be written as $\thidSet_b=\thidSet_b^1\uplus \thidSet_b^2$ and $\phidSet_b=\phidSet_b^1\uplus \phidSet_b^2$ with
  \item $\taskMp:\thidSet_b^1\to\thidSet_a$ is a surjection  and   $\phaserMp:\phidSet_b^1\to\phidSet_a$ is a bijection, and
  \item $(\pcMp_b(\thid_b)\neq\pcMp_a(\thid_a))\implies(\pcMp_b(\thid_b)=*)$ for each
    $\thid_b \in \thidSet_b^1$ with $\thid_a=\taskMp(\thid_b)$, and
  \item for each phaser $\phid_a=\phaserMp(\phid_b)$ in $\phidSet_a$:
    \begin{enumerate}
    \item if $\envMp_a(\phid_a)=(\elwbw_a, \elwbs_a)$  and $\envMp_b(\phid_b)=(\elwbw_b, \elwbs_b)$ then $\elwbw_a \leq \elwbw_b$ and $\elwbs_a\leq \elwbs_b$
    \item  for each $\thid_b\in\thidSet_b^1$ with $\thid_a=\taskMp(\thid_b)$ and
      $\relgapMp_a(\thid_a)(\phid_a)=(var_a,val_a)$, and $\relgapMp_b(\thid_b)(\phid_b)=(var_b,val_b)$, it is the case that:
      \begin{enumerate}
      \item $(var_b \neq var_a) \implies (var_a=*)$, and
      \item $(val_b=\nreg) \iff (val_a=\nreg)$, and
      \item if $val_a=(\lwbw_a,\lwbs_a,\upbw_a,\upbs_a)$ and
        $val_b=(\lwbw_b,\lwbs_b,\upbw_b,\upbs_b)$, then $(\lwbw_a \leq
        \lwbw_b)$, $(\lwbs_a \leq \lwbs_b)$, $(\upbw_b \leq \upbw_a)$
        and $(\upbs_b \leq \upbs_a)$.
      \end{enumerate}
    \item for each $\thid_b\in\thidSet_b^2$ with
      $\relgapMp_b(\thid_b)(\phid_b)=(var,(\lwbw_a,\lwbs_a,\upbw_a,\upbs_a))$,
      with $\envMp_a(\phid_a)=(\elwbw_a,\elwbs_a)$, both $(\elwbw_a
      \leq \lwbw_b)$ and $(\elwbs_a \leq \lwbs_b)$ hold.
    \end{enumerate}
\end{enumerate}
We say in this case that $\taskMp$ and $\phaserMp$ witness the entailment of $\sConf_a$
by $\sConf_b$.

\begin{lemma}[Constraint entailment]
  \label{lem:cstr:entailment}
  $\sConf_a\entailedBy\sConf_b$ implies $\denotationOf{\sConf_b}\subseteq\denotationOf{\sConf_a}$
\end{lemma}
\begin{proof}
  Assume a configuration $\cConf=\tuple{\thidSet_\cConf,\phidSet_\cConf,\boolMp_\cConf,\pcMp_\cConf,\relMp_\cConf}$
  is denoted 
  by $\sConf_b=\tuple{\thidSet_b,\phidSet_b,\boolMp_b,\pcMp_b,\relgapMp_b,\envMp_a}$ with
  $\sConf_a\entailedBy\sConf_b$ and $\sConf_a=\tuple{\thidSet_a,\phidSet_a,\boolMp_a,\pcMp_a,\relgapMp_a,\envMp_a}$.
  We show $\cConf$ is also denoted by $\sConf_a$.

  By assumption, we can write $\thidSet_\cConf$ as a partition
  $\thidSet_\cConf^m\uplus \thidSet_\cConf^e$ and
  $\phidSet_\cConf$ as a partition
  $\phidSet_\cConf^m\uplus \phidSet_\cConf^e$ such that a surjection
  $\taskMp_\cConf:\thidSet_\cConf^m\to\thidSet_b$ and a bijection
  $\phaserMp_\cConf:\phidSet_\cConf^m\to\phidSet_b$ witness the
  denotation of $\cConf$ by $\sConf_b$.
  Also, we can write $\thidSet_b$ as a partition
  $\thidSet_b^m\uplus \thidSet_b^e$ and
  $\phidSet_b$ as a partition
  $\phidSet_b^m\uplus \phidSet_b^e$ such that a surjection
  $\taskMp_\sConf:\thidSet_b^m\to\thidSet_a$ and a bijection
  $\phaserMp_\sConf:\phidSet_b^m\to\phidSet_a$ witness the entailment of
  $\sConf_a$ by $\sConf_b$.
  Let us write $\thidSet_\cConf^m$ as the partition
  $\thidSet_\cConf^{m,m}\uplus\thidSet_\cConf^{m,e}$
  where $\thidSet_\cConf^{m,m}=\taskMp_\cConf^{-1}(\thidSet_b^m)$
  and $\phidSet_\cConf^m$ as the partition
  $\phidSet_\cConf^{m,m}\uplus\phidSet_\cConf^{m,e}$
  where $\phidSet_\cConf^{m,m}=\phaserMp_\cConf^{-1}(\phidSet_b^m)$.
  We define $\taskMp$ to be the restriction of
  $\taskMp_\sConf\circ\taskMp_\cConf$ to $\thidSet_\cConf^{m,m}$,
  i.e., $\taskMp:\thidSet_\cConf^{m,m} \to \thidSet_a$ with
  $\taskMp(\thid)=\taskMp_\sConf(\taskMp_\cConf(\thid))$ for each
  $\thid$ in $\thidSet_\cConf^{m,m}$.  Observe that $\taskMp$ is a
  well defined surjection.
  In addition, we write $\phaserMp$ to mean the restriction of 
  $\phaserMp_\sConf \circ \phaserMp_\cConf$ to $\phidSet_\cConf^{m,m}$.
  Observe that $\phaserMp$ is a well defined bijection.
  We show that $\taskMp$ and $\phaserMp$ witness the denotation of
  $\cConf$ by $\sConf_a$.

  \begin{enumerate}
  \item we have that, for each $\bvar\in\bVarSet$, both 
    $(\boolMp_b(\bvar)\neq\boolMp_\cConf(\bvar))\implies(\boolMp_b(\bvar)=*)$
    and 
    $(\boolMp_a(\bvar)\neq\boolMp_b(\bvar))\implies(\boolMp_a(\bvar)=*)$ hold.
    Hence,
    $(\boolMp_a(\bvar)\neq\boolMp_\cConf(\bvar))\implies(\boolMp_a(\bvar)=*)$ also holds.
  \item $\thidSet_\cConf=\thidSet_\cConf^{m,m}\uplus(\thidSet_\cConf^{m,e}\uplus \thidSet_\cConf^e)$
    and $\phidSet_\cConf=\phidSet_\cConf^{m,m}\uplus(\phidSet_\cConf^{m,e}\uplus \phidSet_\cConf^e)$, with
  \item $\taskMp:\thidSet_\cConf^{m,m} \to \thidSet_a$ is a surjection and
    $\phaserMp: \phidSet_\cConf^{m,m} \to \phidSet_a$ is a bijection, such that
  \item we have that
    $(\pcMp_b(\thid_b)\neq\pcMp_\cConf(\thid_\cConf))\implies(\pcMp_b(\thid_b)=*)$
    for each
    $\thid_\cConf \in \thidSet_\cConf^{m}$ with $\thid_b=\taskMp_\cConf(\thid_\cConf)$ in $\thidSet_b$, and
    $(\pcMp_a(\thid_a)\neq\pcMp_b(\thid_b))\implies(\pcMp_a(\thid_a)=*)$
    for each
    $\thid_b \in \thidSet_b^{m}$ with $\thid_a=\taskMp_\sConf(\thid_b)$ in $\thidSet_a$.
    Since $\thidSet_b^{m,m}\subseteq\thidSet_b^{m}$ and the
    surjection $\taskMp:\thidSet_\cConf^{m,m}\to\thidSet_b$ is the
    restriction of $\taskMp_\sConf\circ\taskMp_\cConf$ to
    $\thidSet_\cConf^{m,m}$,
    we deduce:
    $(\pcMp_a(\thid_a)\neq\pcMp_\cConf(\thid_\cConf))\implies(\pcMp_a(\thid_a)=*)$ for each
    $\thid_\cConf \in \thidSet_\cConf^{m,m}$ with $\thid_a=\taskMp(\thid_\cConf)$ in $\thidSet_a$.
  \item for each $\phid_a=\phaserMp_\sConf(\phid_b)=\phaserMp_\sConf(\phaserMp_\cConf(\phid_\cConf))=\phaserMp(\phid_\cConf)$,
    there is a $\lev:0\leq \lev$ s.t.: 
      \begin{enumerate}
  \item if $\thid_\cConf \in \thidSet_\cConf^{m,m}$ with $\thid_a=\taskMp_\sConf(\taskMp_\cConf(\thid_\cConf))$
    with $\relMp_\cConf(\thid_\cConf)(\phid_\cConf)=(var_\cConf,val_\cConf)$,
    $\relgapMp_b(\thid_b)(\phid_b)=(var_b,val_b)$
    and $\relgapMp_a(\thid_a)(\phid_a)=(var_a,val_a)$, then: 
    \begin{enumerate}
    \item we have $(var_a \neq var_b) \implies (var_a=*)$ and $(var_b \neq var_\cConf) \implies (var_b=*)$.
      Hence, $(var_a \neq var_\cConf) \implies (var_a=*)$.
    \item $(val_\cConf=\nreg) \iff (val_b=\nreg) \iff (val_a=\nreg)$, and
    \item if $(val_\cConf=(\waitVal{}{},\sigVal{}{}))$ and
      $(val_b=(\lwbw_b,\lwbs_b,\upbw_b,\upbs_b))$ and $(val_a=(\lwbw_a,\lwbs_a,\upbw_a,\upbs_a))$,
      then
      $\lwbw_a \leq \lwbw_b \leq \lev - \waitVal{}{} \leq \upbw_b \leq \upbw_a$
      and $\lwbs_a \leq \lwbs_b \leq \sigVal{}{} - \lev \leq \upbs_b \leq \upbs_a$.
    \end{enumerate}
      \item if $\thid_\cConf \in \thidSet_\cConf^{m,e}$ with $\thid_b=\taskMp_\cConf(\thid_\cConf)$
    with $\relMp_\cConf(\thid_\cConf)(\phid_\cConf)=(var_\cConf,val_\cConf)$,
    $\relgapMp_b(\thid_b)(\phid_b)=(var_b,val_b)$
    and $\envMp_a(\phid_a)=(\elwbw_a,\elwbs_a)$, then: 
    \begin{enumerate}
    \item if $(val_\cConf=(\waitVal{}{},\sigVal{}{}))$ and
      $(val_b=(\lwbw_b,\lwbs_b,\upbw_b,\upbs_b))$ and $\envMp_a(\phid_a)=(\elwbw_a,\elwbs_a)$
      then
      $\elwbw_a \leq \lwbw_b \leq \lev - \waitVal{}{}$
      and $\elwbs_a \leq \lwbs_b \leq \sigVal{}{} - \lev $.
    \end{enumerate}   
  \item if $\thid_\cConf\in\thidSet_\cConf^e$ with
    $\relMp_\cConf(\thid_\cConf)(\phid_\cConf)=(var_\cConf,(\waitVal{}{},\sigVal{}{}))$ 
    and $\envMp_b(\phid_b)=(\elwbw_b,\elwbs_b)$ and $\envMp_a(\phid_a)=(\elwbw_a,\elwbs_a)$, then:
    $\elwbs_a \leq \elwbs_b \leq \sigVal{}{} - \lev$ and
    $\elwbw_a \leq \elwbw_b \leq \lev - \waitVal{}{}$
  \end{enumerate}
  \end{enumerate}
\end{proof}

The remaining part of this section aims to establish the following theorem:

\begin{theorem}
  \label{thm:cstrs:wqo}
  $(\sConfSet,\entailedBy)$ is \wqo\ if $\sConfSet$ is $K$-dimension-bounded and $B$-good for some
  pre-defined $K,B\in\nats$. 
\end{theorem}

\noindent The idea is to propose an encoding
for each constraint $\sConf=\sConfTuple$ wrt. some arbitrary 
total orders $<_{\phidSet}$ and $<_{\thidSet}$.
We write $\encodingOf{\sConf}{<_{\thidSet}}{<_{\phidSet}}$ for the encoding of $\sConf$.
We also define an entailment relation $\epreceq$ on encodings.
Then, we show in Lem.~\ref{lem:epre-cpre} that 
$\encodingOf{\sConf}{<_{\thidSet}}{<_{\phidSet}}\epreceq\encodingOf{\sConf'}{<_{\thidSet'}}{<_{\phidSet'}}$
implies $\sConf\entailedBy\sConf'$.
Finally, we show in Lem.~\ref{lem:enc:wqo} that $\epreceq$ is
\wqo\ if the encoded $K$-dimension-bounded constraints are $B$-good for some pre-defined $K,B\in\nats$.
%
%
We start with the named gaps. It is not difficult to show the following lemma:
\begin{lemma}
  \label{lem:named-pairs}
  If $\gapSet$ is a good set of named gaps then $(\gapSet,\bpreceq)$
  is \wqo.
\end{lemma}

A {\em task state} of dimension $K$ is any tuple in
$(\usuff\times\gapSet^K)$ where $K$ is a natural in $\nats$
(corresponding to the number of phasers in the constraint to be encoded).
We write $(\seqVal,\gapElem_1,\ldots,\gapElem_K) \preceq
(\seqVal',\gapElem_1',\ldots,\gapElem_{K'}')$ for two task
states to mean that they have the same dimension (i.e.,
$K=K'$), that $\left(\seqVal\neq \seqVal'\Rightarrow \seqVal=*\right)$, and that
$\gapElem_k\bpreceq \gapElem_k'$ for each $k:1\leq k\leq K$.
Using Higman's lemma \cite{higman1952ordering} and
Lem.~\ref{lem:named-pairs}, we can show the following:

\begin{lemma}
    \label{lem:task-pairs}
    $((\usuff\times\gapSet^K),\apreceq)$ is \wqo\ in case $\gapSet$ is good. 
\end{lemma}

Let $\multisetsOf{\usuff\times\gapSet^K}$ be the set of finite multisets over
$(\usuff\times\gapSet^K)$.
We write $A\aepreceq B$, for $A$ and $B$ two multisets in
$\multisetsOf{\usuff\times\gapSet^K}$, if each element $a \in A$
can be mapped to an element $b\in B$ for which $a \apreceq b$.
By adapting Higman's lemma \cite{higman1952ordering} and using
Lem.~\ref{lem:task-pairs}, we can show the following lemma:

\begin{lemma}
  \label{lem:forall:exists}
  $(\multisetsOf{\usuff\times\gapSet^K},\aepreceq)$ is \wqo\ if $\gapSet$ is good. 
\end{lemma}

We write $A\eapreceq B$, for $A$ and $B$ finite multisets in
$\multisetsOf{\usuff\times\gapSet^K}$, to mean that each $b\in B$ can be mapped to 
some $a\in A$ for which $a \apreceq b$.
Rado's structure~\cite{Jancar99,Marcone2001} shows that
$(\multisetsOf{S},\eapreceq)$ need not be \wqo\ just
because $\preceq$ is $\wqo$ over $S$.
Still, we establish the following result:


\begin{lemma}
  \label{lem:exists:forall}
  $(\multisetsOf{\usuff\times\gapSet^K},\eapreceq)$ is \wqo\ if $\gapSet$ is good. 
\end{lemma}
\begin{proof}
We proceed by contradiction.
Assume, without loss of generality, an infinite sequence $\sqset{A_1,
  A_2, \ldots}$ of $\apreceq$-minimal multisets in $\multisetsOf{\usuff\times\gapSet^K}$ such
that $A_j \neapreceq A_i$ for all $1\leq j < i$.
Notice that: 
\begin{enumerate}
\item for each $i:1\leq j < i$, we can identify an element $\jnoti{a}{i}{j}\in A_i$ such that
$a_j \not\apreceq \jnoti{a}{i}{j}$ for any $a_j \in A_j$. We sometimes write
$\jnoti{a}{i}{j}$ as $(\seqVal^{\neg j}_i,{\jnoti{a}{i}{j}}[1], {\jnoti{a}{i}{j}}[2], \ldots, {\jnoti{a}{i}{j}[K]})$,
where $\jnoti{a}{i}{j}[k]$ is the $k^{th}$ component of $\jnoti{a}{i}{j}$.
\item by the definition of $(\usuff\times\gapSet^K,\apreceq)$ and the
  fact that it is a \wqo, we can extract a subsequence $\sqset{A_{i_1},
    A_{i_2}, \ldots}$ of $\sqset{A_1,A_2,\ldots}$ associated to a sequence of partial mappings
  $(b_{j}:\set{1, \ldots K}\to \nats)$ s.t. for each ${j:1\leq j}$:
\begin{enumerate}
\item the $j$-sequence $\jnoti{a}{i_1}{j} \apreceq \jnoti{a}{i_2}{j} \apreceq \jnoti{a}{i_3}{j} \apreceq \ldots$ is $\apreceq$-monotone,
\item each $j$-sequence has a constant control, i.e., $\seqVal^{\neg j}=\seqVal^{\neg j}_{i_1}=\seqVal^{\neg j}_{i_2}=\ldots$, and
\item for each dimension $k:1\leq k\leq K$, the projection of a $j$-sequence $\jnoti{a}{i_1}{j} \apreceq \jnoti{a}{i_2}{j} \apreceq \jnoti{a}{i_3}{j} \apreceq \ldots$
  on  $k$ is either constant or strictly increasing, i.e.,:
  \begin{enumerate}  
  \item $k\in \domainOf{b_j}$ implies
    $\jnoti{a}{i}{j}[k]=b_j(k)$ for all $i:j<i$
  \item $k\not\in\domainOf{b_j}$ implies
    $\jnoti{a}{i}{j}[k] \bpreceq \jnoti{a}{i'}{j}[k]$ but $\jnoti{a}{i'}{j}[k]\not\!\!\bpreceq \jnoti{a}{i}{j}[k]$ for any $i,i':1\leq j < i < i'$
    \end{enumerate}
\end{enumerate}
\end{enumerate}

We rename the sequence $\sqset{A_{i_1}, A_{i_2}, \ldots}$ into $\sqset{A_{1}, A_{2}, \ldots}$.
We obtain a $\eapreceq$-bad sequence $\sqset{A_1,A_2,\ldots}$ of $\apreceq$-minimal sets
satisfying the constraints depicted in the following figure:

\begin{center}
\begin{tikzpicture}
\scriptsize
\foreach \i in {2, ..., 7}
{
   \pgfmathtruncatemacro{\x}{(\i-1)}
   \node at (\x , -0.25) {$A_{\i}$};
}

\foreach \j in {1, ..., 6}
{
   \pgfmathtruncatemacro{\y}{- \j }
   \node at (0.3 , 0.75*\y) {$\neg A_{\j}$};
}

\foreach \j in {1, ..., 6}
{
  \pgfmathtruncatemacro{\istart}{\j+1}
  \foreach \i in {\istart, ..., 7}
  {
     \pgfmathtruncatemacro{\x}{(\i-1)}
     \pgfmathtruncatemacro{\y}{-\j}  
     \node at (\x , 0.75*\y) {$\jnoti{a}{\i}{\j}$};
     \node at (\x + 0.5 , 0.75*\y) {$\apreceq$};       
  }
}

\node at (7 , -0.25) {$\ldots$};
\node at (8 , -0.25) {$\ldots$};   

\foreach \j in {1, ..., 7}
{
  \foreach \i in {8.5,9.5}
  {
    \pgfmathtruncatemacro{\x}{(\i-1)}
    \pgfmathtruncatemacro{\y}{-\j}  
    \node at (\x , 0.75*\y) {$\ldots$};
  }
}

\node at (7.5 , -0.25) {$A_{i}$};
\foreach \j in {1, ..., 6}{
  \foreach \i in {8}
  {
     \pgfmathtruncatemacro{\x}{(\i-1)}
     \pgfmathtruncatemacro{\y}{-\j}  
    \node at (\x + 0.5 , 0.75*\y) {$\jnoti{a}{i}{\j}$};

  }
}

\node at (7.5 , -5.25) {$\jnoti{a}{i}{j}$};
\node at (0.3 , -5.25) {$\neg A_{j}$};
\end{tikzpicture}
\end{center}

Some observations about the
$\jnoti{a}{i}{j}$ elements:

\begin{enumerate}
\item For any $i,j:1\leq j < i $, $\jnoti{a}{i}{j}\in A_i$  and 
$a_j\not\npreceq \jnoti{a}{i}{j}$ for any $a_j \in A_j$.

\item For any $i,i',j:1\leq j < i < i' $, $\jnoti{a}{i}{j}\apreceq \jnoti{a}{i'}{j}$ with
$\jnoti{a}{i}{j}[k]=\jnoti{a}{i'}{j}[k]$ iff $k\in \domainOf{b_j}$
\item For any $i,j,j':1 \leq j' < j < i$ and $i':j \leq i' $:
$\jnoti{a}{i'}{j'} \not\apreceq \jnoti{a}{i}{j}$. 
\end{enumerate}

Observations (1) and (2) are obtained by construction.  Suppose
observation (3) does not hold, i.e., suppose $\jnoti{a}{i'}{j'}
\apreceq \jnoti{a}{i}{j}$ for some $i,j,j':1\leq j' < j < i$ and $i':j
\leq i'$.  Observation (2) and reflexivity of $\apreceq$ ensure
$\jnoti{a}{j}{j'} \apreceq \jnoti{a}{i'}{j'}$.  By transitivity of
$\apreceq$, we get $\jnoti{a}{j}{j'} \apreceq \jnoti{a}{i}{j}$. This
contradicts observation (1).  Hence observation (3) also holds.

Now, observe that the domain $D_j$ of $b_j$, for any $j:1\leq j$, is a
subset of $\set{1,\ldots,K}$.
Since this domain is finite, there is an infinite number of indices
$j:1\leq j$ with the same domain $D\subseteq\set{1,\ldots K}$.
In other words, an infinite number of $j$-sequences are constant (with
possibly different values) on the same dimensions $D$.
We can therefore extract a sequence $\sqset{j_1,j_2,\ldots}$ of
$j$-sequences that is increasing on each dimension in $D$ (i.e.,
$j_a\leq j_b \implies \textrm{ for each } k\in D. b_{j_a}(k)\leq
b_{j_b}(k)$).
In order for observation (3) to hold (i.e., $\jnoti{a}{i'}{j'} \not\apreceq \jnoti{a}{i}{j}$ for
any $i,j,j':1 \leq j' < j < i$ and $i':j \leq i'$), we need to have some dimensions
on which the $j$-sequences $(\jnoti{a}{i}{j})_{j_1,j_2,\ldots}$ do not increase, these would be
the dimensions in $\set{1,\ldots K} \setminus D$.
But for these dimensions, $(\jnoti{a}{i}{j})_{i:1,2,\ldots}$ is strictly increasing with $i$.
This again contradicts observation (3). 
\end{proof}

Lem.~\ref{lem:exists:forall} will be used in Lem.~\ref{lem:enc:wqo} to
show an entailement relation on encodings of $K$-dimension-bounded and
$B$-good constraints corresponds to a stronger relation than
$\entailedBy$ and is WQO, hence establishing Thm. \ref{thm:cstrs:wqo}.
First, we introduce constraints encodings and an entailement relation
on them.

\paragraph{Encodings of constraints.}

  Given a finite set $Q$ and an associated total order $<_Q$, we write
  $\indexOf{i}{Q}\in Q$ to mean the element of $Q$ with $<_Q$-index
  $i\in\set{1,\ldots |Q|}$\footnote{the $<_Q$-index of an element $q$
    in $Q$ is $1+|\setcomp{q'}{q'\in Q \textrm{ and } q' <_Q q}|$}.
  For instance, given a finite set of phasers $\phidSet$ and
  an associated total order $<_\phidSet$, 
  we write $\indexOf{k}{\phidSet}$ to mean the phaser with $<_\phidSet$-index
  $k$ in $\phidSet$. 
%
%
The encoding of a constraint $\sConf=\sConfTuple$ with respect to
total orders $<_{\phidSet}$ and $<_{\thidSet}$,
written $\encodingOf{\sConf}{<_{\thidSet}}{<_{\phidSet}}$, is a tuple
$\encTuple$ where:

\begin{enumerate}
\item $\boolMp:\bVarSet\to\supscript{\bools}{*}$ is the same as in $\sConf$,
\item $\acc:\set{1,\ldots,|\thidSet|}\to(\usuff\times \gapSet^{|\phidSet|})$
  associates each task $\indexOf{i}{\thidSet}$ to
  a tuple 
  $\acc(i)=(\seqVal,\gapElem_1,\ldots,\gapElem_{|\phidSet|})$
  where $\pcMp(\indexOf{i}{\thidSet})=\seqVal$ and $\relgapMp(\indexOf{i}{\thidSet})(\indexOf{j}{\phidSet})=\gapElem_j$
  for each phaser $\indexOf{j}{\phidSet}$ with index  $j$ in $\phidSet$,
\item $\envAcc:\set{1,\ldots |\phidSet|}\to\nats^2$ associates to each
  phaser $\indexOf{j}{\phidSet}$ in $\phidSet$ the
  pair $\envAcc(j)=\envMp(\indexOf{j}{\phidSet})$.
\end{enumerate}

%
Observe that if two constraints result in the same encoding, then they
can be obtained from each other by renaming the tasks and the phasers.
As a consequence, if a constraint is free (resp. $B$-gap-bounded or $B$-good),
then all constraints resulting in the same encoding will
also be free (resp., $B$-gap-bounded or $B$-good).
We define the dimension of an encoding $\encTuple$ to be the size of
the domain of $\envAcc$ (i.e., the dimension of an encoded constraint).
A (possibly infinite) set of encodings $E$ is said to be free
(resp. $B$-gap-bounded or $B$-good) if all constraints encoded by any
of its elements are free (resp. $B$-gap-bounded or $B$-good).
The set is said to be $K$-dimension-bounded if there is natural $K$ in
$\nats$ that is larger than the dimension of any of its elements.

\paragraph{Entailment of encodings.}
Assume two encodings $\encTuple$ and $\pencTuple$
with
$\acc:\set{1,\ldots,L}\to(\usuff\times \gapSet^{M})$,
$\envAcc:\set{1,\ldots M}\to\nats^2$,
$\acc':\set{1,\ldots,L'}\to(\usuff\times \gapSet^{M'})$,
$\envAcc:\set{1,\ldots M'}\to\nats^2$.
Write $\encTuple \epreceq \pencTuple$ iff: 

\begin{enumerate}
\item for each $\bvar\in\bVarSet$, $(\boolMp(\bvar)\neq\boolMp'(\bvar))\implies(\boolMp(\bvar)=*)$, and

\item $M'=M$ and there is a surjection $h:\set{1,\ldots,L'}\to\set{1,\ldots,L}$
  such that:
  \begin{enumerate}
  \item $\acc(h(i)) \bpreceq \acc'(i)$ for each index $i\in\set{1,\ldots,L'}$,
  \item $\envAcc(j) \leq_2 \envAcc'(j)$ for each index $j\in\set{1,\ldots,M'}$,
  \end{enumerate}
\end{enumerate}


\begin{lemma}
  \label{lem:epre-cpre}
  Let $\sConf=\sConfTuple$ and $\sConf'=\psConfTuple$.
  If $\encodingOf{\sConf}{<_\thidSet}{<_\phidSet}=(\boolMp,\acc,\envAcc)$ and
  $\encodingOf{\sConf'}{<_{\thidSet'}}{<_{\phidSet'}}=(\boolMp',\acc',\envAcc')$,
  then
$(\boolMp,\acc,\envAcc) \epreceq~(\boolMp',\acc',\envAcc')$ implies
$\sConf\entailedBy\sConf'$.
\end{lemma}

\begin{proof}
  From $(\boolMp,\acc,\envAcc) \epreceq~(\boolMp',\acc',\envAcc')$ we deduce
  $|\phidSet|=|\phidSet'|$ and
  the existence of
  a surjection $h:\set{1,\ldots,|\thidSet'|} \to
  \set{1,\ldots,|\thidSet|}$, such that
  $\acc(h(i))\bpreceq\acc'(i)$ for each $i\in\set{1,\ldots,|\thidSet'|}$, and
  $\envAcc(j) \leq_2 \envAcc'(j)$ for each $j\in \set{1,\ldots,|\phidSet'|}$.
  
\begin{enumerate}
\item $\boolMp=\boolMp'$, hence, for each $\bvar\in\bVarSet$,
  $(\boolMp(\bvar) \neq \boolMp'(\bvar) \implies \boolMp(\bvar)=*)$.
\item let $\tau:\thidSet'\to\thidSet$ with
  $\tau(\indexOf{i}{\thidSet'})=\indexOf{h(i)}{\thidSet}$ for each
  $\indexOf{i}{\thidSet'}$ in $\thidSet'$. Let
  $\pi:\phidSet'\to\phidSet$ with
  $\pi(\indexOf{j}{\phidSet'})=\indexOf{j}{\phidSet}$ for each
  $\indexOf{j}{\phidSet'}$ in $\phidSet'$.  Observe $\tau$ is surjective and $\pi$ is
  bijective.
\item for each $i\in\set{1,\ldots,|\thidSet'|}$, we have that $\acc(h(i))\bpreceq\acc'(i)$. By definition,
  $\acc(h(i))$ is the tuple $(\pcMp(\indexOf{h(i)}{\thidSet}),\relgapMp(\indexOf{h(i)}{\thidSet})(\indexOf{1}{\phidSet}), \ldots \relgapMp(\indexOf{h(i)}{\thidSet})(\indexOf{|\phidSet|}{\phidSet}))$
  and   $\acc'(i)$ is the tuple $(\pcMp'(\indexOf{i}{\thidSet'}),\relgapMp(\indexOf{i}{\thidSet'})(\indexOf{1}{\phidSet'}), \ldots \relgapMp(\indexOf{i}{\thidSet'})(\indexOf{|\phidSet'|}{\phidSet'}))$
  where $\phidSet=\set{\indexOf{1}{\phidSet},\ldots,\indexOf{|\phidSet|}{\phidSet}}$ and $\phidSet'=\set{\indexOf{1}{\phidSet'},\ldots,\indexOf{|\phidSet'|}{\phidSet'}}$.
  By definition of $\bpreceq$, we get:
  \begin{enumerate}
  \item $\pcMp(\indexOf{h(i)}{\thidSet})\neq\pcMp'(\indexOf{i}{\thidSet'})\implies\pcMp(\indexOf{h(i)}{\thidSet})=*$. Since
  $\tau(\indexOf{i}{\thidSet'})=\indexOf{h(i)}{\thidSet}$, we get $\pcMp(\tau(\indexOf{i}{\thidSet'}))\neq\pcMp'(\indexOf{i}{\thidSet'})\implies\pcMp(\tau(\indexOf{i}{\thidSet'}))=*$
    for  $\indexOf{i}{\thidSet'}$ in $\thidSet'$.
  \item $\relgapMp(\indexOf{h(i)}{\thidSet})(\indexOf{j}{\phidSet}) \bpreceq \relgapMp'(\indexOf{i}{\thidSet'})(\indexOf{j}{\phidSet'})$ for each $j\in\set{1,\ldots,|\phidSet'|}$.
    Since $\indexOf{j}{\phidSet'}=\indexOf{j}{\phidSet}$, and if $\relgapMp(\indexOf{h(i)}{\thidSet})(\indexOf{j}{\phidSet})=(var,val)$ and
    $\relgapMp'(\indexOf{i}{\thidSet'})(\indexOf{j}{\phidSet'})=(var',val')$,
    we deduce that:
    \begin{enumerate}
    \item $(var \neq var') \implies (var=*)$
    \item $(val=\nreg)\iff(val'=\nreg)$
    \item if $val=(\lwbw,\lwbs,\upbw,\upbs)$ and
      $val'=(\lwbw',\lwbs',\upbw',\upbs')$, then
      $\lwbw \leq \lwbw'$ and $\lwbs \leq \lwbs'$ and $\upbw'\leq \upbw$ and $\upbs'\leq\upbs$      
      \end{enumerate}
  \end{enumerate}
\item for each $j\in\set{1,\ldots,|\phidSet'|}$, we have $\envAcc(j) \leq_2 \envAcc'(j)$. By definition,
  $\envAcc(j)=\envMp(\indexOf{j}{\phidSet})$ and $\envAcc'(j)=\envMp'(\indexOf{j}{\phidSet'})$. Since $\pi(\indexOf{j}{\phidSet'})=\indexOf{j}{\phidSet}$, we deduce that
  $\envMp(\indexOf{j}{\phidSet}) \leq_2 \envMp(\indexOf{j}{\phidSet'})$ for each $j\in \set{1,\ldots,|\phidSet'|}$.
\end{enumerate}

\begin{lemma}
  \label{lem:enc:wqo}
  $(E,\epreceq)$ is \wqo\ if the set $E$ of encodings is $K$-dimension-bounded and $B$-good for some
  pre-defined $K,B\in\nats$.
\end{lemma}

\begin{proof}
  Assume a $K$-dimension-bounded set $E$ of
  $B$-good encodings and an infinite sequence
  $S_1=\sqset{(\boolMp_1,\acc_1,\envAcc_1),(\boolMp_2,\acc_2,\envAcc_2),\ldots}$.
  We show the existence of $i,j: 1\leq i < j$ for which
  $(\boolMp_i,\acc_i,\envAcc_i)\epreceq(\boolMp_j,\acc_j,\envAcc_j)$.
Dimension-boundedness of $E$ ensures there are infinitely many
encodings in $S_1$ with the same dimension, say $K$.
We extract the subsequence $S_2$ consisting in all encodings with
dimension $K$ in $S_1$.
In addition, observe that the set of possible valuations of the
Boolean variables is finite.
We can therefore extract from $S_2$ an infinite subsequence $S_3$ where all
elements share the same valuation of the Boolean variables.
Let us rewrite $S_3$, for simplicity, as the sequence
$\sqset{(\boolMp_1,\acc_1,\envAcc_1),(\boolMp_2,\acc_2,\envAcc_2),\ldots}$.
For each $i:1\leq i$, we can represent the mapping $\envAcc_i$ as the tuple
$\tuple{\envAcc_i(1), \envAcc_i(2), ,\ldots, \envAcc_i(K)}$ in $(\nats^{2})^K$.
Using Higman's lemma, we can extract from $S_3$ a subsequence $S_4$, also renamed
to $\sqset{(\boolMp_1,\acc_1,\envAcc_1),(\boolMp_2,\acc_2,\envAcc_2),\ldots}$ for simplicity,
where $\envAcc_i(k)\leq_2\envAcc_j(k)$ for any $i,j: 1\leq i < j$ and $k:1\leq k\leq K$.

For each mapping $\acc_i:\set{1,\ldots,L_i}\to(\usuff\times
\gapSet^{K})$ in $S_4$, we write $m_{\acc_i}$ to mean the multiset over
$(\usuff\times \gapSet^{K})$ where the number of occurrences of an
element $(\seqVal,\gapElem_1,\ldots,\gapElem_{K})$ coincides with the
number of indices $j$ in $\set{1,\ldots, L_i}$ for which
$\acc_i(j)=(\seqVal,\gapElem_1,\ldots,\gapElem_{K})$.
Consider the sequence
$\sqset{m_{\acc_1},m_{\acc_2},m_{\acc_3},\ldots}$ of elements in
$\multisetsOf{\usuff\times \gapSet^{K}}$. Using the fact that $E$ is
$B$-good together with Lem.~\ref{lem:forall:exists} and
\ref{lem:exists:forall}, we deduce the existence of $i,j:i<j$ for which
$m_{\acc_i} \aepreceq m_{\acc_j}$ and $m_{\acc_i} \eapreceq
m_{\acc_j}$.
We can therefore build a surjection $h:\set{1,\ldots,L_j}\to\set{1,\ldots,L_i}$
such that $\acc_i(h(l))\bpreceq\acc_j(l)$ for each $l$ in $\set{1,\ldots,L_j}$.
\end{proof}

\end{proof}

\section{A symbolic verification procedure}
\label{sec:proc}

We use the constraints from Sec.~\ref{sec:symbolic} as a symbolic
representation in the adaptation of the classical working-list based
backward procedure described below.
This procedure corresponds to
an instantiation of the framework
of Well-Structured-Transition-Systems
\cite{coverability,wsts:everywhere}.
The procedure takes as arguments a program $\program$ and a
$\entailedBy$-minimal set $\sConfSet$ of constraints denoting the
targeted set of configurations.
%
%
Such sets can be easily built from the partial configurations
described in Fig.~\ref{fig:semantics-err}.

The predecessor computation rules in Figures \ref{fig:pre:new} and \ref{fig:pre:asynch} that are used by the procedure need to first concretize the given constraint to explicitly contain the required task(s), phaser(s), phaser variable(s), and control sequence(s). For this purpose, they make use of the rules $\conc$ (a wrapper for $\conctask$ and $\concphaser$), $\concseq$, and $\concvar$. Each concretization rule returns a (possibly empty) set of concrete constraints. Intuitively, $\conctask(\sConf,\thid,\uniquely,\fresh)$ makes sure either $\thid$ that is going to be used in the computation is already in $\sConf$, or adds it as a fresh task. $\concphaser(\sConf,\phid,\uniquely,\fresh)$ concretizes the phaser $\phid$. $\concseq(\sConf,\thid,\seqVal)$ makes sure the sequence of the task $\thid$ is $\seqVal$, which is a requirement for the predecessor computation rule. Otherwise, it will return an empty set. Finally, $\concvar(\sConf,\thid,\phid,\phvar)$ ensures $\thid$ uses the variable $\phvar$ to reference $\phid$, which is again required by the predecessor computation rule. 
In the following sections we introduce and then prove some of the characteristics of the concretization functions, predecessor computation functions, and the verification procedure.

\begin{figure*}
\begin{center}
\scriptsize
$$ 
\begin{array}{c}
\infer[\left(\begin{array}{c}\mathtt{concretize}\\\mathtt{seq~1}\end{array}\right)]
      {
        \sConfTuple \in \concseq(\sConfTuple,\thid, \seqVal)
      }
      {
        \begin{array}{c}
          \thid\in\thidSet  \quad  \pcMp(\thid)=\seqVal
        \end{array}  
      }
      \\
      \\
\infer[\left(\begin{array}{c}\mathtt{concretize}\\\mathtt{seq~2}\end{array}\right)]
      {
        \tuple{\thidSet,\phidSet,\boolMp,\pcMp',\relgapMp,\envMp} \in \concseq(\sConfTuple,\thid,\seqVal)
      }
      {
        \begin{array}{c}
          \thid\in\thidSet \quad
          \pcMp(\thid)=* \quad \pcMp'=\mapSubst{\pcMp}{\thid}{\seqVal} 
        \end{array}  
      }
      \\
      \\
\infer[\left(\begin{array}{c}\mathtt{concretize}\\\mathtt{task~1}\end{array}\right)]
      {
        (\sConfTuple,\uniquely,\fresh) \in \conctask(\sConfTuple,\thid,\uniquely,\fresh)
      }
      {
        \begin{array}{c}
          \thid\in\thidSet 
        \end{array}  
      }
      \\
      \\
\infer[\left(\begin{array}{c}\mathtt{concretize}\\\mathtt{task~2}\end{array}\right)]
      {
 (\tuple{\thidSet',\phidSet,\boolMp,\pcMp',\relgapMp',\envMp},\uniquely',\fresh') \in \conctask(\sConfTuple,\thid,\uniquely,\fresh)
      }
      {
        \begin{array}{c}
          \thid\in\fresh \quad \othid\in\thidSet\setminus\uniquely \quad \thidSet'=\thidSet\cup \set{\thid}     
        \\
        \fresh'=\fresh\setminus \set{\thid} \quad \uniquely'=\uniquely\cup \set{\thid}
        \quad
       \pcMp'= \mapSubst{\pcMp}{\thid}{\pcMp(\othid)} 
          \\
         \relgapMp'=\mapSubst{\relgapMp}{\thid}{\mapSubst{\relgapMp(\othid)}{\phid}{\relgapMp(\othid)(\phid)}~|~ \phid\in\phidSet}
       \end{array}
      }
      \\
      \\
\infer[\left(\begin{array}{c}\mathtt{concretize}\\\mathtt{task~3}\end{array}\right)]
      {
 (\tuple{\thidSet',\phidSet,\boolMp,\pcMp',\relgapMp',\envMp},\uniquely',\fresh') \in \conctask(\sConfTuple,\thid,\uniquely,\fresh)
      }
      {
        \begin{array}{c}
          \thid\neq\thid' \quad \set{\thid,\thid'}\subseteq\fresh \quad \othid\in\thidSet\setminus\uniquely \quad \thidSet'=(\thidSet\setminus\set{\othid})\cup \set{\thid,\thid'}     
        \\
        \fresh'=\fresh\setminus \set{\thid,\thid'} \quad \uniquely'=\uniquely\cup \set{\thid,\thid'}
        \quad
       \pcMp'= \mapSubst{\pcMp}{\thid_i}{\pcMp(\othid)~|~ \thid_i\in \set{\thid,\thid'}} 
          \\
         \relgapMp'=\mapSubst{\relgapMp}{\thid_i}{\mapSubst{\relgapMp(\othid)}{\phid}{\relgapMp(\othid)(\phid)~|~ \phid\in\phidSet}~|~  \thid_i\in \set{\thid,\thid'}}
       \end{array}
      }
      \\
      \\
\infer[\left(\begin{array}{c}\mathtt{concretize}\\\mathtt{task~4}\end{array}\right)]
      {
        (\tuple{\thidSet',\phidSet,\boolMp,\pcMp',\relgapMp',\envMp},\uniquely\cup\set{\thid},\fresh\setminus\set{\thid}) \in \conctask(\sConfTuple,\thid,\uniquely,\fresh)
      }
      {
        \begin{array}{c}
          \thid\in\fresh \quad \thid\in\fresh \quad 
          \thidSet'=\thidSet\cup\set{\thid} \\
          \phidSet=\phidSet_1\uplus\phidSet_2 
          \textrm{ for each }
          \phid_2\in\phidSet_2.~\envMp(\phid_2)=(\elwbw_{\phid_2},\elwbs_{\phid_2})
          \\
          \relgapMp'=\mapSubst{\relgapMp}{\thid}{map_1 \cup map_2}
          \quad map_1 = \setcomp{\phid_1 \leftarrow
            (*,(\nreg))}{\phid_1\in\phidSet_1} \\
          map_2 =
          \setcomp{\phid_2 \leftarrow
            (*,(\elwbw_{\phid_2},\elwbs_{\phid_2},\infty,\infty))}
                  {\phid_2\in\phidSet_2} \quad \pcMp'=\mapSubst{\pcMp}{\thid}{*}
        \end{array}  
      }
      \\
      \\
\infer[\left(\begin{array}{c}\mathtt{concretize}\\\mathtt{phaser~1}\end{array}\right)]
      {
        (\sConfTuple,\uniquely,\fresh) \in \concphaser(\sConfTuple,\phid,\uniquely,\fresh)
      }
      {
        \begin{array}{c}
          \phid\in\phidSet
        \end{array}  
      }
      \\
      \\
\infer[\left(\begin{array}{c}\mathtt{concretize}\\\mathtt{phaser~2}\end{array}\right)]
      {
        (\tuple{\thidSet',\phidSet',\boolMp,\pcMp,\relgapMp',\envMp'},\uniquely\cup(\fresh\setminus\fresh'),\fresh\setminus\thidSet') \in \concphaser(\sConfTuple,\phid,\uniquely,\fresh)
      }
      {
        \begin{array}{c}
          \phid\not\in\phidSet \quad          
 \phidSet' = \phidSet\cup\set{\phid}\quad
 \envMp'=\mapSubst{\envMp}{\phid}{(0,0)}
 \\
 \thidSet=\thidSet_1\cup\thidSet_2 \text{ with } \thidSet_1\cap\thidSet_2 \text{ possibly non-empty and } \thidSet_1\cap\thidSet_2\cap\uniquely=\emptyset
 \\
 \thidSet'=\thidSet_1'\cup\thidSet_2'\quad 
 \thidSet_1'\cap\thidSet_2'=\emptyset  \\
  \taskMp_1: \thidSet_1\rightarrow\thidSet_1' \quad \taskMp_2: \thidSet_2\rightarrow\thidSet_2'\quad \taskMp_1,\taskMp_2 \text{ are bijective functions}
 \\        
 \forall\thid\in\thidSet_1\setminus\thidSet_2.~\taskMp_1(\thid)=\thid \quad \forall\thid\in\thidSet_1\cap\thidSet_2.~\taskMp_1(\thid)\in\set{(\thid,1)}\cup\fresh
 \\
 \forall\thid\in\thidSet_2\setminus\thidSet_1.~\taskMp_2(\thid)=\thid \quad \forall\thid\in\thidSet_1\cap\thidSet_2.~\taskMp_2(\thid)\in\set{(\thid,2)}\cup\fresh
 \\
 \taskMp_1(\thid)=\thid_1 \quad \taskMp_2(\thid)=\thid_2 \\
 \forall\thid\in\thidSet_1,\forall\ophid\in\phidSet. ~\relgapMp'(\thid_1)(\ophid)=\relgapMp(\thid)(\ophid) \\
 \forall\thid\in\thidSet_2,\forall\ophid\in\phidSet. ~\relgapMp'(\thid_2)(\ophid)=\relgapMp(\thid)(\ophid) \\
 \forall\thid\in\thidSet_1.~ \relgapMp'(\thid_1)(\phid)=(*,\nreg) \\
 \forall\thid\in\thidSet_2.~  \relgapMp'(\thid_2)(\phid)=(*,(0,0,\infty,\infty)) 
        \end{array}  
      }
      \\
      \\
      \infer[\left(\begin{array}{c}\mathtt{concretize}\\\mathtt{phaser~var~1}\end{array}\right)]
      {
        \sConfTuple \in \concvar(\sConfTuple,\thid,\phid,\phvar)
      }
      {
        \begin{array}{c}  
          \thid\in\thidSet \quad \phid\in\phidSet \quad
          \relgapMp(\thid)(\phid)=(\phvar,\val)
        \end{array}
      } 
      \\
      \\
\infer[\left(\begin{array}{c}\mathtt{concretize}\\\mathtt{phaser~var~2}\end{array}\right)]
      {
        \tuple{\thidSet,\phidSet,\boolMp,\pcMp,\relgapMp',\envMp} \in \concvar(\sConfTuple,\thid,\phid,\phvar)
      }
      {
        \begin{array}{c}  
          \thid\in\thidSet \quad \phid\in\phidSet \quad
          \relgapMp(\thid)(\phid)=(*,\val)
          \\
          \forall~\ophid\in\phidSet.~ \left(
          \relgapMp(\thid)(\ophid)=(var,\val)
          \Rightarrow var\neq \phvar\right)
          \\
          \relgapMp'(\thid)(\phid)=
          \mapSubst{\relgapMp}{\thid}{\mapSubst{\relgapMp(\thid)}{\phid}{(\phvar,\val)}}
        \end{array}
      } 
\end{array}
$$
\end{center}
\caption[Auxiliary functions used in the predecessor computation of the parameterized Phaser instructions (Part I)]{Auxiliary functions used in the predecessor computation of the parameterized Phaser instructions (Part I). The function $\conc(\sConf,A,B,\uniquely,\fresh)$ concretizes the set of phasers $A$ and the set of tasks $B$ in the constraint $\sConf$. Those  phasers or tasks in $A$ or $B$ that are already in $\sConf$ will be preserved. $\uniquely$ is a subset of $B$, which is already concretized to unique tasks. $\fresh$ is a set of fresh task identifiers that does not intersect $B$.}
\label{fig:pre:aux}
\end{figure*}

\begin{figure*}
\begin{center}
\scriptsize
$$ 
\begin{array}{c}
\infer[\left(\begin{array}{c}\mathtt{1}\end{array}\right)]
      {
        (\sConf'',\uniquely'',\fresh'') \in \conc(\sConf,A\uplus\set{\phid},B,\uniquely,\fresh)
      }
      {
        \begin{array}{c}  
          (\sConf',\uniquely',\fresh')\in\conc(\sConf,A,B,\uniquely,\fresh)\\
          (\sConf'',\uniquely'',\fresh'')\in\concphaser(\sConf',\phid,\uniquely',\fresh')
        \end{array}
      } 
\\
\\
\infer[\left(\begin{array}{c}\mathtt{2}\end{array}\right)]
      {
        (\sConf'',\uniquely'',\fresh'') \in \conc(\sConf,A,B\uplus\set{\thid},\uniquely,\fresh)
      }
      {
        \begin{array}{c}  
          (\sConf',\uniquely',\fresh')\in\conc(\sConf,A,B,\uniquely,\fresh)\\
          (\sConf'',\uniquely'',\fresh'')\in\conctask(\sConf',\thid,\uniquely',\fresh')
        \end{array}
      } 
 \\
 \\
\infer[\left(\begin{array}{c}\mathtt{3}\end{array}\right)]
      {
        (\sConf,\uniquely,\fresh) \in \conc(\sConf,\emptyset,\emptyset,\uniquely,\fresh)
      }
      {
      }
\\\\
\infer[\left(\begin{array}{c}\mathtt{4}\end{array}\right)]
      {
        \conc(\sConf,A,B,\uniquely,\fresh)
      }
      {
        \begin{array}{c}          
          \sConf = \sConfTuple\quad \conc(\sConf,A,B)\\
          B=\uniquely\uplus\fresh \quad \uniquely\subseteq\thidSet\quad \fresh\cap\thidSet=\emptyset 
        \end{array}
      }
       
\end{array}
$$
\end{center}
\caption[Auxiliary functions used in the predecessor computation of the parameterized Phaser instructions (Part II)]{Auxiliary functions used in the predecessor computation of the parameterized Phaser instructions (Part II). The function $\conc(\sConf,A,B,\uniquely,\fresh)$ concretizes the set of phasers $A$ and the set of tasks $B$ in the constraint $\sConf$. Those phasers or tasks in $A$ or $B$ that are already in $\sConf$ will be preserved.}
\label{fig:pre:aux2}
\end{figure*} 

\subsection{Concretization}

In this section, we introduce the concretization functions and prove some properties about them. 
%
Assume a constraint $\sConf=\sConfTupleInd{\sConf}$. We say a set  $\uniquely$ is in $\sConf$ if $\uniquely\subseteq\thidSet_\sConf$ and a set $\fresh$  is \textit{fresh} for $\sConf$ if $\fresh\cap\thidSet_\sConf=\emptyset$. 
We say $\concphaser(\sConf,\phid_\sConf,\uniquely,\fresh)$ (Figure \ref{fig:pre:aux}) concretizes the phaser $\phid_\sConf$ in $\sConf$. Rule \texttt{concretize phaser 1} is used when phaser $\phid_\sConf$ is already concrete and the input constraint will be returned without any modification. Rule \texttt{concretize phaser 2} adds a new phaser $\phid_\sConf$ to the constraint $\sConf$ and non-deterministically registers a set of tasks $\thidSet_1'$ in the concrete constraint that map to $\thidSet_1\subseteq\thidSet$ with $\phid_\sConf$. Moreover, it non-deterministically chooses a set of tasks $\thidSet_2'$ that map to $\thidSet_2\subseteq\thidSet$ to not be registered with $\phid_\sConf$. The task mappings $\taskMp_1$ and $\taskMp_2$ are used to map the tasks from the concrete constraint to constraint $\sConf$.

$\conctask(\sConf,\thid_\sConf,\uniquely,\fresh)$ (Figure \ref{fig:pre:aux}) concretizes the task $\thid_\sConf$ in $\sConf$. Rule \texttt{concretize task 1} is used when task $\thid_\sConf$ is already concrete and the input constraint will be returned without any modification. Rules \texttt{concretize task 2} and \texttt{3} consider the case when a new task $\thid_\sConf$ will be added that copies a task $\othid\in\thidSet$. The difference between \texttt{concretize task 2} and \texttt{3} is in the fact that \texttt{concretize task 3},  unlike \texttt{concretize task 2} concretizes $\othid$ (after being renamed to $\thid'$) as well as $\thid$.  
Rule \texttt{concretize task 4} concretizes a task that is non-deterministically registered with a subset of the phasers in $\phidSet_2\subseteq\phidSet_\sConf$ and not registered to others. Such task needs to have phase bounds that respect the environment of the phasers in $\phidSet_2$.

$\conc(\sConf,A_\sConf,B_\sConf,\uniquely,\fresh)$ (Figure \ref{fig:pre:aux2}) concretizes the phasers $A_\sConf$ and the tasks $B_\sConf$ in $\sConf$ using $\concphaser(\sConf,\phid_\sConf,\uniquely,\fresh)$ and $\conctask(\sConf,\thid_\sConf,\uniquely,\fresh)$. 

Rule \texttt{concretize seq} concretizes the control sequence of a task and rule \texttt{concretize var} concretizes the phaser name a task uses for referencing a phaser. 
Some facts about the the concretizations will follow.

\begin{lemma}
\label{lem:phaser:greater}
For a given constraint $\sConf=\sConfTupleInd{\sConf}$, a set $\uniquely$ that is in $\sConf$, a set $\fresh$ that is fresh for $\sConf$, and a phaser $\phid$ that is possibly in $\phidSet_\sConf$, $\concphaser(\sConf,\phid,\uniquely,\fresh)$ always terminates. Each tuple $(\sConf',\uniquely',\fresh')$ in $\concphaser(\sConf,\phid,\uniquely,\fresh)$ satisfies the following:
\begin{itemize} 
\item $\sConf\preceq\sConf'$,
\item $\fresh'\subseteq\fresh$ and $\fresh'\cap\thidSet_{\sConf'}=\emptyset$,
\item $\uniquely'=\uniquely\cup(\fresh\setminus\fresh')$ and $\uniquely'\subseteq\thidSet_{\sConf'}$,
\end{itemize}  
\end{lemma}

\begin{proof}
By definiton of \texttt{concretize phaser 1} and \texttt{2} in Figure \ref{fig:pre:aux}.
\end{proof}

\begin{lemma}
\label{lem:task:greater}
For a given constraint $\sConf=\sConfTupleInd{\sConf}$, a set $\uniquely$ that is in $\sConf$, a set $\fresh$ that is fresh for $\sConf$, and a task $\thid\in\uniquely\cup\fresh$, $\conctask(\sConf,\thid,\uniquely,\fresh)$ always terminates. Each tuple $(\sConf',\uniquely',\fresh')$ in $\conctask(\sConf,\phid,\uniquely,\fresh)$ satisfies the following:
\begin{itemize} 
\item $\sConf\preceq\sConf'$,
\item $\fresh'\subseteq\fresh$ and $\fresh'\cap\thidSet_{\sConf'}=\emptyset$,
\item $\uniquely'=\uniquely\cup(\fresh\setminus\fresh')$ and $\uniquely'\subseteq\thidSet_{\sConf'}$,
\end{itemize}   
\end{lemma}

\begin{proof}
By definition of \texttt{concretize task 1}, \texttt{2}, \texttt{3}, and \texttt{4} in Figure \ref{fig:pre:aux}.
\end{proof}

Observe that $\concvar(\sConf,\thid,\phid,\phvar)$ and $\concseq(\sConf,\thid,\seqVal)$ in Figure \ref{fig:pre:aux} do not change the tasks, phasers, and phases. One can show that after calling them, the identity relations on $\thidSet$ and $\phidSet$ witness $\sConf\preceq\sConf'$.

\begin{lemma}
\label{lem:var:seq:greater}
For a given constraint $\sConf=\sConfTupleInd{\sConf}$, a set $\uniquely$ that is in $\sConf$, a set $\fresh$ that is fresh for $\sConf$, a task $\thid\in\thidSet_\sConf$, a phaser variable $\phvar$, a phaser $\phid\in\phidSet_\sConf$, and $\seqVal$, $\concvar(\sConf,\thid,\phid,\phvar)$ and $\concseq(\sConf,\thid,\seqVal)$ always terminate and will respectively generate a singleton $\set{\sConf'}$ such that the constraint $\sConf'$ satisfies $\sConf\preceq\sConf'$. 
\end{lemma}

\begin{proof}
By definition of $\concvar(\sConf,\thid,\phid,\phvar)$ and $\concseq(\sConf,\thid,\seqVal)$ in Figure \ref{fig:pre:aux}.
\end{proof}

\begin{lemma}
\label{lem:conc:greater}
Assume $\sConf=\sConfTupleInd{\sConf}$, $A_\sConf$ that is a set of phaser identifiers possibly intersecting $\phidSet_\sConf$, a set $\uniquely$ that is in $\sConf$, a set $\fresh$ that is fresh for $\sConf$, and  $B_\sConf\subseteq\uniquely\cup\fresh$. $\conc(\sConf,A_\sConf, B_\sConf,\uniquely,\fresh)$ always terminates and returns a finite set of tuples. Each tuple $(\sConf',\uniquely',\fresh')$ in $\conc(\sConf,A_\sConf, B_\sConf,\uniquely,\fresh)$ satisfies 
\begin{itemize} 
\item $\sConf\preceq\sConf'$,
\item $\fresh'\subseteq\fresh\setminus B_\sConf$ and $\fresh'\cap\thidSet_{\sConf'}=\emptyset$,
\item $\uniquely'=\uniquely\cup(\fresh\setminus\fresh')$ and $\uniquely'\subseteq\thidSet_{\sConf'}$,
\end{itemize}  
\end{lemma}

\begin{proof}
By Lemmas \ref{lem:phaser:greater} and \ref{lem:task:greater} and induction on $|A_\sConf|+|B_\sConf|$.
\end{proof}

\begin{definition}[$\taskMp$-uniquely-mapped]
For a given configuration $\cConf=\cConfTupleInd{\cConf}$ and constraint $\sConf=\sConfTupleInd{\sConf}$ such that $\taskMp$ and $\phaserMp$ witness $\cConf\in\denotationOf{\sConf}$, a task $\thid\in\thidSet_\cConf$ is said to be $\taskMp$-uniquely-mapped iff $\taskMp(\othid)=\taskMp(\thid)\Longrightarrow \othid=\thid$. 
\end{definition}

\begin{lemma}
\label{unique-phaser}  
Assume a configuration $\cConf=\cConfTupleInd{\cConf}$, a constraint $\sConf=\sConfTupleInd{\sConf}$ such that $\taskMp$ and $\phaserMp$ witness $\cConf\in\denotationOf{\sConf}$, a set $\uniquely$ that is in $\sConf$ such that $\taskMp$ uniquely maps $\taskMp^{-1}(\uniquely)$, and a set $\fresh$ that is fresh for $\sConf$. Let $\phid_\cConf$ be an arbitrary phaser in $\phidSet_\cConf$ and $\phid_\sConf$ be $\phaserMp(\phid_\cConf)$ if $\phid_\cConf$ is mapped by $\phaserMp$ or a fresh phaser, otherwise. A tuple $(\sConf',\uniquely',\fresh')$ exists among $\concphaser(\sConf,\phid_\sConf,\uniquely,\fresh)$ such that:
 
\begin{itemize}
\item $\cConf\in\denotationOf{\sConf'}$ and $\taskMp'$ and $\phaserMp'$ witness the denotation, 
\item $\phaserMp'=\phaserMp[\phid_\cConf\leftarrow\phid_\sConf]$,
\item ${\taskMp'}$ uniquely maps the tasks in ${\taskMp'}^{-1}(\uniquely')$,
\end{itemize}
\end{lemma}

\begin{proof}
The concretization of $\phid_\sConf$ will be performed in one of the following ways:

\begin{itemize}
\item If $\phid_\cConf$ is mapped to $\phid_\sConf$ by $\phaserMp$, the rule (\texttt{concretize phaser 1}) will return $(\sConf,\uniquely,\fresh)$, which is the desired tuple. Observe that the tasks are not altered, hence, the returned tuple satisfies the conditions in the lemma.

\item If $\phid_\cConf$ is not mapped by $\phaserMp$, the rule (\texttt{concretize phaser 2}) generates a set of tuples $(\sConf',\uniquely',\fresh')$ in which the concrete constraints account for all possible registrations of the tasks in $\thidSet_\sConf$ to the new phaser $\phid_\sConf$. 
A constraint $\sConf'=\sConfTupleInd{\sConf'}$ generated by the rule will capture

\begin{itemize}
\item the tasks in $\thidSet_1\subseteq\thidSet_\cConf$ that were mapped by $\taskMp$ but were not registered with $\phid_\cConf$ can be mapped to $\thidSet_1'\subseteq\thidSet_{\sConf'}$ by the task mapping $\taskMp_1(\taskMp)$. 

\item the tasks in $\thidSet_2\subseteq\thidSet_\cConf$ that were mapped by $\taskMp$ and are registered with $\phid_\cConf$ can be mapped to $\thidSet_2'\subseteq\thidSet_{\sConf'}$ by the task mapping $\taskMp_2(\taskMp)$.
\end{itemize}

The task mapping $\taskMp'$ that maps the tasks in $\thidSet_1$ using $\taskMp_1(\taskMp)$ and those in $\thidSet_2$ using $\taskMp_2(\taskMp)$ and the phaser mapping $\phaserMp'=\phaserMp[\phid_\cConf\leftarrow\phid_\sConf]$ witness $\cConf\models\sConf'$. $\taskMp'$ by construction uniquely maps the tasks in ${\taskMp'}^{-1}(\uniquely')$.
\end{itemize}
\end{proof}

\begin{lemma}
\label{unique-task}
Assume a configuration $\cConf=\cConfTupleInd{\cConf}$, a constraint $\sConf=\sConfTupleInd{\sConf}$ such that $\taskMp$ and $\phaserMp$ witness $\cConf\in\denotationOf{\sConf}$, a set $\uniquely$ that is in $\sConf$ such that $\taskMp$ uniquely maps $\taskMp^{-1}(\uniquely)$, a set $\fresh$ that is fresh for $\sConf$, and a set $B_\cConf=(B_{\cConf_1}\uplus B_{\cConf_2})\subseteq\thidSet_\cConf$ such that $\taskMp$ uniquely maps $B_{\cConf_1}$ and does not uniquely map $B_{\cConf_2}$. Let $\thid_\cConf$ be an arbitrary task in $B_\cConf$ and $\thid_\sConf$ be $\taskMp(\thid_\cConf)$ if $\thid_\cConf$ is in $\uniquely$ or a task n $\fresh$, otherwise. $\conctask(\sConf,\thid_\sConf,\uniquely,\fresh)$ will generate a set of tuples among which there is a tuple $(\sConf',\uniquely',\fresh')$ such that 
\begin{itemize}
\item $\cConf\in\denotationOf{\sConf'}$ and $\taskMp'$ and $\phaserMp$ witness the denotation, 
\item $\taskMp'$ uniquely maps the tasks in ${\taskMp'}^{-1}(\uniquely')$,
\end{itemize}
\end{lemma}

\begin{proof}
The concretization of $\thid_\sConf$ will be performed in one of the following ways:
\begin{itemize}
\item If $\thid_\cConf$ is in $B_{\cConf_1}$, it is already uniquely mapped to $\thid_\sConf$, the rule (\texttt{concretize task 1}) will return $(\sConf,\uniquely,\fresh)$, which is the desired tuple. 

\item If $\thid_\cConf$ is mapped by $\taskMp$ but is in $B_{\cConf_2}$, it is mapped to $\taskMp(\thid_\cConf)$ but not uniquely. The rule (\texttt{concretize task 2}) (respectively,  \texttt{concretize task 3}) return in this case a tuple $(\sConf',\uniquely',\fresh')$ for which 
the mappings $\taskMp'=\taskMp[\thid_\cConf\leftarrow\thid_\sConf]$ (respectively, $\taskMp'=\taskMp[\thid_{\cConf}\leftarrow\thid_\sConf][\othid_{\cConf}\leftarrow\othid_\sConf]$ given that $\thid_{\cConf}$ and $\othid_{\cConf}$ are both mapped to $\taskMp(\thid_\cConf)$) and $\phaserMp$ witnesses $\cConf\in\denotationOf{\sConf'}$. The other claims hold by construction. 

\item If $\thid_\cConf$ is not mapped by $\taskMp$, it is in $B_{\cConf_2}$. Let $\phidSet_\cConf^m\subseteq\phidSet_\cConf$ be the set of mapped phasers by $\phaserMp$ and $\phidSet_\cConf^{m,r}\subseteq\phidSet_\cConf^m$ be the set of mapped phasers that $\thid_\cConf$ is registered with. 
In this case, the rule (\texttt{concretize task 4}) returns a set of tuples among which there is a tuple $(\sConf',\uniquely',\fresh')$ in which ${\thid_\sConf}$ is a fresh task that 
is registered with $\phaserMp(\phidSet_\cConf^{m,r})$ and not with $\phaserMp(\phidSet_\cConf^m\setminus\phidSet_\cConf^{m,r})$.  
 Hence, the task and phaser mappings $\taskMp'=\taskMp[\thid_\cConf\leftarrow\thid_\sConf]$ and $\phaserMp$ witnesses $\cConf\models\sConf'$. The other claims hold by construction. 
\end{itemize}
\end{proof}

\begin{definition}[Match]
\label{match}
Assume a tuple $(\cConf,A_\cConf,B_\cConf)$ with $\cConf=\cConfTupleInd{\cConf}$, $A_\cConf\subseteq\phidSet_\cConf$, and $B_\cConf\subseteq\thidSet_\cConf$ and $(\sConf,A_\sConf,B_\sConf,\uniquely,\fresh)$ with $\sConf=\sConfTupleInd{\sConf}$, $A_\sConf$ that is a set of phaser identifiers possibly intersecting $\phidSet_\sConf$, $B_\sConf=\uniquely\cup\fresh$, a set $\uniquely$ that is in $\sConf$, and a set $\fresh$ that is fresh for $\sConf$. We say the tuple $(\cConf,A_\cConf,B_\cConf)$ matches $(\sConf,A_\sConf,B_\sConf,\uniquely,\fresh)$ with respect to $\taskMp$ and $\phaserMp$ if 

\begin{itemize}
\item $\taskMp$ and $\phaserMp$ witness $\cConf\in\denotationOf\sConf$,
\item $\taskMp^{-1}(\uniquely)$ is $\taskMp$-uniquely-mapped,
\item $|A_\cConf|=|A_\sConf|$ and $|B_\cConf|=|B_\sConf|$,
\item $B_\cConf=B_{\cConf_1}\uplus B_{\cConf_2}$,
\item $\taskMp(B_{\cConf_1})\subseteq\uniquely$, i.e., $B_{\cConf_1}$ is $\taskMp$-uniquely-mapped,
\item $\taskMp$ does not uniquely map $B_{\cConf_2}$,
\item $|B_{\cConf_2}|=|\fresh|$.
\end{itemize}

\end{definition}

\begin{lemma}

Given $\cConf=\cConfTupleInd{\cConf}$, $\sConf\sConfTupleInd{\sConf}$, and $\taskMp$ and $\phaserMp$ that witness $\cConf\in\denotationOf{\sConf}$, for any $A_\cConf\subseteq\phidSet_\cConf$ and $B_\cConf\subseteq\thidSet_\cConf$, $(\cConf,A_\cConf,B_\cConf)$ matches $(\sConf,A_\sConf,B_\sConf,\uniquely,\fresh)$ with respect to $\taskMp$ and $\phaserMp$ if 

\begin{itemize}
\item $|A_\cConf|=|A_\sConf|$ and $|B_\cConf|=|B_\sConf|$,
\item $A_\sConf = \phaserMp({A_\cConf}_{|_\phaserMp})\uplus({A_\sConf}\setminus\domainOf{\phaserMp})$,
\item $B_\sConf=\uniquely\uplus\fresh$, $\uniquely\subseteq\thidSet_\sConf$, $\fresh\cap\thidSet_\sConf=\emptyset$,
\item $\taskMp^{-1}(\uniquely)$ is $\taskMp$-uniquely-mapped.
\end{itemize}
\end{lemma}

\begin{proof}
By Definition \ref{match} and construction of $\cConf\in\denotationOf\sConf$.
\end{proof}

\begin{example}
\normalfont
Let $\cConf\in\denotationOf\sConf$ such that $\cConf=\cConfTupleInd{\cConf}$, $\phid_\cConf\in\phidSet_\cConf$, $\thid_\cConf\in\thidSet_\cConf$, $\sConf=\sConfTupleInd{\sConf}$, and $\taskMp$ and $\phaserMp$ witness the denotation. $(\cConf,\set{\phid_\cConf},\set{\thid_\cConf})$ will match some $(\sConf,\set{\phid_\sConf},\set{\thid_\sConf},\uniquely,\fresh)$ for $\uniquely\cup\fresh=\set{\thid_\sConf}$.
\end{example}

\begin{lemma}
\label{unique-conc}  
Assume a configuration $\cConf=\cConfTupleInd{\cConf}$, a constraint $\sConf=\sConfTupleInd{\sConf}$, the mappings $\taskMp$ and $\phaserMp$, and the sets $A_\cConf$, $B_\cConf$, $A_\sConf$, and $B_\sConf$  such that 
$(\cConf,A_\cConf,B_\cConf)$ matches $(\sConf,A_\sConf,B_\sConf,\uniquely,\fresh)$ with respect to $\taskMp$ and $\phaserMp$.  For all the tuples $(\sConf',\uniquely',\fresh')$ in $\conc(\sConf,A_\sConf,B_\sConf,\uniquely,\fresh)$ we have that $\uniquely'=B_\sConf$ and $\fresh'=\emptyset$ and there exists one such tuple that: 
 
\begin{itemize}
\item $\phaserMp':\domainOf{\phaserMp}\cup A_\cConf\rightarrow \phidSet_{\sConf'}$,
\item $\taskMp':\domainOf{\taskMp}\cup B_\cConf\rightarrow \thidSet_{\sConf'}$,
\item $\taskMp'$ uniquely maps $B_{\cConf}$,
\item $\phaserMp'$ and $\taskMp'$ witness $\cConf\in\denotationOf{\sConf'}$.
\end{itemize}
\end{lemma}

\begin{proof}
By Lemmas \ref{unique-phaser} and \ref{unique-task} and induction on $|A_\sConf|+|B_\sConf|$.
\end{proof}

\begin{lemma}
\label{lem:denote:var-seq}
Assume a configuration $\cConf=\cConfTupleInd{\cConf}$ and a constraint $\sConf=\sConfTupleInd{\sConf}$ such that $\taskMp$ and $\phaserMp$ witness $\cConf\in\denotationOf{\sConf}$. Assume also that for some task $\thid_\cConf$ and phaser $\phid_\cConf$, $\taskMp(\thid_\cConf)=\thid_\sConf$, $\phaserMp(\phid_\cConf)=\phid_\sConf$, $\pcMp_\cConf(\thid_\cConf)=\seqVal$, and $\relMp(\thid_\cConf)(\phid_\cConf)=(\phvar,\val)$. $\concvar(\sConf,\thid_\sConf,\phid_\sConf,\phvar)$ and $\concseq(\sConf,\thid_\sConf,\seqVal)$ will respectively generate a singleton $\set{\sConf'}$ such that the constraint $\sConf'$ satisfies $\cConf\in\denotationOf{\sConf'}$.
\end{lemma}

\begin{proof}
By definition of $\concvar(\sConf,\thid,\phid,\phvar)$ and $\concseq(\sConf,\thid,\seqVal)$ in Figure \ref{fig:pre:aux}.
\end{proof}

\subsection{Predecessor Computation}
In this section, we formally define the predecessor computation functions in Figures \ref{fig:pre:new} and \ref{fig:pre:asynch} and prove their soundness and relative completeness. 

\begin{figure*}[t]
  \begin{center}
    \scriptsize
    $$
    \begin{array}{c}
\infer[\mathtt{(newPhaser)}]
      {
        \sConfTuple
        \preReducesTo{\phvar:=\newp}{\thid}
\tuple{\thidSet',\phidSet'\setminus\set{\phid},\boolMp,\pcMp'',\relgapMp'',\envMp'\setminus \set{\phid}}
      }
      {
        \begin{array}{c}
       (\sConf_1,\uniquely',\fresh') \in \conc(\sConfTuple,\set{\phid},\set{\thid})  \\
          \sConf_2 \in \concseq(\sConf_1,\thid,\seqVal)  \quad
          \sConf_\cnc \in \concvar(\sConf_2,\thid,\phid,\phvar)\\
          \sConf_\cnc=\psConfTuple \\
          \seqVal',\seqVal \in \seqSet \quad
          \seqVal'=\phvar:=\newp;\seqVal
          \quad \pcMp''=\mapSubst{\pcMp'}{\thid}{\seqVal'} \\
          (0,0) \models \relgapMp'(\thid)(\phid) \\
          \left(\forall \othid\in\thidSet'.
          ~\relgapMp'(\othid)(\phid)=(\phvar_\othid,\val_\othid)\quad \phvar_\othid=\phvar \text{ or } \val_\othid\neq\nreg\Rightarrow \othid=\thid\right)\\
          \relgapMp'' = \substSet{\relgapMp'}{\setcomp{\othid\leftarrow\relgapMp'(\othid)\setminus \set{\phid}}{\othid\in\thidSet'}}
          \\                 
        \end{array}
      }
      \\
      \\
\infer[\mathtt{(drop)}]
      {
        \sConfTuple
        \preReducesTo{\dereg{\phvar}}{\thid}
        \tuple{\thidSet',\phidSet',\boolMp,\pcMp'',\relgapMp'',\envMp''}
      }
      {
        \begin{array}{c}          
          (\sConf_1,\uniquely',\fresh') \in \conc(\sConfTuple,\set{\phid},\set{\thid})  \\
          \sConf_2 \in \concseq(\sConf_1,\thid,\seqVal)  \quad
          \sConf_\cnc \in \concvar(\sConf_2,\thid,\phid,\phvar)\\
          \sConf_\cnc=\psConfTuple\quad \relgapMp'(\thid)(\phid)=(\phvar,\nreg) \\
          \seqVal',\seqVal \in \seqSet \quad
          \seqVal'=\dereg{\phvar};\seqVal
          \quad \pcMp''=\mapSubst{\pcMp'}{\thid}{\seqVal'} \\
          \othidSet = \setcomp{\othid}{\othid \in \thidSet' \textrm{ and } \relgapMp'(\othid)(\phid)=(val_\othid,(\lwbw_\othid,\lwbs_\othid,\upbw_\othid,\upbs_\othid)}        
          \\
          \uwmin = min \setcomp{\upbw_\othid}{\othid \in \othidSet} \quad 
          \usmin = min \setcomp{\upbs_\othid}{\othid \in \othidSet} \\
     \lwmax = max \setcomp{\lwbw_\othid}{\othid\in \othidSet} \quad
     \lsmax = max \setcomp{\lwbs_\othid}{\othid\in \othidSet} \\
          (\uwmin< \infty \wedge \usmin<\infty) \implies -\uwmin \leq \delta \leq \usmin \\ 
          (\uwmin=\usmin=\infty) \implies -\lwmax \leq \delta \leq \lsmax \\
          map_\delta = \setcomp{\othid\leftarrow\mapSubst{\relgapMp'(\othid)}{\phid}{(val_\othid,(\plusOf{\lwbw_\othid+\delta},\plusOf{\lwbs_\othid-\delta},(\upbw_\othid+\delta),(\upbs_\othid-\delta))}}
          {\othid\in\othidSet} \\
map_\thid = \set{\thid\leftarrow\mapSubst{\relgapMp'(\thid)(\phid)}{\phid}{(\phvar,(0,0,\infty,\infty))}} \quad
          \relgapMp'' = \substSet{\relgapMp'}{map_\thid \cup map_\delta 
          }\\
           \envMp'(\phid)=(\elwbw,\elwbs)\quad\envMp''=\mapSubst{\envMp'}{\phid}{((\elwbw+\delta)^+,(\elwbs-\delta)^+))}
        \end{array}
      }
      \\
      \\ 
\infer[\mathtt{(exit)}]
      {
        \sConfTuple
        \preReducesTo{\exit}{\thid}
        \tuple{\thidSet',\phidSet',\boolMp,\pcMp'',\relgapMp'',\envMp''}
      }
      {
        \begin{array}{c}
          \thid\not\in\thidSet  \quad  \phidSet=\phidSet_1\cup\phidSet_2 
           \\
(\psConfTuple,\uniquely',\fresh') \in \conc(\sConf,\set{},\set{\thid}) 
\\ \forall \phid_i\in\phidSet_1.\isRegistred{\relgapMp'}{\othid_i}{\phid_i}
\quad \pcMp''=\mapSubst{\pcMp'}{\thid}{\exit} 
          \\
          \forall \phid\in\phidSet_1.
          \othidSet^\phid = \setcomp{\othid}{\othid \in \thidSet \textrm{ and } \relgapMp'(\othid)(\phid)=(val^\phid_\othid,(\lwbw^\phid_\othid,\lwbs^\phid_\othid,\upbw^\phid_\othid,\upbs^\phid_\othid)} \\
\uwmin^\phid = min \setcomp{\upbw^\phid_\othid}{\othid \in \othidSet^\phid} \quad 
          \usmin^\phid = min \setcomp{\upbs^\phid_\othid}{\othid \in \othidSet^\phid} \\
          \lwmax^\phid = max \setcomp{\lwbw^\phid_\othid}{\othid\in \othidSet^\phid} \quad
 \lsmax^\phid = max \setcomp{\lwbs^\phid_\othid}{\othid\in \othidSet^\phid} \\
          map_\thid = \setcomp{\phid\leftarrow(*,(0,0,\infty,\infty))}{\phid\in\phidSet_1} \cup \setcomp{\phid\leftarrow(*,\nreg)}{\phid\in\phidSet_2}\\
          \forall \phid\in\phidSet_1.
          (\uwmin^{\phid}< \infty \wedge \usmin^{\phid}<\infty) \implies -\uwmin^{\phid} \leq \delta^{\phid} \leq \usmin^{\phid} \\
           \forall \phid\in\phidSet_1. 
          (\uwmin^{\phid}=\usmin^{\phid}=\infty) \implies -\lwmax^{\phid} \leq \delta^{\phid} \leq \lsmax^{\phid} \\
           \forall \phid\in\phidSet_1, \othid\in\othidSet^\phid. \relgapMp'(\othid)(\phid)+\delta^\phid=
          (val^\phid_\othid,(\plusOf{\lwbw^\phid_\othid+\delta^\phid},\plusOf{\lwbs^\phid_\othid-\delta^\phid},(\upbw^\phid_\othid+\delta^\phid),(\upbs^\phid_\othid-\delta^\phid)))\\
           \relgapMp''= \substSet{\relgapMp'}{(\thid\leftarrow map_\thid) \cup
             \setcomp{\othid\leftarrow \mapSubst{\relgapMp'(\othid)}{\phid}{\relgapMp'(\othid)(\phid)+\delta^\phid}}{\othid\in\othidSet^\phid \textrm{ and } \phid\in\phidSet_1}}\\
             \envMp''=\mapSubst{\envMp'}{\phid}{((\elwbw+\delta^\phid)^+,(\elwbs-\delta^\phid)^+))~|~{\phid\in\phidSet_1\text{ and }\envMp'(\phid)=(\elwbw,\elwbs)}}
        \end{array}
      }
    \end{array}
    $$
  \end{center}
  \caption[Predecessor computation for parameterized Phaser statements (Part I)]{Derivation rules for computing predecessors wrt. parameterized Phaser instructions $\newp,\dereg{\phvar}$ and $\exit$. For each rule, a task $\thid$ executes the statement. If $\thid$ does not belong to $\thidSet$, it will be added by concretizing $\sConf$. Otherwise, the concretization will preserve it. In $\newp$ and $\dereg{\phvar}$, a phaser $\phid$ is required, which  is either added by concretization or is preserved by it. }
  \label{fig:pre:new}
\end{figure*}

\begin{figure*}[t]
\begin{center}
\scriptsize
$$ 
\begin{array}{c}
  \infer[\mathtt{(asynch)}]
        {
          \sConfTuple
          \preReducesTo{\asynch{\task}{\phvar_1,\ldots \phvar_k}}{\thid}
          \tuple{\thidSet'\setminus\set{\othid},\phidSet',\boolMp,\pcMp''\setminus\set{\othid},\relgapMp''\setminus\set{\othid},\envMp'}
        }
        {
          \begin{array}{c}
  (\sConf_1,\uniquely',\fresh')\in\conc(\sConfTuple,\set{\phid_1,\ldots \phid_k},\set{\thid,\othid})  \\
          \sConf_2 \in \concseq(\sConf_1,\thid,\seqVal)  \quad \sConf_\cnc \in \concseq(\sConf_2,\othid,\stmt)  \\          
\text{ for each } i:1\leq i\leq k \quad \sConf_{3+i} \in \concvar(\sConf_{2+i},\thid,\phid_i,\phvar_i) \\
\text{ for each } i:1\leq i\leq k \quad \sConf_{3+k+i} \in \concvar(\sConf_{2+k+i},\othid,\phid_i,\ophvar_i)\\
          \sConf_\cnc=\sConf_{3+2k} = \psConfTuple\\
            \seqVal',\seqVal \in \seqSet \quad            
          \parametersOf{\task}=(\ophvar_1,\ldots \ophvar_k) \quad  \seqVal'=\asynch{\task}{\phvar_1,\ldots \phvar_k}\set{\stmt};\seqVal
          \quad \pcMp''=\mapSubst{\pcMp'}{\thid}{\seqVal'} \\
          \textrm{ for each } i:1\leq i\leq k \quad
          \relgapMp'(\thid)(\phid_i)=(\phvar_i,(\lwbw^\thid_{i},\lwbs_i^\thid, \upbw_i^\thid, \upbs_i^\thid))
          ~\quad \relgapMp'(\othid)(\phid_i)=(\ophvar_i,(\lwbw_{i}^\othid,\lwbs_i^\othid, \upbw_i^\othid, \upbs_i^\othid)) \\
          \relgapMp''=\mapSubst{\relgapMp'}{\thid}
                              {\substSet{\relgapMp'(\thid)}
                                {\setcomp{\phid_i\leftarrow(\phvar_i,(\lwbw_{i}^{{\thid}},\lwbs_i^{{\thid}}, \upbw_i^{{\thid}}, \upbs_i^{{\thid}})\sqcap (\lwbw_{i}^{{\othid}},\lwbs_i^{{\othid}}, \upbw_i^{{\othid}}, \upbs_i^{{\othid}}))}
                                  {i:{1}\leq i\leq {k}}}}\\
          \relgapMp'(\othid)(\phid) \not\in \set{-,*}\times\set{\nreg} \implies \phid \in \set{\phid_1,\ldots \phid_k} 
          \end{array}
 } 
\\
\\
\infer[\mathtt{(signal~I)}]
      {\sConfTuple
        \preReducesTo{\sig{\phvar}}{\thid}
        \tuple{\thidSet',\phidSet',\boolMp,\pcMp'',\relgapMp''',\envMp''}
      }
      {
        \begin{array}{c}
       (\sConf_1,\uniquely',\fresh') \in \conc(\sConfTuple,\set{\phid},\set{\thid})  \\
          \sConf_2 \in \concseq(\sConf_1,\thid,\seqVal)  \quad
          \sConf_\cnc \in \concvar(\sConf_2,\thid,\phid,\phvar)\\
          \sConf_\cnc = \psConfTuple\\
          \seqVal',\seqVal \in \seqSet \quad
          \seqVal'=\sig{\phvar};\seqVal
          \quad \pcMp''=\mapSubst{\pcMp'}{\thid}{\seqVal'} \\
          \othidSet = \setcomp{\othid}{\othid \in \thidSet' \textrm{ and } \relgapMp'(\othid)(\phid)=(var_\othid,(\lwbw_\othid,\lwbs_\othid,\upbw_\othid,\upbs_\othid))} \quad
          \textrm{for each } \othid\in\othidSet.~\upbw_\othid \geq 1 \\ 
          \relgapMp''=\substSet{\relgapMp'}
                   {\setcomp{\othid\leftarrow\mapSubst{\relgapMp'(\othid)}{\phid}{(var_\othid,((\lwbw_{\othid}-1)^+,\lwbs_\othid+1, \upbw_\othid-1, \upbs_\othid+1))}}{\othid\in\othidSet\setminus\set{\thid}}}
                   \\
                   \relgapMp'''= \mapSubst{\relgapMp''}{\thid}{\mapSubst{\relgapMp''}{\phid}{(var_\thid,((\lwbw_{\thid}-1)^+,\lwbs_\thid, \upbw_\thid-1, \upbs_\thid))}} \\
                   \envMp''=\mapSubst{\envMp'}{\phid}{(\elwbw-1)^+,\elwbs+1}
        \end{array}
      }
\\
\\
\infer[\mathtt{(signal~II)}]
      {\sConfTuple
        \preReducesTo{\sig{\phvar}}{\thid}
        \tuple{\thidSet',\phidSet',\boolMp,\pcMp'',\relgapMp'',\envMp'}
      }
      {
        \begin{array}{c}         
       (\sConf_1,\uniquely',\fresh') \in \conc(\sConfTuple,\set{\phid},\set{\thid})  \\
          \sConf_2 \in \concseq(\sConf_1,\thid,\seqVal)  \quad
          \sConf_\cnc \in \concvar(\sConf_2,\thid,\phid,\phvar)\\
          \sConf_\cnc = \psConfTuple\\
          \seqVal',\seqVal \in \seqSet \quad
          \seqVal'=\sig{\phvar};\seqVal
          \quad \pcMp''=\mapSubst{\pcMp'}{\thid}{\seqVal'} \\ 
    \relgapMp'(\thid)(\phid)=(var_\thid,(\lwbw_\thid,\lwbs_\thid,\upbw_\thid,\upbs_\thid)) \quad
          \upbs_\thid \geq 1 \\ 
          \relgapMp''=\mapSubst{\relgapMp'}{\thid}{\mapSubst{\relgapMp'(\thid)}{\phid}{(var_\thid,(\lwbw_\thid,(\lwbs_\thid-1)^+,\upbw_\thid,\upbs_\thid-1))}}
        \end{array}
      }
\\
\\
\infer[\mathtt{(wait)}]
      {\sConfTuple
        \preReducesTo{\wait{\phvar}}{\thid}
        \tuple{\thidSet',\phidSet',\boolMp,\pcMp'',\relgapMp'',\envMp'}
      }
      {
        \begin{array}{c}\          
          (\sConf_1,\uniquely',\fresh') \in \conc(\sConfTuple,\set{\phid},\set{\thid}) \\
          \sConf_2 \in \concseq(\sConf_1,\thid,\seqVal)  \quad
          \sConf_\cnc, \in \concvar(\sConf_2,\thid,\phid,\phvar)\\
        \sConf_\cnc = \psConfTuple\\
          \seqVal',\seqVal \in \seqSet \quad
          \seqVal'=\wait{\phvar};\seqVal
          \quad \pcMp''=\mapSubst{\pcMp'}{\thid}{\seqVal'} \\
       \relgapMp'(\thid)(\phid)=(var_\thid,(\lwbw_\thid,\lwbs_\thid,\upbw_\thid,\upbs_\thid)) \\
          \relgapMp''=\mapSubst{\relgapMp'}{\thid}{\mapSubst{\relgapMp'(\thid)}{\phid}{(var_\thid,(\lwbw_\thid+1,\lwbs_\thid,\upbw_\thid+1,\upbs_\thid))}}
        \end{array}
      }

              \end{array}
$$
\end{center}
  \caption[Predecessor computation for parameterized Phaser statements (Part II)]{Derivation rules for predecessor computation wrt. parameterized Phaser instructions $\asynch{\task}{\phvar_1,\ldots \phvar_k}\set{\stmt}, \sig{\phvar}$ and $\wait{\phvar}$.  For each rule, a task $\thid$ executes the statement. If $\thid$ does not belong to $\thidSet$, it will be added by concretizing $\sConf$. Otherwise, the concretization will preserve it. In the given statements, one or more phasers are required, which are either added by concretization or are preserved by that.}
  \label{fig:pre:asynch}
\end{figure*}

\begin{theorem}
\label{th:tacas:sound-pre}
Each predecessor computation rule $\preReducesTo{\stmt}{}$ in Figures \ref{fig:pre:asynch} and \ref{fig:pre:new} is sound with respect to the semantic rules of Figure \ref{fig:semantics}.
\end{theorem}

\begin{proof}
Assume a configuration $\cConf$ and a constraint $\sConf$ such that $\cConf\in\denotationOf\sConf$.

\paragraph{\textbf{newPhaser.}}
Assume $\cConf'\reducesTo{\phvar:=\newp}{\thid_\cConf}\cConf$ for some $\cConf'$ where $\phid_\cConf$ is the new phaser id in $\cConf$. 
We exhibit $\sConf'$ as well as the task and phaser mappings that witness $\cConf'\in\denotationOf{\sConf'}$ such that  $\sConf\preReducesTo{\phvar:=\newp}{\thid_\sConf}\sConf'$ for some $\thid_\sConf$. 
The rule starts by concretizing $\sConf$ according to $\cConf$ and $A_\cConf=\set{\thid_\cConf}$ and $B_\cConf=\set{\phid_\cConf}$. 
Lemmas \ref{unique-conc} and \ref{lem:denote:var-seq} ensure that after concretizations, a concrete constraint $\sConf_\cnc$ is generated such that $\cConf\in\denotationOf{\sConf_\cnc}$ with respect to some task and phaser mappings $\taskMp$ and $\phaserMp$ so that $\thid_\cConf$ is uniquely mapped to $\thid_\sConf$ by $\taskMp$, and $\phid_\cConf$ is mapped to some $\phid_\sConf$ by $\phaserMp$. 
The task $\thid_\cConf$ in $\cConf'$ creates $\phid_\cConf$, hence, it must be the only task registered with $\phid_\cConf$ or referencing it. This, combined with the definition of $\taskMp$ and $\phaserMp$,  guarantees that $\thid_\sConf$ is the only task in $\sConf$ which is registered with $\phid_\sConf$ or referencing it. 
The new mappings $\taskMp$ and $\phaserMp'=\phaserMp\setminus\set{\phid_\cConf}$ witness $\cConf'\in\denotationOf{\sConf'}$.

\paragraph{\textbf{signal.}}

Assume $\cConf'\reducesTo{\sig{\phvar}}{\thid_\cConf}\cConf$ for some $\cConf'$. 
We exhibit $\sConf'=\tuple{\thidSet_\sConf',\phidSet_\sConf',\boolMp_\sConf',\pcMp_\sConf',\relgapMp_\sConf',\envMp_\sConf'}$ as well as the task and phaser mappings that witness $\cConf'\in\denotationOf{\sConf'}$ such that $\sConf\preReducesTo{\sig{\phvar}}{\thid_\sConf}\sConf'$ for some $\thid_\sConf$.  
The rule starts by concretizing $\thid_\sConf$ and $\phid_\sConf$ in $\sConf$. 
Lemmas \ref{unique-conc} and \ref{lem:denote:var-seq} ensure that a concrete constraint $\sConf_\cnc=\sConfTupleInd{\sConf}$ is generated such that $\cConf\in\denotationOf{\sConf_\cnc}$ with respect to some task and phaser mappings $\taskMp$ and $\phaserMp$ so that $\thid_\cConf$ is uniquely mapped to some $\thid_\sConf$ by $\taskMp$, and $\phid_\cConf$ is mapped to some $\phid_\sConf$ by $\phaserMp$. 
The task $\thid_\cConf$ in $\cConf'$ has just incremented the signal phase of the task $\thid_\cConf$ on the phaser $\phid_\cConf$. %
There is a level $\lev\ge 0$ that separates the signal phases of the tasks registered with $\phid_\conc$ from their wait phases. Incrementing the signal $\sigVal{\thid_\cConf}{}$ of task $\thid_\cConf$ on $\phid_\cConf$ satisfies the phase bounds in $\sConf_\cnc$. The case in which $\lev<\sigVal{\thid_\cConf}{}$ can be captured by $\mathtt{signal~II}$ (here, $\upbs>0$ and $\lev'=\lev$). The case where $\lev=\sigVal{\thid_\cConf}{}$ can be captured with $\mathtt{signal~I}$ (here, $\upbw>0$ and $\lev'=\lev-1$).

\paragraph{\textbf{wait.}}

Assume $\cConf'\reducesTo{\wait{\phvar}}{\thid_\cConf}\cConf$ for some $\cConf'$. 
We exhibit $\sConf'$ as well as the task and phaser mappings that witness $\cConf'\in\denotationOf{\sConf'}$ such that $\sConf\preReducesTo{\wait{\phvar}}{\thid_\sConf}\sConf'$ for some $\thid_\sConf$.   
Similar to the \texttt{signal} rule, this rule starts by concretizing $\thid_\cConf$ and $\phid_\cConf$ in $\cConf$ and uniquely maps them to $\thid_\sConf$ and $\phid_\sConf$ in $\sConf_\cnc$.
The task $\thid_\cConf$ in $\cConf'$ has just incremented the wait phase of the task $\thid_\cConf$ on the phaser $\phid_\cConf$. Let $\relgapMp_\sConf(\thid_\sConf)(\phid_\sConf)=(\phvar,\gaptuple)$. 
A level $\lev> 0$ exists that shows the signal and wait phases of the tasks registered with $\phid_\cConf$ respect the phase bounds in $\sConf_\cnc$. The same level can be used to show that phases of $\cConf'$ respect the phase bounds in $\sConf'$ that is generated by the rule.

\paragraph{\textbf{drop.}}
Assume $\cConf'\reducesTo{\dereg{\phvar}}{\thid_\cConf}\cConf$ for some $\cConf'$. 
We exhibit $\sConf'=\tuple{\thidSet_\sConf',\phidSet_\sConf,\boolMp_\sConf,\pcMp_\sConf',\relgapMp_\sConf',\envMp_\sConf}$ such that $\cConf'\in\denotationOf{\sConf'}$ and $\sConf\preReducesTo{\dereg{\phvar}}{\thid_\sConf}\sConf'$ for some task $\thid_\sConf$.
Similar to the \texttt{signal} rule, this rule starts by concretizing $\thid_\cConf$ and $\phid_\cConf$ in $\cConf$ and uniquely maps them to $\thid_\sConf$ and $\phid_\sConf$ in $\sConf_\cnc$.
Since $\cConf\in\denotationOf{\sConf_\cnc}$, there exists a level $\lev$ that separates signal and wait phases of the tasks registered with $\phid_\sConf$. 
The task $\thid_\cConf$ in $\cConf'$ has just dropped the phaser $\phid_\cConf$. By existence of $\cConf'$, a $\lev'$ exists that takes into account the phases of $\thid_\sConf$ as well. 
If $\delta<-\lwmax$ (respectively, $\delta>\lsmax$), we can show $\lev-\lwmax$ (respectively, $\lev+\lsmax$) also separates the phases in $\cConf'$.  Otherwise, $\lev+\delta$ with $-\lwmax\leq\delta\leq\lsmax$ separates the phases in $\cConf'$. We adapt the bounds for each such $\delta$ in this case. In a similar manner, if $\usmin<\infty$ and $\uwmin<\infty$ for some task, then a level $\lev+\delta$ with $-\uwmin\leq\delta\leq\usmin$ must separate the phases in $\cConf'$. We adapt the phase bounds to this case.
\paragraph{\textbf{exit.}}
Assume $\cConf'\reducesTo{\exit{}}{\thid_\cConf}\cConf$ for some $\cConf'$. Proof of soundness of the rule $\mathtt{exit}$ has a similar approach to that of the \texttt{drop} rule. The difference is that \texttt{exit} intuitively has to iterate through all the phasers in $\phidSet_\cConf$ with which $\thid_\cConf$ is registered and drop them.  The mappings $\taskMp'=\taskMp\cup\set{\thid_\cConf\rightarrow\thid_\sConf}$ and $\phaserMp$ witness the denotation.

\paragraph{\textbf{asynch.}}
Assume $\cConf'\reducesTo{\asynch{\task}{\phvar_1,\ldots \phvar_k}\{\seqVal\}}{\thid_\cConf}\cConf$ for some $\cConf'$ where $\parametersOf{\task}=(\ophvar_1,\ldots \ophvar_k)$.
The task $\thid_\cConf$ has just spawned $\othid_\cConf$ in $\cConf$. 
We exhibit $\sConf'=\tuple{\thidSet_\sConf',\phidSet_\sConf',\boolMp_\sConf',\pcMp_\sConf',\relgapMp_\sConf',\envMp_\sConf'}$ such that $\cConf'\in\denotationOf{\sConf'}$ and $\sConf\preReducesTo{\asynch{\task}{\phvar_1,\ldots \phvar_k}\{\seqVal\}}{\thid_\sConf}\sConf'$. 
The rule starts by concretizing $\thid_\sConf$, $\othid_\sConf$, and $\phid_{i_\sConf}$ for every $i:1\leq i \leq k$ in $\sConf$. 
Lemmas \ref{unique-conc} and \ref{lem:denote:var-seq} ensure that a concrete constraint $\sConf_\cnc=\sConfTupleInd{\sConf}$ is generated such that $\cConf\in\denotationOf{\sConf_\cnc}$ with respect to some task and phaser mappings $\taskMp$ and $\phaserMp$ so that $\thid_\cConf$ and $\othid_\cConf$ are uniquely mapped to some $\thid_\sConf$ and $\othid_\sConf$ by $\taskMp$, and  $\phid_{i_\cConf}$ is mapped to some $\phid_{i_\sConf}$ by $\phaserMp$ for each $i:1\le i\le k$.  
The task $\thid_\cConf$ in $\cConf'$ has just spawned the task $\othid_\cConf$. 
 $\thid_\cConf$ and $\othid_\cConf$ are registered with each $\phid_{i_\cConf}$ for $i:1\le i\le k$ and have the same phases in $\relgapMp_\cConf$.
$\cConf\in\denotationOf{\sConf_\cnc}$ ensures that for each $i:1\le i\le k$, $\relgapMp_\sConf$ can be constrained so that $\thid_\sConf$ and $\othid_\sConf$ are registered with each $\phid_{i_\sConf}$ in $\relgapMp_\sConf$ and have the same phases. 
%
Hence,  the meet of the signal and wait gaps of $\thid_\sConf$ and $\othid_\sConf$ in $\relgapMp_\sConf$ is not empty. This meet will actually be the phase of $\thid_\sConf$ on phasers $\phid_{i_\sConf}$ for $i:1\leq i\leq k$. 
$\relMp_\cConf'$ is then obtained from $\relMp_\cConf$ by removing the phases of $\othid_\cConf$. 
$\relgapMp_\sConf'$ is also obtained from $\relgapMp_\sConf$ by removing $\othid_\sConf$.  
The task and phaser mappings $\taskMp'=\taskMp\setminus\set{\othid}$ and $\phaserMp$ witness that $\sConf'$ denotes $\cConf'$. 

\end{proof}

We define $\cConf_0\reducesTo{\stmt}{\thidSet}\cConf_n$ to mean a sequence $\cConf_0 \reducesTo{\stmt}{\thid_1} \cConf_1\ldots\cConf_{n-1}\reducesTo{\stmt}{\thid_n} \cConf_n$ where $\thidSet=\set{\thid_1,\ldots \thid_n}$ are the tasks in $\cConf_0$. 
Observe that this is in the transitive closure of $\reducesTo{\stmt}{}$.

\begin{theorem}
\label{th:tacas:complete-pre}

Each predecessor computation rule $\preReducesTo{\stmt}{}$ in Figures \ref{fig:pre:asynch} and \ref{fig:pre:new} except for the rule $\mathtt{newPhaser}$ is complete with respect to the semantic rules of those in Figure \ref{fig:semantics} and $\reducesTo{\stmt}{\uniquely}$ defined above. The rule $\mathtt{newPhaser}$ is complete only when the task \texttt{main} executes it.
\end{theorem}

\begin{proof}

Assume $\sConf\preReducesTo{\stmt}{\thid_\sConf}\sConf'$ according to Figures \ref{fig:pre:asynch} and \ref{fig:pre:new}. For any configuration $\cConf'=\cConfTupleInd{\cConf'}$ where $\cConf'\in\denotationOf{\sConf'}$, we exhibit a configuration $\cConf=\cConfTupleInd{\cConf}$ such that $\cConf'\reducesTo{\stmt}{\taskMp^{-1}(\thid_\sConf)}\cConf$ and $\cConf\in\denotationOf{\sConf}$. 
Note that any predecessor computation rule starts by concretizing $\sConf$ to some concrete constraint $\sConf_\cnc=\sConfTupleInd{\sConf}$. Then, the predecessor $\sConf'$ is obtained from the $\sConf_\cnc$. Lemmas  \ref{lem:var:seq:greater} and \ref{lem:conc:greater} ensure $\sConf\preceq\sConf_\cnc$ for any concrete constraint $\sConf_\cnc$. Hence, for every rule we show $\cConf\in\denotationOf{\sConf_\cnc}$. This implies $\cConf\in\denotationOf{\sConf}$.

Assume $\taskMp$ and $\phaserMp$ witness $\cConf'\in\denotationOf{\sConf'}$. We show that using $\taskMp^{-1}(\thid_\sConf)$ for $\cConf'\reducesTo{\stmt}{\taskMp^{-1}(\thid_\sConf)}\cConf$ generates the desired configuration $\cConf$.
The intuition is that if \textit{all} tasks in $\thidSet_{\cConf'}$ that are associated to $\thid_\sConf$ by $\taskMp$ execute $\stmt$, we obtain a configuration $\cConf$ that is denoted by some concrete constraint $\sConf_\cnc$, hence, denoted by $\sConf$.

\paragraph{\textbf{newPhaser.}}
This rule is only complete when $\taskMp^{-1}(\thid_\sConf)$ is a singleton. To simplify the presentation, we show completeness when \texttt{main} executes \texttt{newPhaser}, because then we are sure $\taskMp^{-1}(\thid_\sConf)$ is a singleton.   
Assume $\thid_\sConf$ is the task \texttt{main} in some $\sConf$ and $\sConf\preReducesTo{\phvar:=\newp}{\thid_\sConf}\sConf'$. Assume also $\phid_\sConf$ is the new phaser id. 
Let $\phid_\cConf\not\in\phidSet_{\cConf'}$ and  $\phidSet_\cConf=\phidSet_{\cConf'}\cup\set{\phid_\cConf}$.  $\relMp_\cConf$ is obtained from $\relMp_{\cConf'}$ by assigning $\relMp_{\cConf}(\thid_\cConf)(\phid_\cConf)=(\phvar,(0,0))$ where $\thid_\cConf$ is the task \texttt{main} in $\cConf'$. 
Since $(0,0)\models\relgapMp_{\sConf}(\thid_\sConf)(\phid_\sConf)$, $\thid_\cConf$ can again be associated with $\thid_\sConf$ in $\sConf_\cnc$ and $\relMp_{\cConf}(\othid_\cConf)(\phid_\cConf)=(-,\nreg)$ for all other tasks $\othid_\cConf\in\thidSet_\cConf'\setminus\set{\thid_\cConf}$. Moreover, $\relMp_{\sConf}(\othid_\sConf)(\phid_\sConf)=(-,\nreg)$ for all tasks $\othid_\sConf\in\thidSet_\sConf'\setminus\set{\thid_\cConf}$.   
Hence, for $\cConf=\tuple{\thidSet_{\cConf},\phidSet_\cConf,\boolMp_{\cConf}, \pcMp_\cConf, \relMp_\cConf}$, we have $\cConf'\reducesTo{\phvar:=\newp}{\thid_\cConf}\cConf$ and the task and phaser mappings $\taskMp$ and $\phaserMp'=\phaserMp\cup\set{\phid_\cConf\rightarrow\phid_\sConf}$ witness $\cConf\in\denotationOf{\sConf_\cnc}$. 

\paragraph{\textbf{signal}.}
Assume $\sConf\preReducesTo{\sig{\phvar}}{\thid_\sConf}\sConf'$ for some $\sConf$. 
%
By definition of $\cConf'\in\denotationOf{\sConf'}$, there exists $\phid_{\cConf}$ such that $\phaserMp(\phid_{\cConf})=\phid_\sConf$ and all tasks in $\taskMp^{-1}(\thid_\sConf)$ are associated with $\thid_\sConf$ by $\taskMp$.
$\cConf'$ is denoted by $\sConf'$ and is obtained via \textbf{signal~I} or \textbf{signal~II}. There is therefore a level $\lev\ge 0$  which shows that the signal and wait phases of the tasks registered in $\cConf'$ respect the phase bounds in $\sConf'$. We can show that $\lev+1$ captures that the signal and wait phases of the tasks registered in $\cConf$ respect the phase bounds in $\sConf$ if $\sConf'$ is obtained via \textbf{signal~I}, and $\lev$ shows that the signal and wait phases of the tasks registered in $\cConf$ respect the phase bounds in $\sConf$ if $\sConf'$ is obtained via \textbf{signal~II}.

In both cases, the task and phaser mappings $\taskMp$ and $\phaserMp$ should be used for the denotation.

\paragraph{\textbf{wait.}} Assume $\sConf\preReducesTo{\wait{\phvar}}{\thid_\sConf}\sConf'$ for some $\sConf$. 
By definition of $\cConf'\in\denotationOf{\sConf'}$, there exists $\phid_{\cConf}$ such that $\phaserMp(\phid_{\cConf})=\phid_\sConf$ and each task $\thid\in\taskMp^{-1}(\thid_\sConf)$ is associated with $\thid_\sConf$ by $\taskMp$.
%
If $\lev$ witnesses $\cConf'\in\denotationOf{\sConf'}$, then the same level witnesses $\cConf\in\denotationOf{\sConf}$ with the task and phaser mappings $\taskMp$ and $\phaserMp$. 
\paragraph{\textbf{drop.}}
Assume $\sConf\preReducesTo{\dereg{\phvar}}{\thid_\sConf}\sConf'$ for some $\sConf$. 
By definition of $\cConf'\in\denotationOf{\sConf'}$, there exists  $\phid_{\cConf}$ such that $\phaserMp(\phid_{\cConf})=\phid_\sConf$ and each task $\thid\in\taskMp^{-1}(\thid_\sConf)$ is associated with $\thid_\sConf$ by $\taskMp$.
 $\relMp_\cConf$ is obtained from $\relMp_{\cConf'}$ by 1) not modifying the phases of the tasks that are not associated with $\thid_\sConf$, 2) assigning $\relMp_{\cConf}(\thid)(\phid_{\cConf})=(-,\nreg)$ for each $\thid\in\taskMp^{-1}(\thid_\sConf)$. 
A level $\lev\ge 0$ witnesses these tasks respect the phase bounds in $\sConf'$ (which is obtained for a certain $\delta$). We can show $\lev-\delta$ witnesses denotation of $\cConf$ by $\sConf_\cnc$. 
The third group of tasks to consider are those in $\thidSet_\cConf\setminus\taskMp^{-1}(\thid_\sConf)$ which are registered with $\phid_\cConf$. The phases for these tasks are not modified. 
Hence, the task and phaser mappings $\taskMp$ and $\phaserMp$ witness $\cConf\in\denotationOf{\sConf_\cnc}$. 

\paragraph{\textbf{exit.}}
This proof is very similar to the proof of completeness of $\mathtt{drop}$. The only difference is that the proof of $\mathtt{drop}$ needs to be extended by iterating through all phasers dropped in $\cConf'$. The mappings $\taskMp'=\taskMp\setminus\taskMp^{-1}(\thid_\sConf)$ and $\phaserMp$ witness $\cConf\in\denotationOf{\sConf_\cnc}$. 

\paragraph{\textbf{asynch.}}
Assume $\sConf\preReducesTo{\asynch{\task}{\phvar_1,\ldots \phvar_k}\{\seqVal\}}{\thid_\sConf}\sConf'$ for some $\sConf$. 
Let $\othid_\sConf$ be the newly spawned task, and $\phid_{1_\sConf},\ldots \phid_{k_\sConf}$ be the phasers passed to $\othid_\sConf$. 
By definition of $\cConf'\in\denotationOf{\sConf'}$, there exist  $\phid_{1_\cConf},\ldots \phid_{k_\cConf}$ that $\phaserMp(\phid_{i_\cConf})=\phid_{i_\sConf}$ for each $i:1\leq i\leq k$  and  $\taskMp$ associates each task $\thid\in\taskMp^{-1}(\thid_\sConf)$ to $\thid_\sConf$, and each task $\othid_\thid$ spawned by $\thid$ to $\othid_\sConf$. 
We obtain $\relMp_\cConf$ from $\relMp_{\cConf'}$ by copying $\relMp_{\cConf'}(\thid)(\phid_{i_\cConf})$ to $\relMp_{\cConf}(\othid_\thid)(\phid_{i_\cConf})$  for each $\thid\in\taskMp^{-1}(\thid_\sConf)$ and $i:1\leq i\leq k$. 
Therefore, each $\thid\in\taskMp^{-1}(\thid_\sConf)$ and $\othid_\thid$ can respectively be associated to $\thid_\sConf$ and $\othid_\sConf$ in $\sConf$. 
As a result, for $\cConf=\tuple{\thidSet_\cConf,\phidSet_{\cConf'},\boolMp_{\cConf'}, \pcMp_\cConf, \relMp_\cConf}$,  we have $\cConf'\reducesTo{\asynch{\task}{\phvar_1,\ldots \phvar_k}\{\seqVal\}}{\taskMp^{-1}(\thid_\sConf)}\cConf$ and the task and phaser mappings $\taskMp'=\taskMp\cup\set{\thid\rightarrow\thid_\sConf~|~\thid\in\taskMp^{-1}(\thid_\sConf)}\cup\set{\othid_\thid\rightarrow\othid_\cConf~|~\thid\in\taskMp^{-1}(\thid_\sConf)}$ and $\phaserMp$ witness $\cConf\in\denotationOf{\sConf_\cnc}$.

\end{proof}

\subsection{Verification Procedure}
The procedure makes use of a predecessors computation (line 7) that
results, for a constraint $\sConf$ and a statement $\stmt$, in a
finite set
$\mathtt{pre_\stmt}=\setcomp{\sConf'}{\sConf\preReducesTo{}{\stmt}{}\sConf'}$.
We write $\mathtt{pre}$ for the union of $\mathtt{pre_\stmt}$ for all $\stmt$.
These computations are described in Figures \ref{fig:pre:new} and \ref{fig:pre:asynch}.
Intuitively, the program statement for which the predecessors set is
being computed can be executed by a
task captured by the constraint $\sConf=\sConfTuple$ explicitly (i.e., $\thid\in\thidSet$)
or implicitly (i.e., $\othid\not\in\thidSet$ but satisfying the
environment gaps).
%
%
%
%
For all but atomic statements (i.e., $\nextblock{\phvar}{\seqVal}$),
the set $\mathtt{pre_\stmt}=\setcomp{\sConf'}{\sConf\preReducesTo{}{\stmt}{}\sConf'}$
is exact in the sense that 
$\setcomp{\cConf'}{\cConf\in\denotationOf{\sConf} \textrm{ and }
  \cConf'\reducesTo{}{\stmt}{}\cConf} \subseteq
\bigcup_{\sConf'\in \mathtt{pre_\stmt}}\denotationOf{\sConf'}
\subseteq
\setcomp{\cConf'}{\cConf\in\denotationOf{\sConf} \textrm{ and }
  \cConf'\reducesTo{}{\stmt}^+\cConf}$.
Intuitively, the predecessors calculation for
the atomic $\nextblock{\phvar}{\stmt}$ statement is
only an over-approxiamation because such an instruction
can encode a test-and-set operation.
Such an operation can be made to only be carried by exactly one task.
Our representation allows for more tasks (and larger gaps), but the additional tasks
may not be able to carry the atomic operation.
We would therefore obtain a non-exact over-approximation and avoid
this issue by only applying the procedure to non-atomic programs.
In fact, we show in Sect.~\ref{sec:limits} that allowing atomic
instructions results in the undecidability of the problems addressed by
Thm. \ref{thm:decidable:control} and \ref{thm:decidable:plain}.
Using Lem.~\ref{lem:cstr:entailment} and the exactness of
$\mathtt{pre}$ we can show by induction partial correctness.

\begin{procedure}
    \caption{check($\program$,$\sConfSet$), a
      simple working list procedure for checking constraints
      reachability.}
    \label{proc:check}
    
    \SetKwFunction{KwContinue}{continue}

    \KwIn{$\program=\programTuple$ and a $\entailedBy$-minimal target set $\sConfSet$}
    
    \KwOut{A symbolic run to $\sConfSet$ or the value
      \texttt{unreachable}}

    Initialize both $\sConfSetWorking$ and $\sConfSetVisited$ to
    $\setcomp{\tuple{\sConf,\sConf}}{\sConf\in\sConfSet}$\;

    \While{there exists $\tuple{\sConf,trace}\in\sConfSetWorking$}
          {
            remove $\tuple{\sConf,trace}$ from $\sConfSetWorking$\;

            let $\sConfTuple=\sConf$\;

            
            \lIf{$\cConfInit\models\sConf$}{\Return $trace$}
            
            \ForEach{$\thid\in\thidSet\cup\set{\othid}$ where $\othid\not\in\thidSet$}
                    {
                      \ForEach{$\sConf'\textrm{ s.t. }
                        \sConf \preReducesTo{\thid}{\stmt}\sConf'$}
                              {
                                \If{$\osConf\not\entailedBy\sConf'$ for all $\tuple{\osConf,\_}\in\sConfSetVisited$}
                                   {
                                     Remove from $\sConfSetWorking$ and $\sConfSetVisited$ each $\tuple{\osConf,\_}$
                                     s.t.  $\sConf'\entailedBy\osConf$\;
                                     
                                     Add $\tuple{\sConf', \sConf'\cdot\stmt\cdot trace}$ to
                                     both $\sConfSetWorking$ and $\sConfSetVisited$\;
                                   }
                              }
                    }
          }
          \Return {unreachable} \;
\end{procedure}

\begin{lemma}[Soundness]
\label{lem:soundness}
If ``\ref{proc:check}''($\program$,$\sConfSet$)
returns \texttt{unreachable}, then 
$\cConf_{strt}\notmovesto{*}{}\denotationOf{\sConfSetBad}$.
\end{lemma}

\begin{proof}
By soundness of $\preReducesTo{}{}$ (Theorem \ref{th:tacas:sound-pre}), the procedure ``\ref{proc:check}''($\program$,$\sConfSet$) is sound.
\end{proof}

\begin{lemma}[Relative completeness]
\label{lem:completeness}
If the procedure ``\ref{proc:check}''($\program$,$\sConfSet$) returns a trace
$\sConf_0;\stmt_0;\ldots;\sConf_{n-1};\stmt_{n-1},\sConf_n$ in which only the task $\mathtt{main}$ has executed $\mathtt{newPhaser}$,  there are $\cConf_0,\ldots\cConf_n$ with $\cConf_0=\cConf_{strt}$, $\cConf_n\in\denotationOf{\sConfSetBad}$ and $\cConf_{i-1}\reducesTo{\stmt_{i-1}}{}^+\cConf_i$ for $i:1<i\leq n$.
\end{lemma}

\begin{proof} 
By relative completeness of $\preReducesTo{}{}$ (Theorem \ref{th:tacas:complete-pre}), the procedure ``\ref{proc:check}''($\program$,$\sConfSet$) is complete relative to the runs is which only the task \texttt{main} is allowed to execute \texttt{newPhaser}. 
\end{proof}

  We can also show the procedure to terminate if we only manipulate $K$-bounded-dimension
  and $B$-good constraints.
  
\begin{lemma}[Termination]
\label{lem:termination}
\textup{check(}$\program$\textup{,}$\sConfSet$\textup{)}
terminates if there are $K,B\in\nats$  s.t.
all constraints in $\mathtt{Visited}$ are $K$-dimension-bounded and $B$-good.
\end{lemma}
\paragraph{Proof sketch.}
Suppose the procedure does not terminate. 
The infinite sequence of constraints passing the test at line 8 violates
Thm.~\ref{thm:cstrs:wqo}.

\begin{theorem}
\label{thm:decidable:control}
Control reachability for non-atomic phaser programs generating a finite
number of phasers is decidable. 
\end{theorem}
\paragraph{Proof sketch.}
  Systematically drop, in the backward procedure, constraints
  violating $K$-dimension-boundedness (as none of the denoted
  configurations is reachable) ensures $K$-boundedness.
  Also, the set of target constraints is free (since we are checking
  control reachability) and this is preserved by the $\mathtt{pre}$ computation
  in Fig.~\ref{fig:pre:aux},~\ref{fig:pre:new}, and~\ref{fig:pre:asynch}.
  Finally, Lemmas \ref{lem:soundness}, \ref{lem:completeness}, and \ref{lem:termination} ensure soundness, relative completeness, and termination.

\begin{theorem}
\label{thm:decidable:plain}
Plain reachability for non-atomic phaser programs
generating at most $K$ phasers with, for each phaser,
$B$-bounded gaps is decidable.
\end{theorem}
\paragraph{Proof sketch.}
  Systematically drop, in the backward procedure, constraints
  requiring more than $K$ phasers or larger than $B$ gaps-values for some phaser
  gaps (as none of the denoted configurations is reachable) ensures
  $K$-dimension-boundedness and $B$-goodness.
  Finally, Lemmas \ref{lem:soundness}, \ref{lem:completeness}, and \ref{lem:termination} ensure soundness, relative completeness, and termination.

\section{Limitations of deciding reachability}
\label{sec:limits}

Assume a program $\program=\programTuple$ and its initial configuration
$\cConfInit$. 
We show  a number of parameterized reachability
problems to be undecidable.
First, we address checking control reachability when restricting
to configurations with at most $K$ task-referenced phasers.
We call this $K$-control-reachability.

\begin{definition}[$K$-control-reachability]
  Given a partial control configuration $\cConf$, we write
  $\breach{\program}{\cConf}{K}$, and say $\cConf$ is
  $K$-control-reachable, to mean there are $n+1$ configurations
  $(\cConf_i)_{i:0\leq i\leq n}$, each with at most $K$ reachable
  phasers (i.e., phasers referenced by at least a task variable) s.t.
  $\cConfInit =\cConf_0$ and $\cConf_i \movesto{} \cConf_{i+1}$ for
  $i:0 \leq i < n - 1$ with $\cConf_n$ equivalent to a configuration
  that includes $\cConf$.
\end{definition}

\lstset{numbersep=-7pt, basicstyle=\ttfamily\scriptsize}

\begin{figure}
  \begin{center}
    \begin{tabular}{cc}
\begin{minipage}{.45\textwidth}
{
\begin{lstlisting}
   bool s1, s2, ..., sF; 
   bool $\mathtt{xDec}$,$\mathtt{yDec}$;
   main(){
      $\mathtt{xPh}$ = newPhaser();
      $\mathtt{yPh}$ = newPhaser();
      while(true){  
         //(${\color{olive}\mathtt{q_i}}$:inc(x):${\color{olive}\mathtt{q_j}}$)
         if(ndet() $\wedge$ si){
            asynch(xTask,$\mathtt{xPh}$); 
            si$=\false$;
            sj$=\true$;
         }
         //(${\color{olive}\mathtt{q_i}}$:inc(y):${\color{olive}\mathtt{q_j}}$)
         if(ndet() $\wedge$ si){
            asynch(yTask,$\mathtt{yPh}$);
            si$=\false$;
            sj$=\true$;
         }   
         //(${\color{olive}\mathtt{q_i}}$:dec(x):${\color{olive}\mathtt{q_j}}$)
         if(ndet() $\wedge$ si){
            $\mathtt{xDec}=\true$;  
            $\mathtt{xPh}$.next();  
            $\mathtt{xPh}$.next();  
            assert(!$\mathtt{xDec}$);
            si$=\false$;
            sj$=\true$;
         }    
         //(${\color{olive}\mathtt{q_i}}$:dec(y):${\color{olive}\mathtt{q_j}}$)
         if(ndet() $\wedge$ si){
            $\mathtt{yDec}=\true$;  
            $\mathtt{yPh}$.next();
            $\mathtt{xPh}$.next();
            assert($\mathtt{yDex}$);
            si$=\false$;
            sj$=\true$;
         }       
         //(${\color{olive}\mathtt{q_i}}$:test(x):${\color{olive}\mathtt{q_j}}$)      
         if(ndet() $\wedge$ si){
            $\mathtt{xPh}$.signal();  
            $\mathtt{xPh}$.wait(); 
            si$=\false$;
            sj$=\true$;
         }
\end{lstlisting}
}
\end{minipage}
&
\begin{minipage}{.45\textwidth}
{
\begin{lstlisting}[firstnumber=44]        
    		//(${\color{olive}\mathtt{q_i}}$:test(y):${\color{olive}\mathtt{q_j}}$)
         if(ndet() $\wedge$ si){
            $\mathtt{yPh}$.signal(); 
            $\mathtt{yPh}$.wait();
            si$=\false$;
            sj$=\true$;
         }
       }
   }//($\color{olive}\text{end of main}$)  
   
   
   // ****** xTask *******
   xTask($\mathtt{xPh}$){  
      while(true){
         if($\mathtt{xDec}$){ 
            if(ndet()){ 
               $\mathtt{xDec}=\false$;
               p1 = newPhaser(); 
               $\mathtt{xPh}$.next();
               exit;      
            }    
            $\mathtt{xPh}$.next();  
            $\mathtt{xPh}$.next();  
         }
      }
   }  
    
   // ****** yTask *******
   yTask($\mathtt{yPh}$){ 
      while(true){
         if($\mathtt{yDec}$){
            if(ndet()){
               $\mathtt{!yDec}=\false$;
               p1 = newPhaser(); 
               $\mathtt{yPh}$.next();
               exit;      
            }
         
            $\mathtt{yPh}$.next();
            $\mathtt{yPh}$.next();    
         }
      }
   }
  
\end{lstlisting}
}
\end{minipage}
\end{tabular}
\end{center}

  \caption{For the proof of Thm.\ref{thm:control:bounded}. Encoding a Minsky machine
    with the two counters $\set{x,y}$.
    The value of counter $x$ is represented by the number of instances of $\mathtt{xTask}$
    tasks registred to phaser $\mathtt{xPh}$.
    The construction ensures
    that runs trying to decrement by more than 1 will result in configurations with larger number of phasers.
    }
\label{fig:thm:control:bounded1}
\end{figure}

\begin{theorem}
  \label{thm:control:bounded}
$K$-control-reachability is undecidable in general. 
\end{theorem}
\paragraph{Proof sketch.}
  Encode state reachability of an arbitrary Minsky machine with
  counters $x$ and $y$ using $K$-control-reachability of a suitable
  phaser program.
  The program (see \cite{phasers:param:abs-1811-07142}) has five tasks: $\mathtt{main}$,
  $\mathtt{xTask}$, $\mathtt{yTask}$, $\mathtt{child1}$ and
  $\mathtt{child2}$.
  Machine states are captured with shared variables and
  counter values with phasers $\mathtt{xPh}$ for counter $x$
  (resp. $\mathtt{yPh}$ for counter $y$). Then, (1) spawn an instance
  of $\mathtt{xTask}$ (resp. $\mathtt{yTask}$) and register it to
  $\mathtt{xPh}$ (resp. $\mathtt{yPh}$) for increments, and (2) perform a wait on
  $\mathtt{xPh}$ (resp. $\mathtt{yPh}$) to test for zero. 
  Decrementing a counter, say $x$, involves asking
  an $\mathtt{xTask}$, via shared variables, to exit (hence, to
  deregister from $\mathtt{xPh}$).
  However, more than one task might participate in the decrement operation.
  For this reason, each participating task builds a path
  from $\mathtt{xPh}$ to $\mathtt{child2}$ with two phasers.
  If more than one $\mathtt{xTask}$ participates in the decrement, then
  the number of reachable phasers of an intermediary configuration will be at
  least five.
  As a result, the phaser program will reach a configuration
  corresponding to $s_F$ via configurations having at most 4 reachable phasers
  iff the counter machine reaches a configuration with state $s_F$.

\begin{theorem}
  \label{thm:control:atomic:bounded:programs}
 Control reachability of phaser programs generating a finite number of phasers is undecidable if atomic statements are allowed.
\end{theorem}
\paragraph{Proof sketch.}
We encode state reachability problem of an arbitrary Minsky machine
with counters $x$ and $y$ using a phaser program with atomic
statements.
The phaser program (captured in
Fig. \ref{fig:thm:control:atomic:bounded}) has three tasks:
$\mathtt{main}$, $\mathtt{xTask}$ and $\mathtt{yTask}$.
The idea is to associate a phaser $\mathtt{xPh}$ to counter $x$
(resp. $\mathtt{yPh}$ to counter $y$) and to perform a signal followed
by a wait on $\mathtt{xPh}$ (resp. $\mathtt{yPh}$) to test for zero on
counter $x$ (resp. counter $y$).
Incrementing and decrementing is performed by asking spawned tasks
to spawn a new instance (incrementing) or to deregister
(decrementing). Atomic-next statements are used to ensure exactly one
task is spawned or deregistred.
  As a result, the phaser program will reach a configuration sending the variable $\mathtt{s}$
  to $s_F$ 
  iff the counter machine reaches a configuration with state $s_F$.

\lstset{numbersep=-7pt, basicstyle=\ttfamily\scriptsize}

\begin{figure}
  \begin{center}
    \begin{tabular}{ccc}
\begin{minipage}{.45\textwidth}
{
\begin{lstlisting}
   bool xDec,yDec,s1, s2, ...,sF; 
  
   main(){
      xPh = newPhaser();
      yPh = newPhaser();
      while(true){  
         //(${\color{olive}\mathtt{q_i}}$:inc(x):${\color{olive}\mathtt{q_j}}$)
         if(ndet() && si)
         {
            asynch(xTask,$\mathtt{xPh}$);              
            si$=\false$;
            sj$=\true$;
         }
         
         //(${\color{olive}\mathtt{q_i}}$:dec(x):${\color{olive}\mathtt{q_j}}$)
         if(ndet() && si)
         {
            $\mathtt{xDec=\true}$; 
            $\mathtt{xPh}$.next(){   
              $\mathtt{xDec}=\false$;  
              exit;                
            }   
            
            si$=\false$;
            sj$=\true$;
         }  
\end{lstlisting}
}
\end{minipage}
&
\begin{minipage}{.45\textwidth}
{
\begin{lstlisting}[firstnumber=27]    
         //(${\color{olive}\mathtt{q_i}}$:test(x):${\color{olive}\mathtt{q_j}}$)
         if(ndet() && si )
         {
            $\mathtt{xPh}$.next();  
            si$=\false$;
            sj$=\true$;
         }
         //(${\color{olive}\mathtt{q_i}}$:inc(y):${\color{olive}\mathtt{q_j}}$)
         //(${\color{olive}\mathtt{q_i}}$:dec(y):${\color{olive}\mathtt{q_j}}$)
         //(${\color{olive}\mathtt{q_i}}$:test(y):${\color{olive}\mathtt{q_j}}$)
            ...
   }  
    
    
   //**** xTask *****
   xTask($\mathtt{xPh}$){
      while(true){
      
         if($\mathtt{xDec}$){
            $\mathtt{xPh}$.next(){            
              $\mathtt{xDec}=\false$;  
              exit;    
              
           }
        }
   }
\end{lstlisting}
}
\end{minipage}
\end{tabular}
\end{center}

  \caption{In proof of Thm.\ref{thm:control:atomic:bounded:programs} for control reachability of atomic phaser programs. Encoding a Minsky machine with the two $\set{x,y}$.
    The value of counter $x$ is represented by the number of instances of $\mathtt{xTask}$
    tasks registered to phaser $\mathtt{xPh}$.
    The construction ensures increments or decrements involve exactly one task. 
    }
\label{fig:thm:control:atomic:bounded}
\end{figure}

Finally, even with finite numbers of tasks and phasers, but with
arbitrary gap-bounds, we can show \cite{fmcad17} the following.

\begin{theorem}
  \label{thm:class:reachability}
  Plain reachability of of non-atomic phaser programs generating a
  finite number of phasers is undecidable if the generated gaps are not
  bounded.
\end{theorem}

\section{Conclusion}
\label{sec:conc}

We have studied parameterized plain (e.g., deadlocks) and control
(e.g., assertions) reachability problems for phaser programs.
We have proposed an exact verification procedure for non-atomic programs.
The procedure can be used for answering both control and plain
reachability problems.
We summarize our findings in Table~\ref{tab:findings}.
The procedure is guaranteed to terminate, even for programs that may
generate arbitrary many tasks but finitely many phasers, when checking
control reachability or when checking plain reachability with
bounded gaps.
These results were obtained using a non-trivial symbolic
representation for which termination had required showing an
$\eapreceq$ preorder on multisets on gaps on natural numbers to be a \wqo.
We are working on a tool that implements the procedure in order to
verify programs that dynamically spawn tasks and synchronize them with
phasers.
We believe our general decidability results are useful to reason
about synchronization constructs other than phasers.
For instance, a traditional static barrier can be captured with one
phaser and with bounded gaps (in fact one).
Similarly, one phaser with one producer and arbitrary many consumers
can be used to capture futures where a ``get'' instruction can be
modeled with a wait.
We believe our negative results can also be used. For instance, atomic
instructions can be modeled using test-and-set operation and may result in
the undecidability of the reachability problem.
This suggests more general applications of the work are to be investigated.

{\scriptsize
  \begin{table}
    \scriptsize
\centering
  \begin{tabular}{|c|c|c|c|c|}
  \cline{2-5}
  \multicolumn{1}{c}{}&\multicolumn{4}{|c|}{Arbitrary numbers of tasks}\\   \cline{2-5}
  \multicolumn{1}{c}{}& \multicolumn{2}{|c|}{Finite dimension} & $K$-reachability & Arbitrary dimension \\ \hline
  Bounded   & ctrl atomic~\xmark                               &  plain non-atomic~\cmark          &                                    &  \\ 
  gaps      & (Thm.\ref{thm:control:atomic:bounded:programs})  &  (Thm.\ref{thm:decidable:plain})  &   ctrl non-atomic~\xmark           & ctrl non-atomic~\xmark \\ \cline{1-3}
  Arbitrary & ctrl non-atomic~\cmark                           &  plain non-atomic~\xmark          &   (Thm.\ref{thm:control:bounded})  & (Thm.\ref{thm:control:general}) \\ 
  gaps      & (Thm.\ref{thm:decidable:control})                &  (From \cite{fmcad17})            &                                    &   \\ \hline
\end{tabular}
\label{tab:findings}
\caption{Findings summary: $ctrl$ stands for control reachability
  and $plain$ for plain reachability; $atomic$ stands for allowing the
  $\nextblock{\phvar}{\stmt}$ atomic instruction and $non-atomic$ for forbidding
  it (resulting in non-atomic programs). Recall the dimension of a phaser program
  is the number of dynamically generated phasers.
}
\end{table}
}

\bibliographystyle{splncs04}
\bibliography{bibdatabase}




\end{document}